\documentclass[reprint,amssymb,superscriptaddress,nofootinbib]{revtex4-1}

\usepackage{xcolor}

\definecolor{dgreen}{RGB}{0,140,0}

\def\q{\text{q}}

\usepackage{tcolorbox}
\usepackage{amsfonts}
\usepackage{dsfont}
\usepackage{bm}
\usepackage{times}
\usepackage{verbatim}
\usepackage[utf8]{inputenc}
\usepackage[T1]{fontenc}
\usepackage[normalem]{ulem}
\usepackage[bottom]{footmisc}
\usepackage{amsmath}
\usepackage{amssymb}
\usepackage{mathrsfs}
\usepackage{mathtools}
\usepackage{amsthm}
\usepackage{physics}
\usepackage{booktabs}
\usepackage{bbm}

\makeatletter
\def\maketag@@@#1{\hbox{\m@th\normalfont\normalsize#1}} 
\makeatother

\usepackage{graphicx}
\usepackage{float}
\usepackage{subfigure} 
\usepackage{dcolumn}

\usepackage{pgfplots}
\usepgfplotslibrary{groupplots}

\newtheorem{theorem}{Theorem}
\newtheorem{proposition}[theorem]{Proposition}

\newtheorem{lemma}[theorem]{Lemma}
\newtheorem{corollary}[theorem]{Corollary}

\newtheorem*{remark}{Definition}

\DeclarePairedDelimiterXPP{\intercalr}[1]{\mathsf{Tr}}{[}{]}{}{#1}
\DeclarePairedDelimiterXPP{\intercalrAbs}[1]{\mathsf{TrAbs}}{[}{]}{}{#1}
\DeclarePairedDelimiterXPP{\opTr}[1]{\mathrm{Tr}}{[}{]}{}{#1}
\DeclarePairedDelimiterXPP{\bbTr}[1]{\mathbb{T}\mathrm{r}}{[}{]}{}{#1}
\DeclarePairedDelimiterXPP{\opTrAbs}[1]{\mathrm{TrAbs}}{[}{]}{}{#1}

\def \cH{{\cal H}}
\def \Lra{\Leftrightarrow}
\def \Ra{\Rightarrow}
\def \cX{{\cal X}}
\def \cP{{\cal P}}
\def \cI{{\cal I}}

\def\bbP{\mathbb{P}}
\def \cE{{\cal E}}
\def\bbH{\mathbb{H}}
\newcommand{\bbA}{\ensuremath{{\mathbb{B}}}} 
\newcommand{\bbL}{\ensuremath{{\mathbb{L}}}} 
\newcommand{\bbS}{{\mathbb{S}}}
\newcommand{\bbV}{\ensuremath{{\mathbb{V}}}} 
\newcommand{\bbW}{\ensuremath{{\mathbb{W}}}} 
\newcommand{\bbX}{\ensuremath{{\mathbb{X}}}}

\newcommand{\Ree}{\operatorname{Re}}
\newcommand{\Imm}{\operatorname{Im}}
\newcommand{\id}{\mathbb{I}}
\newcommand{\CN}{C_{\mathrm N}}
\newcommand{\CNH}{C_{\mathrm{NH}}}
\newcommand{\CH}{C_{\mathrm H}}

\newcommand{\FNH}{F_{\mathrm{NH}}}
\newcommand{\FH}{F_{\mathrm H}}

\def \lofh{{\cal L}({\cal H})}

\newenvironment{prooftprop4}[1][Proof (Proposition 4)]{\noindent\textbf{#1.} }{\ \rule{0.5em}{0.5em}}

\DeclareMathOperator*{\argmin}{arg\,min}

\def\ANU{Centre for Quantum Computation and Communication Technology, Department of Quantum Science, Australian National University, Canberra, ACT 2601, Australia.}

\def\CQT{Centre for Quantum Technologies (CQT), National University of Singapore, Singapore 117543, Republic of Singapore}

\def\UEC{Graduate School of Informatics and Engineering, The
  University of Electro-Communications, Tokyo 182-8585, Japan}
\begin{document}

\title{The gap persistence theorem for quantum multiparameter estimation}
\author{Lorc\'{a}n O. Conlon}
\email{lorcanconlon@gmail.com}
\affiliation{\ANU}
\affiliation{Quantum Innovation Centre (Q.InC), Agency for Science Technology and Research (A*STAR), 2 Fusionopolis Way, Innovis \#08-03, Singapore 138634, Republic of Singapore}
\affiliation{\CQT}
\author{Jun Suzuki}
\email{junsuzuki@uec.ac.jp}
\affiliation{\UEC}
\author{Ping Koy Lam}
\affiliation{\ANU}
\affiliation{Quantum Innovation Centre (Q.InC), Agency for Science Technology and Research (A*STAR), 2 Fusionopolis Way, Innovis \#08-03, Singapore 138634, Republic of Singapore}
\affiliation{\CQT}
\author{Syed M. Assad}
\email{cqtsma@gmail.com}
\affiliation{\ANU}
\affiliation{Quantum Innovation Centre (Q.InC), Agency for Science Technology and Research (A*STAR), 2 Fusionopolis Way, Innovis \#08-03, Singapore 138634, Republic of Singapore}

\begin{abstract}
One key aspect of quantum metrology, measurement incompatibility, is evident only through the simultaneous estimation of multiple parameters. The symmetric logarithmic derivative Cram{\'{e}}r-Rao bound (SLDCRB), gives the attainable precision, if the optimal measurements for estimating each individual parameter commute with one another. As such, when the optimal measurements do not commute, the SLDCRB is not necessarily attainable. In this regard, the Holevo Cram{\'{e}}r-Rao bound (HCRB) plays a fundamental role, providing the ultimate attainable precisions when one allows simultaneous measurements on infinitely many copies of a quantum state. For practical purposes, the Nagaoka Cram{\'{e}}r-Rao bound (NCRB), or Nagaoka--Hayashi Cram{\'{e}}r-Rao bound (NHCRB), is more relevant, as a lower bound when restricted to measuring quantum states individually. The interplay between these three bounds dictates how rapidly the ultimate metrological precisions can be approached through collective measurements on finite copies of the probe state. In this work we investigate this interplay. For full rank probe states, we prove that if the HCRB cannot be saturated with measurements on a single copy of the probe state, then it cannot be saturated for \textit{any finite} number of copies of the probe state. As an application of this result, we show that it is \textit{impossible} to saturate the HCRB for several physically motivated problems, including the simultaneous estimation of phase and phase diffusion. We provide necessary and sufficient conditions for the attainability of the SLDCRB with separable measurements. We further prove that if the SLDCRB cannot be saturated with measurements on a single copy of a full rank probe state, it cannot be saturated with collective measurements on \textit{any finite} number of copies of the probe state. We demonstrate that these results do not necessarily hold for rank deficient states. Finally, for practical purposes it is more relevant to study the rate at which the NHCRB converges to the HCRB. We prove that given $M$ copies of the probe state, the NHCRB converges to the HCRB at rate of at least $1/\sqrt{M}$. However, we present examples where the convergence can happen arbitrarily fast. We believe that these results comprehensively characterise the role of entangling measurements in quantum multiparameter estimation.
\end{abstract}
\maketitle

\section{Introduction} 

\begin{figure*}[t]
\includegraphics[width=\textwidth]{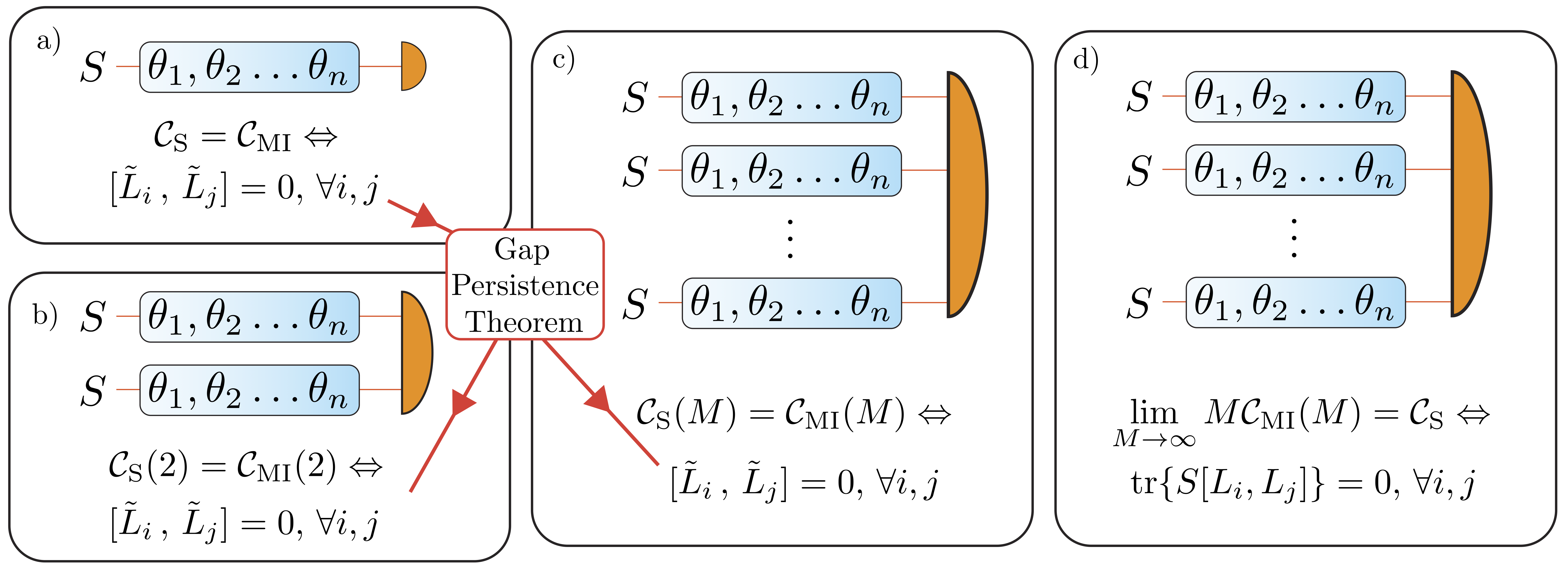}
\caption{\textbf{Attainability of the symmetric logarithmic derivative Cram{\'{e}}r-Rao bound (SLDCRB), $\mathcal{C}_\text{S}$, in quantum multiparameter metrology.} \mbox{a) For} estimating $n$ parameters $(\theta_1\hdots\theta_n)$ with separable measurements on a single copy of the quantum probe state $S$, the single-copy most informative bound, $\mathcal{C}_\text{MI}$, applies. In this scenario, the necessary and sufficient conditions for the attainability of $\mathcal{C}_\text{S}$ are shown in terms of the operators $\tilde{L}_{i}$. The relationship between $\tilde{L}_{i}$ and the corresponding SLD operators $L_{i}$ is provided in the main text (Theorem~\ref{theorem:mainSLD}).
b), c) When performing collective measurements on any finite number of copies, $M$, of $S$, the $M$-copy most informative bound, $\mathcal{C}_\text{MI}(M)$, applies. In this case we prove that the necessary and sufficient conditions for the attainability of the $M$-copy SLDCRB $\mathcal{C}_\text{s}(M)$ are the same as those in the single-copy case. This is done through the gap persistence theorem. A similar result applies for the Holevo Cram{\'{e}}r-Rao bound, $\mathcal{C}_\text{H}$. d) Shown are the necessary and sufficient conditions in terms of the SLD operators $L_{i}$ for saturating $\mathcal{C}_\text{S}$ when allowing for collective measurements on infinitely many copies of the probe state. }
\label{fig:schematic}
\end{figure*}

Quantum metrology is one of the most promising quantum technologies, outperforming classical resources in real world applications~\cite{aasi2013enhanced,casacio2021quantum}. Much of the excitement surrounding quantum metrology stems from the fact that quantum resources can offer improved scaling of the measurement error, with respect to the number of probe states~\cite{leibfried2004toward,higgins2007entanglement,dorner2009optimal,kacprowicz2010experimental,slussarenko2017unconditional,daryanoosh2018experimental,wang2019heisenberg,mccormick2019quantum,pedrozo2020entanglement,marciniak2021optimal,boixo2008quantum,napolitano2011interaction,conlon2026fault} or an energy constraint~\cite{caves1981quantum,jaekel1990quantum,schnabel2010quantum,guo2020distributed}, over classical resources. However, most experimental demonstrations focus on single parameter estimation, whereas the more fundamental quantum nature of metrology is only revealed through multiparameter estimation. The uncertainty principle~\cite{heisenberg1985anschaulichen,robertson1929uncertainty,arthurs1965simultaneous,conlon2026100}, one of the key tenets of quantum mechanics, is only witnessed when multiple non commuting observables are measured simultaneously. Furthermore, multiparameter estimation is well motivated physically~\cite{monras2011measurement,humphreys2013quantum,vidrighin2014joint,yue2014quantum,baumgratz2016quantum,zhang2017quantum,cimini2019quantum,hou2020minimal,yung2024comparison,yung2025saturating} and so, unsurprisingly, has become a prominent research area. For recent reviews on multiparameter estimation see Refs.~\cite{szczykulska2016multi,liu2019quantum,sidhu2020geometric,demkowicz2020multi,conlon2026100}.
%
%
%

To determine limits on how well multiple parameters can be estimated, it is common to turn to variations of quantum Cram{\'{e}}r-Rao bounds~\cite{helstrom1967minimum,helstrom1968minimum,yuen1973,gill2005state,holevo1973statistical,holevo2011probabilistic,nagaoka2005new,nagaoka2005generalization,conlon2021efficient}. In order for a Cram{\'{e}}r-Rao bound to be practically useful, it is necessary to know the measurement which saturates this bound. The symmetric logarithmic derivative Cram{\'{e}}r-Rao bound, $\mathcal{C}_\text{S}$, (SLDCRB)~\cite{helstrom1967minimum,helstrom1968minimum} (note that this is often referred to as the quantum Cram{\'{e}}r-Rao bound or Helstrom bound) is of particular importance, as for single parameter estimation there always exists an optimal measurement saturating this bound~\cite{braunstein1994statistical} with asymptotically many repetitions of the experiment~\cite{barndorff2000fisher,hayashi2005statistical}. Note that when only finite repetitions are allowed, the SLDCRB is not necessarily a tight bound~\cite{genoni2012optical,tsang2012ziv,rubio2019quantum,lu2016quantum}. Nevertheless, in this asymptotic regime, the theoretically optimal measurement is known for single parameter estimation. For estimating multiple parameters however, if the optimal measurements do not commute, they cannot be performed simultaneously. In this case the SLDCRB may not be attainable. Another bound of particular importance, is the Holevo Cram{\'{e}}r-Rao bound, $\mathcal{C}_\text{H}$, (HCRB)~\cite{holevo1973statistical,holevo2011probabilistic}. The reason for its importance is that the HCRB is a tight bound which can be asymptotically approached through collective measurements on infinitely many copies of the quantum state~\cite{kahn2009local, yamagata2013quantum, yang2019attaining}. In a few specific cases, measurements saturating the SLDCRB~\cite{helstrom1968minimum,young1975asymptotically,belavkin2004generalized} and HCRB~\cite{holevo2011probabilistic,matsumoto2002new,bradshaw2017tight,bradshaw2018ultimate} are known. However, general conditions for when the SLDCRB or HCRB can be attained by a measurement capable of generating entanglement across a finite number of copies of the quantum state remain unknown. 

The attainability of the HCRB and SLDCRB are questions of both practical and fundamental significance. From a practical viewpoint, implementing collective measurements on even a finite number of copies of the probe state is difficult~\cite{roccia2017entangling,parniak2018beating,hou2018deterministic,wu2019experimentally,yuan2020direct,conlon2023approaching,conlon2023discriminating,yung2025saturating,yung2026beating}. Additionally, it has recently been shown that both of these bounds can be arbitrarily far from the attainable, finite copy precision~\cite{das2025holevo}. Hence, knowledge of the necessary and sufficient conditions to saturate the HCRB or SLDCRB with collective measurements on a finite number of copies would be very beneficial. From a theoretical viewpoint, the attainability of the SLDCRB is intrinsically related to measurement incompatibility~\cite{lu2021incorporating} and the uncertainty principle~\cite{robertson1929uncertainty,arthurs1965simultaneous,heisenberg1985anschaulichen,ozawa2003universally,ozawa2004uncertainty,branciard2013error,watanabe2011uncertainty}. The necessary and sufficient conditions for saturating the SLDCRB in the infinite-copy limit are known~\cite{ragy2016compatibility}. In spite of some recent progress~\cite{amari2000methods,suzuki2020quantum,pezze2017optimal,yang2019optimal,PhysRevLett.128.250502,PhysRevA.105.062442,conlon2025role}, when the SLDCRB can be saturated with measurements on a finite number of copies of the probe state remains an open problem~\cite{horodecki2022five}.

In this work we introduce the \textit{gap persistence theorem} to solve these problems. Put simply, a gap persistence theorem between two bounds states that the two bounds can be equal for any finite number of copies of the probe state, if and only if they are equal for a single-copy of the probe state. Hence, if the gap persistence theorem can be shown to hold for either the HCRB or SLDCRB and an appropriate separable measurement bound, the question of finite-copy attainability reduces to that of single-copy attainability. Perhaps more importantly, the gap persistence theorem also implies that if the HCRB or SLDCRB cannot be attained with separable measurements on a single-copy of the probe state, then these two bounds can \textit{never} be saturated by any measurement on finite copies of the probe state, i.e. the gap persists.

The gap persistence theorem leads us to the following main results. Note that we shall use the Nagaoka Cram{\'{e}}r-Rao bound, $\mathcal{C}_\text{N}$, (NCRB) and Nagaoka--Hayashi Cram{\'{e}}r-Rao bound, $\mathcal{C}_\text{NH}$, (NHCRB) as a bound on separable measurement precisions when estimating two and more than two parameters respectively.
\begin{enumerate}
    \item We show that the gap persistence theorem holds for the HCRB and the NHCRB for full rank probe states. We also show that the gap persistence theorem holds for the HCRB and the most informative bound, $\mathcal{C}_\text{MI}$, corresponding to the best possible separable measurement for full rank probe states.\footnote{Note that although the NHCRB is a bound on $\mathcal{C}_\text{MI}$, i.e. $\mathcal{C}_\text{MI}\geq\mathcal{C}_\text{NH}$, it is not necessarily a tight bound, i.e. we can find examples where $\mathcal{C}_\text{MI}>\mathcal{C}_\text{NH}$~\cite{hayashi2023tight,conlon2025role}.} Therefore, if the HCRB cannot be reached in the single-copy setting, the HCRB, often referred to as the ultimate attainable limit in quantum metrology, can never be reached for any finite copy, full rank setting. Note that this implies the gap persistence theorem always holds for qubit states.
    \item We prove that in the limit of infinitely many copies of the probe state the NHCRB converges to the HCRB.
    \item We prove that the gap persistence theorem applies for the SLDCRB and the most informative bound for full rank probe states. We also that for general rank-deficient probe states, the gap persistence theorem applies for the SLDCRB and the NHCRB.
    \item We provide necessary and sufficient conditions for saturating the SLDCRB with collective measurements on any finite number of copies of a full rank probe state. These conditions are summarised in Fig.~\ref{fig:schematic}.
    \item We study the rate of convergence of the NHCRB to the HCRB, providing a lower bound on the speed of convergence.
\end{enumerate}

Before proceeding, it will be useful to pause to highlight the importance of these results. Firstly, from a fundamental viewpoint, reducing the question of finite copy attainability of the SLDCRB or HCRB to that of single copy attainability makes the problem much easier to solve. Additionally, from a practical viewpoint, we demonstrate the limitations of the HCRB, namely that no practical measurement can saturate it in many cases. We also note that the gap between the SLDCRB or HCRB and the most informative bound can scale with the dimensions of the system in question, which can be arbitrarily large~\cite{das2025holevo}. The persistence of an arbitrarily large gap is undoubtedly of practical relevance. Finally, the lower bounds we present on the rate of convergence of the NHCRB to the HCRB provides an efficient method for evaluating the benefit of entangling measurements.

The layout of this paper is as follows. We first introduce the notation and all preliminaries necessary in Sec.~\ref{sec:prelim}. We also introduce the gap persistence theorem and present some basic results in this section. In Sec.~\ref{sec:twopar}, we consider two-parameter estimation and show that for full rank probe states the gap persistence theorem holds for the HCRB and the NCRB and also for the HCRB and the most informative bound. Strikingly, we present a rank-deficient probe state where these gap persistence theorems do not hold. We prove that in the infinite-copy limit the NCRB asymptotes to the HCRB. In Sec.~\ref{sec:multipar} we consider multiparameter estimation, proving that for full rank probe states the gap persistence theorem holds for the HCRB and NHCRB, and also that in the infinite-copy limit the NHCRB tends to the HCRB. We also prove that the gap persistence theorem holds for the SLDCRB and the most informative bound for full rank probe states. Intriguingly, we prove that for general rank deficient states a gap persistence theorem holds between the NHCRB and SLDCRB. In Sec.~\ref{sec:convergence}, we study the rate of convergence of the NHCRB to the HCRB, proving a general lower bound on the rate of convergence. In Sec.~\ref{sec:eg} we present several examples that illustrate the importance of our results. We then discuss several important open questions in Sec.~\ref{sec:OP}, before concluding in Sec.~\ref{sec:disc}.

\section{Preliminaries}
\label{sec:prelim}

We consider a density matrix $S_\theta$ in a $d$ dimensional Hilbert space $\mathcal{H}$, which is a function of $n$ parameters $S_\theta(\theta_1,\theta_2,\hdots,\theta_n)$ and is not necessarily full rank. As we are concerned with local estimation we shall drop the dependence on $\theta$ going forward. The parameters could be any physical quantity we wish to learn about, e.g. Gaussian displacements~\cite{bradshaw2018ultimate,bradshaw2017tight,conlon2023verifying} or qubit rotations~\cite{conlon2023multiparameter}. Before proceeding to any technical detail, we clarify some of the terminology we will use throughout this article. Firstly, we consider both separable and collective measurements. In our terminology a separable measurement refers to measuring the probe states $S$ individually, shown in Fig.~\ref{fig:schematic} a) (a precise mathematical definition is presented in the next section). Note that this still allows the use of an ancilla state. Although a separable measurement acts on individual quantum states, it is necessary to consider many repetitions of this process to be in the regime where Cram{\'{e}}r-Rao bounds are applicable~\cite{genoni2012optical,tsang2012ziv,rubio2019quantum,lu2016quantum}. Therefore, when we refer to separable measurements on $S$, we implicitly assume that this measurement is repeated $N\gg1$ times. An $M$-copy collective measurement refers to measuring $M$ copies of $S$, $S^{\otimes M}$, with a possibly entangling measurement, shown in Fig.~\ref{fig:schematic} b)-d). Assuming that we repeat the $M$-copy measurement $N$ times, we use $N\times M$ total copies of $S$. Therefore, an $M$-copy collective measurement uses $M$ times more probe states than a separable measurement and so a direct comparison of the two is not fair. However, this can be remedied by considering rescaled variances, corresponding to the variance per copy of $S$ (more detail is provided in the next section). Secondly, throughout this manuscript we shall often refer to certain Cram{\'{e}}r-Rao bounds as attainable. By this we mean that there exists a physical positive operator valued measure (POVM) which gives the same variance as the bound in question at a known local value of $(\theta_1,\theta_2,...,\theta_n)$. Hence, although the HCRB can be asymptotically approached through collective measurements on infinitely many copies of the quantum state~\cite{kahn2009local, yamagata2013quantum, yang2019attaining}, we will, where appropriate, refer to it as unattainable. This is because, as the gap persistence theorem shows, if the HCRB cannot be saturated in the single-copy setting, there does not exist a POVM which reaches the HCRB with any finite number of copies of the probe state. 

\subsection{Multiparameter estimation preliminary material}
\subsubsection{Local estimation}

In general when estimating $n$ parameters, $(\theta_1,\theta_2,\hdots,\theta_n)$, the aim of multiparameter estimation is to construct an estimate of every $\theta$ value, denoted $\hat{\theta}_i$, with the minimum possible total weighted mean squared error (MSE). The MSE is defined as $v=\text{Tr}[WV(\theta)]$, where $W$ is a $n\times n$ positive definite weight matrix and $V(\theta)$ is the following $n\times n$ matrix:
\begin{equation}
\label{eq:msegeneric}
[V(\theta)]_{i,j}=\mathbb{E}\left[(\hat{\theta}_i-\theta_i)(\hat{\theta}_j-\theta_j)\right]
\end{equation}
(Note that we use $\text{Tr}[A]$ to denote the trace over the classical parameter space $\mathcal{H}_\text{Cl}$, which is an $n$-dimensional real Hilbert space.) Given $N$ copies of the quantum state to measure, $S^{\otimes N}$, we can implement any measurement on $\mathcal{H}^{\otimes N}$. At this stage we do not place any restriction on the type of measurement - it could be separable, adaptive or entangling. To construct the estimates, $\hat{\theta}_i$, one can implement a measurement described by a POVM, which is a set of positive operators $\{\Pi_k\}$ that sum to the identity
\begin{equation}
\label{Eq:BGsumpi}
\sum_{k}\Pi_{k}=\mathbb{I}_{d^N}\;.
\end{equation}
The $k$-th measurement outcome occurs with probability
\mbox{$p_k=\text{tr}\{S^{\otimes N}\Pi_{k}\}$} (We use $\text{tr}\{A\}$ to denote the trace over quantum operators in $\mathcal{H}$). Based on the observed measurement outcome $k$, the experimenter aims to construct unbiased estimators for the parameters of interest as \mbox{$\hat{\theta}_i=\hat{\theta}_i(k)$}. Here $\hat{\theta}_i(k)$ are understood to be pre-determined constants assigned to each measurement outcome (as opposed to $\hat{\theta}_i$ being a random variable, see e.g. Refs.~\cite{hayashi2023tight,hayashi2024finding}). Thus given $N$ copies of $S$ to measure, the estimated values of $\theta$ are determined completely from the measurement result on all $N$ copies of $S$. In this setting, the MSE matrix, $V(\theta)$, in Eq.~\eqref{eq:msegeneric} simplifies to
\begin{equation}
\label{eqMSEmatrix}
[V(\theta)]_{ij}=\sum_{k}(\hat{\theta}_{i}(k)-\theta_{i})(\hat{\theta}_{j}(k)-\theta_{j})p_k\;,
\end{equation}
where the index $k$ is over all possible measurement outcomes for a POVM acting on $S^{\otimes N}$. 

The mean and variance of the estimator is defined as $\langle\hat{\theta}_i\rangle=\sum_k\hat{\theta}_i(k)p_k$ and $\langle\hat{\theta}_i^2\rangle-\langle\hat{\theta}_i\rangle^2=\sum_k\hat{\theta}_i(k)^2p_k-(\sum_k\hat{\theta}_i(k)p_k)^2$ respectively, where the sum is over all possible measurement outcomes $k$. In general, for global estimation, an unbiased estimator must satisfy $\langle\hat{\theta}_i\rangle=\theta_i$ for all $\theta$ and $i$. The most general goal of quantum metrology is to minimise the weighted trace of $V(\theta)$ for a globally unbiased estimator. Accordingly we define the globally unbiased most informative bound as
\begin{equation}
\label{problem1}
\mathcal{C}_\text{MI}^{\text{GUb}}(S^{\otimes N}, W) = \min_{\{\Pi_k\},\hat{\theta}}\text{Tr}[WV(\theta)]\;,
\end{equation}
where the minimisation is over all $\{\Pi_k\},\hat{\theta}$ satisfying the globally unbiased condition. This bound corresponds to the MSE when the optimal POVM and estimator are used, and represents the best precision available on average over all possible
data when using globally unbiased estimators. However, there some problems that arise in this setting. First, for some estimation problems unbiased estimators do not exist. For example, for a fixed POVM, we recover the standard problem in classical statistics where it is well known that unbiased estimators may not exist. Secondly, the optimisation problem is very complex for finite $N$.

To get around these obstacles, we shall relax the requirement of a globally unbiased estimator. In this work we focus on local estimation, where we only require that the estimator’s are unbiased to first order around a known value $\theta_0=(\theta_{0,1},\theta_{0,2},...,\theta_{0,n})$. The local unbiased conditions are that $\langle\hat{\theta}_i\rangle=\theta_{0,i}$ to first order in the Taylor series expansion~\cite{Fujiwara2006Adaptive}:
\begin{equation}
    \begin{split}
&\langle\hat{\theta}_i\rangle=\theta_{0,i}\;,\\
&\frac{\partial }{\partial\theta_j}\langle\hat{\theta}_i\rangle\bigg|_{\theta=\theta_0}=\delta_{ij}\;.
    \end{split}
\end{equation}
In this setting we can define the most informative bound as
\begin{equation}
\mathcal{C}_\text{MI}(S^{\otimes N},W) = \min_{\{\Pi_k\},\hat{\theta}}\text{Tr}[WV(\theta)]\;,
\end{equation}
where the minimisation is over all $\{\Pi_k\},\hat{\theta}$ satisfying the locally unbiased conditions. Note that strictly speaking the minimisation in the above bound should be an infimum~\cite{yamagata2026sufficient}. However, as this does not affect our main results we use minimum throughout. This bound corresponds to the MSE when the optimal locally unbiased POVM and estimator are used. We note that the maximum likelihood estimator, which is asymptotically unbiased, can achieve a MSE that scales as $1/N$, under standard regularity conditions, see e.g. Ref.~\cite{van2000asymptotic}.

It is important to note that in the local estimation setting, the optimal measurement and estimator, $\{\Pi_k^*\},\hat{\theta}^*$, will depend on the unknown parameter $\theta$. 
Nevertheless, $\mathcal{C}_{\text{MI}}(S^{\otimes N})$ presents the best precision available on average over all possible data at each unknown $\theta$. One cannot beat this bound as long as we use a locally unbiased estimator for any finite sample size. In this sense, it sets the fundamental limit within the framework of locally unbiased estimation.


\subsubsection{Most informative bound and entangling measurements}
Thus far, we have considered the most informative bound on $S^{\otimes N}$ as an optimisation problem over $\{\Pi_k\},\hat{\theta}$ acting on $\mathcal{H}^{\otimes N}$. That is the measurements we consider may be separable, adaptive or entangling. We refer to a measurement acting on the quantum states one by one as a separable measurement. That is, a separable measurement is of the form $\Pi_\text{sep}=\{\Pi_{k,1}\}\otimes\{\Pi_{k,2}\}\hdots\otimes\{\Pi_{k,N}\}$, where each set of POVMs $\{\Pi_{k,i}\}$ acts on $\mathcal{H}$. In principle, the POVMs could be adaptive with $\{\Pi_{k,i}\}$ depending on the $i-1$ previous measurement results, see for example Ref.~\cite{li2023optimal}. In these settings, we can denote the overall measurement outcome as $\mathbf{k}=(k_1,k_2|k_1,k_3|k_1k_2,...k_N|k_1...k_{N-1})$, which occurs with probability $p_\mathbf{k}=\text{tr}\{S^{\otimes N}(\Pi_{k_1}\otimes\Pi_{k_2}...\otimes\Pi_{k_N})\}$. Finally an entangling measurement is one in which each $\Pi_k$ acts on $\mathcal{H}^{\otimes N}$.

While it is clear that entangling measurements are more powerful than separable, it is desirable to study the attainable precision under separable or adaptive measurements owing to their experimental attainability. As such we define
\begin{equation}
\mathcal{C}_{\text{MI,separable}}(S^{\otimes N},W) = \min_{\{\Pi_k\},\hat{\theta}}\text{Tr}[W\;V(\theta)]\;,
\end{equation}
where the minimisation is over all $\{\Pi_k\},\hat{\theta}$ satisfying the locally unbiased conditions and also $\{\Pi_k\}$ corresponding to a separable POVM. An important result in this context is the following proposition:
\begin{proposition}
\label{prop1}
For all separable POVMs, we have
\begin{equation}
\mathcal{C}_{\text{MI,separable}}(S^{\otimes N},W) =\frac{1}{N}\mathcal{C}_{\text{MI}}(S,W)
\end{equation}
\end{proposition}
\noindent Note that we do not need to assume any form for estimators, and thus  
they are general, i.e.,
$\hat{\theta}(x_1,x_2,\ldots,x_N)$, where $x_i$ denotes the measurement result from the $i$th copy of $S$. The proof is presented in appendix~\ref{apen:CMI}.

The importance of proposition~\ref{prop1} is that it demonstrates that under separable measurements the most informative bound is additive. This means that when evaluating the most informative bound for separable measurements acting on $S^{\otimes N}$, it is sufficient to analyse only the single-copy setting. The intuitive reason for this result, is that in local estimation because $\theta_0$ is assumed to be known in advance (we are only searching for small changes in this neighbourhood), the locally optimal measurement can be implemented on each copy of $S$. Repeating this POVM independently on each copy is sufficient to achieve the optimal local asymptotic precision, so adaptive POVMs do not improve the leading $1/N$ scaling in this setting. Therefore, for local estimation the optimal separable measurement is of the form $\{\Pi_{k}\}\otimes\{\Pi_{k}\}\hdots\otimes\{\Pi_{k}\}$~\cite{barndorff2000fisher,hayashi2005statistical}. 

While it is not feasible to perform entangling measurements on $S^{\otimes N}$ for $N\to\infty$, it is still feasible to perform entangling measurements on $S^{\otimes M}$ for some finite $M$~\cite{conlon2023approaching,conlon2023discriminating,hou2018deterministic}. In general, the MSE can be reduced by using entangling measurements, whereby $\{\Pi_k\}$ can act on $M$ copies of $S$, $S^{\otimes M}$. Given $S^{\otimes MN}$, we can consider measurements of the form $\Pi_{\text{sep},M}=\{\Pi_{k,1}\}\otimes\{\Pi_{k,2}\}\hdots\otimes\{\Pi_{k,N}\}$, where each set of POVMs $\{\Pi_{k,i}\}$ acts on $\mathcal{H}^{\otimes M}$. We call such a measurement an $M$-copy measurement. In this case, the $M$-copy separable measurement most informative bound would become\footnote{Separable here refers to separable with respect to $N$, not $M$}:
\begin{equation}
    \mathcal{C}_{\text{MI,separable},M}(S^{\otimes MN},W) =\frac{1}{N}\mathcal{C}_{\text{MI}}(S^{\otimes M},W)\;.
\end{equation}
Once again this simply serves to illustrate that when considering entangling measurements that act on $\mathcal{H}^{\otimes M}$ given $S^{\otimes MN}$, it is sufficient to evaluate $\mathcal{C}_{\text{MI}}(S^{\otimes M},W)$.

Given, the above discussion, one may wonder why we bother to introduce $N$ at all, as in the context of separable or $M$-copy entangling measurements, it results in a $1/N$ factor. However, it is important to repeat the experiment asymptotically many times to ensure certain statistical properties~\cite{genoni2012optical,tsang2012ziv,rubio2019quantum,lu2016quantum}. It is worth pausing for a moment to discuss the additivity of the most informative bound. Owing to the power of entangling measurements, in general we will have $\mathcal{C}_{\text{MI}}(S^{\otimes M},W)\leq\mathcal{C}_{\text{MI}}(S,W)/M$~\cite{conlon2021efficient}, i.e. the most informative bound is sub-additive in $M$. The sub-additivity corresponds to the power of entangling measurements, whereas the additivity with respect to $N$ is a classical effect from repeating the same optimal measurement procedure $N$ times (corresponding to $\{\Pi^*\},\hat{\theta}^*$).

Repeating both a separable measurement and an $M$-copy measurement $N$ times requires using $M$ times as many copies of $S$ in the latter setting. In accordance with proposition~\ref{prop1}, if we were to allocate $M$ times as many copies to the separable measurement the MSE would decrease by a factor of $M$, i.e. using $S^{\otimes MN}$ rather than $S^{\otimes N}$, while still restricting to measurements on $\mathcal{H}$. Therefore, going forward we shall multiply variances and Cram{\'{e}}r-Rao bounds associated with an $M$-copy measurement by a factor of $M$ and we shall drop any explicit reference to $N$. This normalises the resources used in different settings.

As such, we can focus from here on, on the quantity $\mathcal{C}_{\text{MI}}(S^{\otimes M},W)$, with $M$ any integer greater than or equal to 1. This optimisation problem is non-convex and is extremely difficult to solve in all but the simplest cases (see Ref.~\cite{yung2025most} for a solution for two parameter estimation with pure states). Efficient methods for finding the optimal POVM are known~\cite{kimizu2024adaptive,zhang2024qestoptpovm}, however, these methods are not guaranteed to converge to the true optimal point. Along this direction, Hayashi and Ouyang have derived an alternative form of $\mathcal{C}_{\text{MI}}(S^{\otimes M},W)$ as~\cite{hayashi2023tight}
\begin{equation}
\mathcal{C}_{\text{MI}}(S^{\otimes M},W)=\mathcal{C}_{\text{conic}}(S^{\otimes M},W)\;,
\end{equation}
where the right hand side is a conic programming optimisation problem under the constraint of a separable cone. While this is mathematically interesting, it is difficult to compute in practice. As such there has been significant interest in introducing efficiently computable bounds on the most informative bound, as introduced in the next subsection.

Going forward we shall use the notation $\mathcal{C}_\text{MI}(S^{\otimes M})=\mathcal{C}_\text{MI}(M)$.

\subsection{Known Cram{\'{e}}r-Rao bounds - two parameter estimation}
We define the Holevo function for two parameters as~\cite{sidhu2021tight}
  \begin{align}
  \label{eq:holfun}
F_\text{H}(\mathbf{X}) \coloneqq \tr{S (X^2+Y^2)} + \abs{\tr{A(\mathbf{X})}}\,,
  \end{align}
  where we introduce the notation $A(\mathbf{X})\coloneqq\mathrm{i}\sqrt{S}[X,Y]\sqrt{S}$ and $\mathbf{X}=(X,Y)$. More generally, the Holevo function can incorporate a positive definite square weight matrix $W$, however for now we shall set this to be the identity matrix. The extension of our results to include a weight matrix will be discussed later in the manuscript. The HCRB for two parameter estimation is given by 
  \begin{align}
  \label{HCRB1}
    \mathcal{C}_\text{H} = \min_\mathbf{X} F_\text{H}(\mathbf{X})\,,    
  \end{align}
  where the optimisation is over $\mathbf{X}=(X,Y)$, where $X$ and $Y$ are Hermitian matrices satisfying the
  locally unbiased conditions:
  \begin{equation}
  \label{equnb}
  \begin{split}
  \tr{SX}=0, &\tr{S_xX}=1\text{ and } \tr{S_yX}=0\\
  \tr{SY}=0, &\tr{S_xY}=0\text{ and } \tr{S_yY}=1\;,
  \end{split}
  \end{equation}
where $S_x=\partial S/\partial\theta_1$ and $S_y=\partial S/\partial\theta_2$. We note that the minimum of Eq.~\eqref{HCRB1} always exists for finite dimensional systems~\cite{suzuki2020quantum}. Let us denote the vector of matrices which optimise the HCRB as 
\begin{equation}
\mathbf{X}_\text{H}=\argmin F_\text{H}(\mathbf{X})\;.
\end{equation}
Note that in some cases there are multiple solutions to the HCRB, and we will address this situation later in the manuscript. The numerical tractability of this minimisation problem has been investigated recently~\cite{albarelli2019evaluating, sidhu2021tight}. 

The HCRB acts as an lower bound on the attainable MSE when we allow collective measurements on infinitely many copies of the quantum state, i.e. \mbox{$\mathcal{C}_\text{H}\leq\lim_{M\to\infty}M\mathcal{C}_\text{MI}(M)$.} As mentioned in the introduction, \mbox{$\mathcal{C}_\text{H}=\lim_{M\to\infty}M\mathcal{C}_\text{MI}(M)$}~\cite{kahn2009local, yamagata2013quantum, yang2019attaining}. Specifically, this means that for sufficiently large $N$ and $M$ there exists a POVM $\Pi_{\text{sep},M}$ such that
\begin{equation}
   \lim_{N\to\inf} N \text{Tr}[WV(\Pi_{\text{sep},M})]=\mathcal{C}_\text{MI}(M)= \frac{1}{M} \mathcal{C}_\text{H} +o(M^{-1})
\end{equation}
For any finite $M$, we have  \mbox{$\mathcal{C}_\text{H}\leq M\mathcal{C}_\text{MI}(M)$.}

Next we define the Nagaoka function for two parameters as~\cite{nagaoka2005new,nagaoka2005generalization}
  \begin{align}
  \label{eq:nagfun}
   F_\text{N}(\mathbf{X}) \coloneqq \tr{S (X^2+Y^2)} +   \text{tr}\{\abs{A(\mathbf{X})}\}\,,
  \end{align}
where $\abs{A}=\sqrt{A^\dagger A}$ such that $\text{tr}\{\abs{A}\}$ is the sum of the absolute values of the eigenvalues of $A$. The NCRB for two parameter estimation is
  \begin{align}
  \label{eq:nagbound}
    \mathcal{C}_\text{N} = \min_\mathbf{X} F_\text{N}(\mathbf{X})\,,    
  \end{align}
  where the optimisation is over $\mathbf{X}=(X,Y)$ satisfying the same
  unbiased conditions as for the HCRB, Eq.~\eqref{equnb}. As the absolute value in the second term of the Nagaoka function is inside the summation, $F_\text{H}(\mathbf{X}) \leq  F_\text{N}(\mathbf{X})$ for all $\mathbf{X}$ with equality if and only if all non-zero eigenvalues of $A(\mathbf{X})$ are either all positive or all negative. Similarly to the HCRB, we shall define the matrices which optimise the NCRB as 
 \begin{equation}
 \label{eq:Nagoptimiser}
\mathbf{X}_\text{N}=\argmin F_\text{N}(\mathbf{X})\;.
\end{equation}
 We stress that the NCRB is a mathematical bound, not related to any estimator. It satisfies $\mathcal{C}_\text{N}\leq\mathcal{C}_\text{MI}$. From proposition~\ref{prop1}, we therefore know that $\mathcal{C}_\text{N}\leq N\mathcal{C}_{\text{MI,separable}}(N)$. This can be interpreted as the statement that given $S^{\otimes N}$ and restricted to separable measurements the NCRB gives a bound on the attainable precision. This is the sense in which the NCRB is a bound on the precision attainable with separable measurements.

The NCRB can be computed based on $S^{\otimes M}$, and the $M$-copy derivatives, denoted as $\mathcal{C}_\text{N} (M)$. As mentioned in the introduction, the NCRB satisfies $\mathcal{C}_\text{N} (M)\leq\mathcal{C}_\text{MI}(M)\qquad\forall M$. Similarly, this implies $\mathcal{C}_\text{N}(M)\leq N\mathcal{C}_{\text{MI,separable}}(N\times M)$.

\subsection{Known Cram{\'{e}}r-Rao bounds - multiparameter estimation}
We now generalise the above bounds to the multiparameter setting. The Holevo function can be written as
\begin{align}
\label{eq:Hfn}
F_\text{H}[\mathbf{X}|W]= \text{Tr}[ W\Re Z[\mathbf{X}]] +\text{Tr} [|\sqrt{W}\Im Z[\mathbf{X}]\sqrt{W}|]\;,
\end{align}
where
\begin{align}
\label{eq_zmat}
Z[\mathbf{X}]_{jk} \coloneqq  \text{tr}\{ S X_j X_k\}\;,
\end{align}
 As above, $\mathbf{X}=(X_1,X_2,\hdots X_n)$ is a vector of Hermitian matrices $X_i$. The HCRB can then be written as
 \begin{align}
  \label{HCRB2}
    \mathcal{C}_\text{H}[W] = \min_\mathbf{X} F_\text{H}[\mathbf{X}|W]\,,    
  \end{align}
  where the optimisation is over $\mathbf{X}$ matrices that are subject to the locally unbiased conditions: $\tr{SX_i}=0$ and $\tr{S_iX_j}=\delta_{i,j}$, where $S_i=\partial S/\partial\theta_i$. 
  

We next define the Nagaoka--Hayashi (NH) function for a fixed $\mathbf{X}$ as~\cite{conlon2021efficient}
\begin{align}
\begin{split}
\label{eq:NHfunctionmain}
F_\text{NH}[\mathbf{X}\,|W]&:= \min_{\bbL}\{\mathbb{T}\mathrm{r}\{\bbW\bbS \bbL\}\,|\, \bbL\ge \mathbf{X}\mathbf{X}^{\intercal},\\
&\hspace{0.4cm}\qquad\qquad\,\bbL_{ij}=\bbL_{ji}\mbox{: Hermitian} \}\\
\end{split}
\end{align}
where $\bbS=I\otimes S$, and $\bbW=W\otimes I$ is a weight matrix, both acting on the extended space, $\mathcal{H}_{\rm{Cl}} \otimes \mathcal{H}$. $\mathbb{T}\mathrm{r}\{\}$ denotes trace over both classical and quantum systems. $\bbL$ is a block matrix with the individual blocks denoted $\bbL_{ji}$. The NHCRB is then expressed as 
\begin{equation}
\label{eq:NHCRBmain}
\mathcal{C}_\text{NH}[W]=\min_{\mathbf{X}}\{F_\text{NH}[\mathbf{X}|W] \,|\, X_i:\,\mathrm{l.u.}\text{ } \mathrm{and}\text{ } \mathrm{Hermitian} \}\;.
\end{equation}
As we show in Appendix~\ref{apen:NHSLD}, the NH function can be split into two parts as 
\begin{equation}
\label{eq:NHCRBdecomp}
F_\text{NH}[\mathbf{X}|W]= \text{Tr}[W\Re Z[\mathbf{X}]] + A^{\text{NH}}_{-}[\mathbf{X}|W].
\end{equation}
The second term of the NH function $A^{\text{NH}}_{-}[\mathbf{X}|W]$ is given by
\begin{equation}
\label{NHfunsecondterm}
\begin{split}
A^{\text{NH}}_{-}[\mathbf{X}|W]=\min_{\bbV}\{&\mathbb{T}\mathrm{r}\{\bbV\}\,|\, \bbV\ge 0, \\
&\,a(\bbV)+(\bbW\bbS)^{1/2}a(\mathbf{X}\mathbf{X}^{\intercal})(\bbW\bbS)^{1/2}
=0 \}\;,
\end{split}
\end{equation} 
where $a(\bbX)$ represents the antisymmetrized matrix with respect to the first Hilbert space. 
Explicitly, for $\bbX=[\bbX_{ij}]$ with $\bbX_{ij}\in\lofh$, $a(\bbX)=\frac12[\bbX_{ij}-\bbX_{ji}]$. 

 Let $(\bbL_{*},\mathbf{X}_{*})$ be an optimiser for the NHCRB. Note that when $\mathcal{C}_{\text{MI}}=\mathcal{C}_{\text{NH}}$, we have
\begin{equation}
\bbL_{*,ij}=\sum_{k}(\hat{\theta}_{i}(k)-\theta_i)(\hat{\theta}_{j}(k)-\theta_j)\Pi_k\;,
\end{equation}
and
\begin{equation}
\mathbf{X}_{*,i}=\sum_{k}(\hat{\theta}_{i}(k)-\theta_i)\Pi_k\;.
\end{equation}
This follows directly from Ref.~\cite{conlon2021efficient}. In appendix~\ref{apen:McopyNHCRB} we show $\mathcal{C}_\text{NH}(M)\le \mathcal{C}_\text{MI}(M)$ (see also Ref.~\cite{conlon2021efficient}). From proposition~\ref{prop1}, we therefore have $\mathcal{C}_\text{NH}(M)\le N \mathcal{C}_\text{MI,separable}(N\times M)$, i.e. the NHCRB is a bound on the attainable precision using $M$-copy measurements on $S^{\otimes MN}$.


The final function we introduce is the symmetric logarithmic derivative (SLD) function for estimating $n$ parameters, defined as
\begin{equation}
\label{equation:slddefinition}
F_\text{S}[\mathbf{X}|W]=\text{Tr}[W\text{Re}Z[X]]
\end{equation}
The SLDCRB can then be expressed as~\cite{nagaoka2005new}
\begin{equation}
\label{eq:SLDCRB}
 \mathcal{C}_\text{S}[W]=\min_\mathbf{X} F_\text{S}[\mathbf{X}|W]\,,   
\end{equation}
where the minimisation is over $\mathbf{X}=(X_1,X_2\hdots X_n)$, where the $X_i$ are Hermitian matrices satisfying the unbiased conditions, $\tr{SX_i}=0$ and $\tr{S_iX_j}=\delta_{i,j}$. The matrices which optimise the SLDCRB are
 \begin{equation}
 \label{eq:SLDoptimisers}
\mathbf{X}_\text{S}=\argmin F_\text{S}[\mathbf{X}]\;.
\end{equation}
$\mathbf{X}_\text{S}$ is a vector of matrices $X_{\text{S},i}$, which are commonly denoted $L^i$. The SLD operators, $L_{i}$, satisfy $S_i=(SL_i+L_iS)/2$ and the two matrices are related through $L^{i}:=\sum_{j} ({J_{\text{S}}}^{-1})_{ji}L_{j}$ where $J_{\text{S}}$ is the SLD Fisher information matrix, $J_{\text{S},ij}=\text{tr}\{S(L_iL_j+L_jL_i)/2\}$. Note that the assumption that the partial derivatives $S_i$ are linearly independent ensures $J_{\text{S}}$ is invertible. For consistency with the HCRB and NCRB, we shall primarily use the notation $X_{\text{S},i}$ instead of $L^i$.

When the weight matrix is set to the identity, we simply denote the above quantities without $W$. 
For example, the NH function is $F_\text{NH}[\mathbf{X}]$.

Throughout this manuscript we will mainly evaluate the two parameter HCRB and NCRB for $M$ copies of the quantum state, $S^{\otimes M}$, denoted as $\mathcal{C}_\text{H}(M)$ and $\mathcal{C}_\text{N}(M)$ respectively. When the argument $M$ is dropped we refer to the single-copy bound. The corresponding optimal matrices shall be denoted $\mathbf{X}_{\text{H},M}=(X_{\text{H},M},Y_{\text{H},M})$ and $\mathbf{X}_{\text{N},M}=(X_{\text{N},M},Y_{\text{N},M})$. Similarly, for estimating any number of parameters, the $M$-copy SLDCRB, HCRB, and NHCRB shall be denoted $\mathcal{C}_\text{x}(M)$, and the corresponding optimal matrices are $\mathbf{X}_{\text{x},M}=(X_{\text{x},i,M})$ with x as S, H and NH respectively. We denote the most informative bound as $\mathcal{C}_\text{MI}$, which is the minimum variance attainable with separable measurements. When considering $M$ copies of $S$, $\mathcal{C}_\text{MI}(M)$ denotes the most informative bound optimised over all possible collective measurements on $M$ copies of $S$. 
The relation $F_\text{S}[\mathbf{X}]\leq F_\text{H}[\mathbf{X}]\leq F_\text{N}[\mathbf{X}]$ gives rise to the following orderings
\begin{equation}
\mathcal{C}_\text{S}\leq\mathcal{C}_\text{H}\leq\mathcal{C}_\text{N}\leq\mathcal{C}_\text{MI}\;,
\end{equation}
and
\begin{equation}
\mathcal{C}_\text{S}\leq\mathcal{C}_\text{H}\leq\mathcal{C}_\text{NH}\leq\mathcal{C}_\text{MI}\;.
\end{equation}
We note that these inequalities hold in the $M$-copy setting (see appendix~\ref{apen:McopyNHCRB} for a detailed discussion on $\mathcal{C}_\text{NH}(M)$). In order to provide greater physical intuition for what these different Cramér-Rao bounds mean, we consider some sample cases. When $\mathcal{C}_\text{S}=\mathcal{C}_\text{H}=\mathcal{C}_\text{NH}=\mathcal{C}_\text{MI}$, then all the bounds introduced above are tight bounds, and there exists a measurement saturating all the bounds. When $\mathcal{C}_\text{S}\leq\mathcal{C}_\text{H}<\mathcal{C}_\text{NH}=\mathcal{C}_\text{MI}$, only the NHCRB is tight. This is the case for several physically motivated examples~\cite{conlon2023approaching,conlon2021efficient,conlon2023multiparameter}. However, there do exist examples where $\mathcal{C}_\text{NH}<\mathcal{C}_\text{MI}$\cite{hayashi2023tight,conlon2025role}, in which case the NHCRB is a closer bound to the most informative bound, but it is not a tight bound.

\subsection{Gap persistence property}
We next introduce the concept of gap persistence between two bounds. For two bounds $\mathcal{C}_1$ and $\mathcal{C}_2$ satisfying $\mathcal{C}_1\geq\mathcal{C}_2$, we define the difference between them as
\begin{equation}
\label{eq:deltaGPT}
\Delta_{1-2}(M)\coloneqq \mathcal{C}_1(M)-\mathcal{C}_2(M)\;.
\end{equation}
For $M=1$, we shall simply denote this quantity as $\Delta_{1-2}$. We first define weak gap persistence
\begin{remark}[Weak gap persistence]
The two bounds, $\mathcal{C}_1$ and $\mathcal{C}_2$ are said to exhibit weak gap persistence if the following property holds.
For all models, if $\Delta_{1-2}\neq0$, then $\Delta_{1-2}(M)\neq0$ for all finite $M$.
\end{remark}
\noindent This means that if there is a gap between the two bounds for a single copy of the probe state, the two bounds will never become equal by performing collective measurements on $M$ copies of the state. However, weak gap persistence does not require $\Delta_{1-2}=0\Leftrightarrow\Delta_{1-2}(M)=0$. A stronger version is
\begin{remark}[Strong gap persistence]
The two bounds, $\mathcal{C}_1$ and $\mathcal{C}_2$ are said to exhibit strong gap persistence if the following property holds.
For all models, if $\Delta_{1-2}=0$, then $\Delta_{1-2}(M)=0$ for all finite $M$ and if $\Delta_{1-2}\neq0$, then $\Delta_{1-2}(M)\neq0$ for all finite $M$.
\end{remark}
\noindent Hence, strong gap persistence includes the weak version. As a simple example, strong gap persistence holds between the HCRB and SLDCRB, due to the additivity of the two bounds. When gap persistence can be shown to hold between two bounds, we shall refer to this as a gap persistence theorem. Hence, there exists a strong gap persistence theorem between the HCRB and SLDCRB. Note that although Eq.~\eqref{eq:deltaGPT} refers to a specific model, the gap persistence theorem applies for all models. Gap persistence theorems can be extended to include weight matrices, as the Cram{\'{e}}r-Rao bounds in general include weight matrices. In what follows, we shall mainly discuss the case of the identity weight matrix.

\subsection{Basic results}
We now present some basic results which will aid the proof of the main results. We first exclude the locally quasi-classical model, where $A(\mathbf{X}_{\text{S}})=\mathrm{i}\sqrt{S}[X_{\text{S}},Y_{\text{S}}]\sqrt{S}=0$. We shall address this situation when discussing multiparameter estimation in Sec.~\ref{sec:multipar}.

The first basic result concerns estimation with $M$ copies of the quantum state, $S^{\otimes M}$. For two parameter estimation, if
  $\mathbf{X}=(X_{1},Y_{1})$ satisfy the locally unbiased conditions, then
  we can construct locally unbiased estimator operators in the $M$-copy setting $\mathbf{X}_{M}=(X_{M},Y_{M})$. The required estimator operators are
  \begin{equation}
  \label{eq:twocopyest}
  \begin{split}
   X_{M}&=\sum_{k=1}^M\frac{1}{M}X^{(k)}\;,\\    
     Y_{M}&=\sum_{k=1}^M\frac{1}{M}Y^{(k)}\,,    
  \end{split}
  \end{equation}
  where $X^{(k)}=\mathbb{I}\otimes\mathbb{I}\otimes\ldots\otimes X_{1}\otimes\ldots\otimes\mathbb{I}$, where $X_{1}$ is in the $k$th position and $\mathbb{I}$ is the identity matrix. This gives $F_\text{H}(\mathbf{X}_{M})=F_\text{H}(\mathbf{X})/M$, verified by direct computation in Appendix~\ref{apen:hol2copy}. The following lemma proves the additivity of the HCRB.
    \begin{lemma} [Lemma 4 in Ref.~\cite{hayashi2008asymptotic}]
 \label{lemma:mcopyhol} 
  $\mathcal{C}_\text{H}(M)=\mathcal{C}_\text{H}/M$ and an optimiser for $\mathcal{C}_\text{H}(M)$ is $\mathbf{X}_{\text{H},M}=\sum_{k=1}^M\frac{1}{M}\mathbf{X}_\text{H}^{(k)}$. 
\end{lemma}
\noindent The extension to the $n$ parameter case with
  $\mathbf{X}=(X_{1},X_{2}...X_n)$ follows in a similar way with the unbiased estimator operators in the $M$-copy setting denoted as $\mathbf{X}_{M}$.

Lemma~\ref{lemma:mcopyhol} does not necessarily apply to the NCRB. In general, $\mathcal{C}_\text{N}(M)\leq \mathcal{C}_\text{N}/M$ (see Appendix~\ref{apen:subadditivity}). By direct calculation, we can show that the SLDCRB is also additive. Hence, we get the following lemma
\begin{lemma}
 \label{lemma:mcopysld} 
  $\mathcal{C}_\text{S}(M)=\mathcal{C}_\text{S}/M$ and the optimiser for $\mathcal{C}_\text{S}(M)$ is $\mathbf{X}_{\text{S},M}=\sum_{k=1}^M\frac{1}{M}\mathbf{X}_\text{S}^{(k)}$. 
\end{lemma}
  
 Our next basic result concerns additive bounds.
 \begin{proposition}
 \label{prop:additivebounds}
 For an additive bound $\mathcal{C}$, $\mathcal{C}_\text{MI}=\mathcal{C}$ implies $\mathcal{C}_\text{MI}(M)=\mathcal{C}(M)=\mathcal{C}/M$.
 \end{proposition} 
  \begin{prooftprop4}
By definition, the most informative bound implies subadditivity $\mathcal{C}_\text{MI}(M)\leq\mathcal{C}_\text{MI}/M$. The assumption of additivity of $\mathcal{C}$ implies $\mathcal{C}_\text{MI}(M)\leq\mathcal{C}_\text{MI}/M=\mathcal{C}/M=\mathcal{C}(M)$. However, $\mathcal{C}_\text{MI}(M)\geq\mathcal{C}(M)$ must hold, hence $\mathcal{C}_\text{MI}(M)=\mathcal{C}(M)$.
  \end{prooftprop4}
 
  Note that this proposition also holds between the NCRB (or NHCRB) and the HCRB or SLDCRB, due to the subadditivity of the NCRB (NHCRB), i.e. \mbox{$\mathcal{C}_\text{N}=\mathcal{C}_\text{H}\implies\mathcal{C}_\text{N}(M)=\mathcal{C}_\text{H}(M)$.}
    
For our next basic result we decompose $A(\mathbf{X})$ into positive and negative parts as \mbox{$A(\mathbf{X})=A_+(\mathbf{X})-A_-(\mathbf{X})$}, where $A_\pm(\mathbf{X})=(\abs{A}\pm A)/2\geq0$. To make this decomposition unique, $A_\pm(\mathbf{X})$ are defined as the positive and negative parts of $A(\mathbf{X})$ when written in its eigenbasis respectively. As mentioned in the preliminaries, $F_\text{H}(\mathbf{X}) \leq  F_\text{N}(\mathbf{X})$ for all $\mathbf{X}$. With this decomposition of $A(\mathbf{X})$, we can now give the necessary and sufficient condition for equality between $F_\text{H}(\mathbf{X})$ and $F_\text{N}(\mathbf{X})$.
\begin{lemma} 
\label{lemm:equality_cond}
$F_\text{H}(\mathbf{X})=F_\text{N}(\mathbf{X})$ iff $\text{tr}\{A_+(\mathbf{X})\}=0$ or $\text{tr}\{A_-(\mathbf{X})\}=0$. Furthermore, for $S>0$, this is equivalent to $\mathrm{i}[X,Y]\geq0$ or $\mathrm{i}[X,Y]\leq0$.
\end{lemma}
\noindent Note that the condition $\text{tr}\{A_+(\mathbf{X})\}=0$ or $\text{tr}\{A_-(\mathbf{X})\}=0$ is equivalent to the statement that 
$A(\mathbf{X})=\mathrm{i}\sqrt{S}[X,Y]\sqrt{S}$ is a semidefinite matrix, i.e., $A(\mathbf{X})\geq0$ or $A(\mathbf{X})\leq0$. 

  
We now consider the case when there are multiple solutions to the HCRB and NCRB/NHCRB. (See Appendix~\ref{subsec:MultipleSols} for the importance of considering multiple solutions.) Let us denote the set of optimisers for the HCRB and NCRB/NHCRB as $\mathcal{X}_\text{H}$ and $\mathcal{X}_\text{N}$/$\mathcal{X}_\text{NH}$ respectively. 
    \begin{lemma}
  \label{lemm:ch=cnManySol} 
The necessary and sufficient condition for $\mathcal{C}_\text{H}=\mathcal{C}_\text{N}$ is the existence of an estimator operator $\mathbf{X}_*$ in $\mathcal{X}_\text{H}$ such that $F_\text{H}(\mathbf{X}_*)=F_\text{N}(\mathbf{X}_*)$.
  \end{lemma}
  \begin{proof}[\textbf{Proof (Lemma~\ref{lemm:ch=cnManySol})}]
  The necessary part follows from two equality conditions for the following inequalities. 
  For any $\mathbf{X}_{\text{N}}\in \mathcal{X}_\text{N}$, which is an optimiser for the Nagaoka function, we have 
  \begin{equation}
  \mathcal{C}_\text{N}= F_\text{N}(\mathbf{X}_{\text{N}})\geq F_\text{H}(\mathbf{X}_{\text{N}})\geq 
  F_\text{H}(\mathbf{X}_{\text{H}})=\mathcal{C}_\text{H}. 
  \end{equation}
  To have $\mathcal{C}_\text{H}=\mathcal{C}_\text{N}$, the first inequality requires 
  $F_\text{N}(\mathbf{X}_{\text{N}})= F_\text{H}(\mathbf{X}_{\text{N}})$. 
  Additionally, $\mathbf{X}_{\text{N}}$ must be in the set $\mathcal{X}_\text{H}$ from the second inequality. 
  The existence of such optimiser immediately proves the sufficient part.
  \end{proof}
  
  Having the necessary and sufficient condition, it is also useful to express it as follows.
  \begin{equation}
  \label{cond:finite_gap}
  \mathcal{C}_\text{H}\neq\mathcal{C}_\text{N}\ \iff\ \forall \mathbf{X}\in\mathcal{X}_\text{H},F_\text{N}(\mathbf{X})>F_\text{H}(\mathbf{X}).
  \end{equation}
This result also applies when replacing the NCRB with the NHCRB.

  
  
Our next basic result concerns the convexity of the HCRB. In Appendix~\ref{apen:strictconvex} we introduce an equivalence relation for linear operators, which allows us to define a quotient space $\mathcal{L}^{\sim_S}(\mathcal{H})$ of $\mathcal{L}(\mathcal{H})$, the set of all matrices. 
  \begin{proposition}
  \label{prop:quotient}
  Within the quotient space $\mathcal{L}^{\sim_S}(\mathcal{H})$ the Holevo function $F_\text{H}$ is convex. For $S\succ0$, $F_\text{H}$ is strictly convex and the solution is unique. \end{proposition}
  This is proven in appendix~\ref{apen:strictconvex}. Interestingly, as the Holevo function is not strictly convex when $S$ is rank deficient we are able to find examples where the solution to the HCRB is not unique, see Appendix~\ref{subsec:MultipleSols}. The following lemma also follows.
  \begin{lemma}
  \label{prop:quotientSLD}
  Within the quotient space $\mathcal{L}^{\sim_S}(\mathcal{H})$ the SLD function $F_\text{S}$ is strictly convex and has a unique solution.
  \end{lemma}
  
As we shall see below, the existence of a unique solution is central to whether or not a gap persistence theorem holds. We present two more basic results that are only applicable when there is a unique solution to the HCRB up to the equivalence relation, Eq.~\eqref{eqequiv}. 
  \begin{lemma}
  \label{lemm:ch=cn} 
 If there is a unique solution to the HCRB and $\mathcal{C}_\text{H}=\mathcal{C}_\text{N}$, then
  $\mathbf{X}_\text{H}=\mathbf{X}_\text{N}$ up to the equivalence relation, Eq.~\eqref{eqequiv}.
  \end{lemma}
  \begin{proof}[\textbf{Proof (Lemma~\ref{lemm:ch=cn})}]
  $\mathcal{C}_\text{H}=\mathcal{C}_\text{N}$ leads to $F_\text{H}(\mathbf{X}_\text{N})\geq F_\text{H}(\mathbf{X}_\text{H})=F_\text{N}(\mathbf{X_\text{N}})\geq F_\text{H}(\mathbf{X}_\text{N})$. The first inequality follows from the definition of $\mathbf{X}_\text{H}$, the equality follows from $\mathcal{C}_\text{H}=\mathcal{C}_\text{N}$ and the last inequality holds as the Nagaoka function cannot be smaller than the Holevo function. Hence, $F_\text{H}(\mathbf{X}_\text{H})=F_\text{H}(\mathbf{X}_\text{N})$. As we assume there is a unique solution to the HCRB, $\mathbf{X}_\text{H}=\mathbf{X_\text{N}}$ follows.
  \end{proof}
  
   By this lemma, we know that $\mathcal{C}_\text{H}=\mathcal{C}_\text{N}$ implies $F_\text{H}(\mathbf{X}_\text{H})=F_\text{N}(\mathbf{X}_\text{H})$. Taking the contraposition of this gives the following corollary
   \begin{corollary}
     \label{cor:gapHN} 
   If there is a unique solution to the HCRB and $F_\text{H}(\mathbf{X}_\text{H})\neq F_\text{N}(\mathbf{X}_\text{H})$, then $\mathcal{C}_\text{N}>\mathcal{C}_\text{H}$, i.e. there is a gap between the NCRB and the HCRB.
   \end{corollary}
Using the form of the $M$-copy estimator operator, Eq.~\eqref{eq:twocopyest}, and the above two results it is easy to show that $\tr{A_+(\mathbf{X}_\text{H})}\neq0$ implies $\tr{A_+(\mathbf{X}_{\text{H},M})}\neq0$, and similarly for $A_-$. We note again that from proposition~\ref{prop:quotient}, there is a unique solution to the HCRB for full rank probe states.

For convenience, we provide a table summarising our notation (Table~\ref{tab:notation}) and also our main results (Table~\ref{tab:results}).

\begin{table*}[t]
\centering
\caption{Notation and definitions.}
\label{tab:notation}
\begin{tabular}{@{}ll@{}}
\toprule
\textbf{Notation} & \textbf{Definition} \\
\midrule
$S$ & Density Matrix \\
$d$ & Dimensions of $S$ \\
$\mathcal{H}$ & $d$-dimensional Hilbert space that $S$ lives on \\
$\mathcal{H}_\text{Cl}$ & $n$-dimensional real Hilbert space \\
$\mathcal{H}_\text{Cl}\otimes \mathcal{H}$ & Joint classical-quantum Hilbert space\\
$\text{Tr}[A]$ & Trace over $\mathcal{H}_\text{Cl}$\\
$\text{tr}\{A\}$ & Trace over $\mathcal{H}$\\
$\mathbb{T}\mathrm{r}[A]$ & Trace over $\mathcal{H}_\text{Cl}\otimes \mathcal{H}$\\
$n$ & Number of parameters \\
$W$ & Weight matrix\\
$\theta_i$ & Unknown parameters \\
$V$ & Covariance matrix \\
$\mathcal{C}_\text{MI}$ & Most informative Cramer-Rao bound \\
$F_\text{H}(\mathbf{X})$ & Holevo function (two parameter, Eq.~\eqref{eq:holfun}\\
$F_\text{H}[\mathbf{X}|W]$ & Holevo function ($n$ parameter, Eq.~\eqref{eq:Hfn}\\
$\mathcal{C}_\text{H}$ & Holevo Cramer-Rao bound (Eqs.~\eqref{HCRB1} and \eqref{HCRB2}) \\
$\mathbf{X}_\text{H}$&Optimiser for the HCRB\\
$\mathbf{X}_{\text{H},M}$ & $M$-copy optimiser for the HCRB\\
$F_\text{N}(\mathbf{X})$ & Nagaoka function (two parameter, Eq.~\eqref{eq:nagfun}\\
$\mathcal{C}_\text{N}$ & Nagaoka Cramer-Rao bound (Eq.~\eqref{eq:nagbound}) \\
$\mathbf{X}_\text{N}$ & Optimiser for the NCRB (Eq.~\eqref{eq:Nagoptimiser})\\
$F_\text{NH}[\mathbf{X}|W]$ & Nagaoka--Hayashi function ($n$ parameter, Eq.~\eqref{eq:NHfunctionmain}\\
$A_-^\text{NH}[\mathbf{X}|W]$ & Second term in Nagaoka--Hayashi function (Eq.~\eqref{NHfunsecondterm}\\
$\mathcal{C}_\text{NH}$ & Nagaoka--Hayashi  Cramer-Rao bound (Eq.~\eqref{eq:NHCRBmain}) \\
$\mathbf{X}_\text{NH}$ & Optimiser for the NHCRB \\
$F_\text{S}[\mathbf{X}|W]$ & Symmetric logarithmic derivative function ($n$ parameter, Eq.~\eqref{equation:slddefinition}\\
$\mathcal{C}_\text{S}$ & Symmetric logarithmic derivative Cramer-Rao bound (Eq.~\eqref{eq:SLDCRB}) \\
$\mathbf{X}_\text{S}$ & Optimiser for the SLDCRB (Eq.~\eqref{eq:SLDoptimisers}) \\
$\mathbb{S}=I\otimes S$ & Extended density matrix operator. Acts on $\mathcal{H}_\text{Cl}\otimes \mathcal{H}$
\\
$\mathbb{W}=W\otimes I$ & Extended weight matrix. Acts on $\mathcal{H}_\text{Cl}\otimes \mathcal{H}$\\
$\Delta_{1-2}(M)=\mathcal{C}_1(M)-\mathcal{C}_2(M)$ & Difference between two $M$-copy Cramer-Rao bounds\\
$A(\mathbf{X})=\mathrm{i}\sqrt{S}[X,Y]\sqrt{S}$ &Two parameter commutator function\\
$\bbA(\mathbf{X})$ &$n$ parameter commutator function satisfying $\bbA(\mathbf{X})_{ij}=\mathrm{i}\sqrt{S}[X_i,X_j]\sqrt{S}$ \\
$\mathcal{L}^{\sim S}$ &Quotient space
\end{tabular}
\end{table*}

\section{Two parameter estimation}
\label{sec:twopar}
Before presenting results for general multiparameter estimation, we first present some results for two-parameter estimation to provide the reader some intuition going forward. We first consider full rank density matrices, where the result is almost trivial, before moving onto the more interesting case of rank deficient density matrices. 
\subsection{Full rank density matrix}
For full rank density matrices, $S>0$, we have the following proposition. 
\begin{proposition}
\label{prop:fullrank}
For $S>0$ in any finite dimensional system, if the optimisers for the HCRB do not commute, \mbox{$[X_H,Y_H]\neq0$}, a finite gap exists between the NCRB and HCRB, i.e. $\mathcal{C}_\text{N}(M)>\mathcal{C}_\text{H}(M)$ for all finite $M$.
\end{proposition}
\begin{proof}[\textbf{Proof (Proposition~\ref{prop:fullrank})}]
{\color{black} $[X_H,Y_H]\neq0$ implies $\mathcal{C}_\text{N}>\mathcal{C}_\text{H}$ from Lemma~\ref{lemm:equality_cond} and Proposition~\ref{prop:quotient}. For $M$ copies, we know
$\mathcal{C}_{\rm N}(M)
=
F_{\rm N}(\mathbf X_{{\rm N},M})
\geq
F_{\rm H}(\mathbf X_{{\rm N},M})
\geq
F_{\rm H}(\mathbf X_{{\rm H},M})
=
\mathcal{C}_{\rm H}(M).$ We now prove that at least one of the two inequalities is strict. If both were
equalities, then $\mathbf X_{{\rm N},M}$ would also be an optimiser
of the $M$-copy HCRB, and such an optimiser would commute ($[X_{\rm N},Y_{\rm N}]=0$.). Since $S^{\otimes M}>0$, the HCRB optimiser
is unique, so $\mathbf X_{{\rm N},M}=\mathbf X_{{\rm H},M}$. The unique $M$-copy HCRB optimisers are then given by Eq.~\eqref{eq:twocopyest}. Taking the partial trace
over copies $2,\ldots,M$ in $[X_{{\rm H},M},Y_{{\rm H},M}]$ gives an expression proportional to $[X_{\rm H},Y_{\rm H}]$. As $[X_{\rm H},Y_{\rm H}]\neq0$, the $M$-copy Holevo optimiser is noncommuting, which contradicts our starting assumption. Therefore, at least one of the inequalities is strict and $\mathcal{C}_{\rm N}(M)>\mathcal{C}_{\rm H}(M).$}
\end{proof}

We note that Proposition~\ref{prop:fullrank} does not hold for infinite dimensional systems. This is because $\mathrm{i}[X,Y]$ can have a definite sign in infinite dimensional systems. For example, when $X$ and $Y$ are canonical conjugate operators, we have \mbox{$\mathrm{i}[X,Y]\propto\mathbb{I}$}. Hence, when estimating Gaussian displacements, the HCRB can be attained by a separable measurement~\cite{holevo2011probabilistic,bradshaw2017tight,bradshaw2018ultimate}. Proposition~\ref{prop:fullrank} then implies that there exists a gap persistence theorem between the NCRB and HCRB for full rank density matrices. 

We now extend this to a gap persistence theorem between the most informative bound and the HCRB with the following theorem
\begin{proposition}
\label{prop:fullrankMI}
For $S>0$ in any finite dimensional system, if the optimisers for the HCRB do not commute, \mbox{$[X_H,Y_H]\neq0$}, a finite gap exists between the most informative bound and HCRB, i.e. $\mathcal{C}_\text{MI}(M)>\mathcal{C}_\text{H}(M)$ for all finite $M$.
\end{proposition}

{\color{black}
\begin{proof}
By Proposition~\ref{prop:fullrank}, $\mathcal{C}_{\rm N}(M)>\mathcal{C}_{\rm H}(M)$
for every finite $M$. As
$\mathcal{C}_{\rm MI}(M)\geq \mathcal{C}_{\rm N}(M)$, it follows immediately that $\mathcal{C}_{\rm MI}(M)>\mathcal{C}_{\rm H}(M)$.
\end{proof}
}

This implies that the gap persistence theorem always holds for qubit states. If the qubit state is pure then $\mathcal{C}_{\rm H}=C_{\rm MI}$ for any number of copies. On the other hand, if the qubit state is mixed then it is full rank.

\subsection{Rank deficient density matrix}
We now consider the case when $S$ is not necessarily full rank.
\begin{theorem}
\label{theorem:main}
A strong gap persistence theorem does not hold between the NCRB and HCRB. Specifically, we can find an example, such that $\Delta_{\text{N}-\text{H}}\neq0$, but \mbox{$\Delta_{\text{N}-\text{H}}(2)=0$}.
\end{theorem}
The specific counter-example to the gap persistence theorem is presented in appendix~\ref{apen:SGPmiHCRB}. Building on this counter-example, we are able to prove that a strong gap persistence theorem does not hold between the most informative bound and the HCRB. This is stated as the following theorem
\begin{theorem}
\label{theorem:mainHCRBMI}
A strong gap persistence theorem does not hold between the most informative bound and HCRB. Specifically, we can find an example, such that $\Delta_{\text{MI}-\text{H}}\neq0$, but \mbox{$\Delta_{\text{MI}-\text{H}}(2)=0$}.
\end{theorem}
This example is presented in appendix~\ref{apen2parMIHCRB}.

Finally, it is worth noting that these results hold, even for non-identity weight matrices. The Holevo and Nagaoka functions, as defined in Eqs.~\eqref{eq:holfun} and \eqref{eq:nagfun} respectively, do not include a positive definite weight matrix $W$. However, a positive definite weight matrix can be accounted for by a reparameterisation of the model (see \mbox{Sec. V} of Ref.~\cite{fujiwara1999estimation}). All of our results then hold for the reparameterised model with the identity matrix as the weight matrix.

\subsection{Infinitely many copies of the probe state}
\label{subsec:inf1}
Regardless of the rank of $S$, we expect $\displaystyle\lim_{M\to\infty}M\mathcal{C}_\text{N}(M)=\mathcal{C}_\text{H}$. This statement is contained in the following theorem. 
\begin{theorem}
\label{theorem:infinite}
For estimating two parameters with any density matrix $S$, we have $\displaystyle\lim_{M\to\infty}M\mathcal{C}_\text{N}(M)=\mathcal{C}_\text{H}$.
\end{theorem}
We provide a sketch of the proof here and the complete proof is presented in Appendix~\ref{apen:asymptotic}. Using a lemma from Ref.~\cite{hayashi1999}, we can show the equivalence between $MF_\text{N}(\mathbf{X}_M)$ and $F_\text{H}(\mathbf{X})$ in the limit $M\to\infty$. This allows us to show that $\lim\limits_{M\to\infty}M\mathcal{C}_\text{N}(M)\leq\mathcal{C}_\text{H}$, but as $M\mathcal{C}_\text{N}(M)\geq M\mathcal{C}_\text{H}(M)=\mathcal{C}_\text{H}$, Theorem~\ref{theorem:infinite} is proven.

\section{Multiparameter estimation}
\label{sec:multipar}
We now turn to estimating more than two parameters. In this scenario, as the NCRB no longer applies, we shall use the NHCRB, $\mathcal{C}_\text{NH}$, to bound the precision attainable with separable measurements~\cite{conlon2021efficient} (as before, separable with respect to $N$, not $M$). We first present gap persistence theorems between 1) the NHCRB and the SLDCRB, and 2) the most informative bound and the SLDCRB. We then prove a gap persistence theorem holds between the NHCRB and the HCRB, as well as proving the asymptotic equivalence of the NHCRB and the HCRB.

\subsection{Gap persistence theorems for the SLDCRB}

We first examine the attainability of the SLDCRB in the single-copy setting, which has recently been recognised as one of five open problems in quantum information theory~\cite{horodecki2022five}. Partial solutions to this problem have been given by \citet{pezze2017optimal}, \citet{yang2019optimal} and~\citet{PhysRevLett.128.250502,PhysRevA.105.062442} in the view of quantum metrology. We note that necessary and sufficient conditions for the full rank case were first stated by Amari and Nagaoka, see Sec. 7.4 of Ref.~\cite{amari2000methods}. Furthermore, Suzuki, Yang and Hayashi (see Appendix B1 of Ref.~\cite{suzuki2020quantum}) gave the necessary and sufficient condition for the more general rank-deficient case. In this sense, the open problem seems to be solved as far as the single-copy setting is concerned. However, a more ambitious goal is to provide constructive conditions for the attainability of the SLDCRB. This problem remains open~\cite{conlon2025role,imai2026hierarchy}.

Additionally, this open problem may be extended beyond the single-copy scenario to any finite-copy setting. Using the gap persistence theorem, we provide the necessary and sufficient conditions for the attainability of the SLDCRB with collective measurements on any finite number of copies of the probe state. To do this, we first present a simple and alternative proof for the necessary and sufficient conditions in the single-copy setting, stated as the following theorem:
\begin{theorem}
\label{theorem:mainSLD}
For any $n$-parameter model $S$, which is not necessarily full rank, the following conditions are equivalent. 
\begin{itemize}
\item[i)] There exists a POVM and a locally unbiased estimator whose MSE matrix is equal to the inverse of the SLD Fisher information matrix.
\item[ii)] $\mathcal{C}_{\text{MI}}[W]=\mathcal{C}_{\text{NH}}[W]=\mathcal{C}_{\text{S}}[W],\,\forall W>0$. 
\item[iii)] $\mathcal{C}_{\text{MI}}[\mathbb{I}]=\mathcal{C}_{\text{NH}}[\mathbb{I}]=\mathcal{C}_{\text{S}}[\mathbb{I}]$. 
\item[iv)] There exists a Hilbert space $\cH_{0}$, a state on it, $S_{0}$, and a unitary on $\cH\otimes \cH_{0}$ 
such that the SLD operators $\tilde{L}_{i}$ for the extended model $U(S\otimes S_{0})U^{\dagger}$ commute with each other, i.e.,  
$[\tilde{L}_{i}\,,\,\tilde{L}_{j}]=0,\,\forall i,j$. 
\end{itemize}
\end{theorem}

Note that in condition ii), we have introduced an explicit dependence on a positive weight matrix. When condition iv) is satisfied, we shall call the model locally quasi-classical. 

We provide a sketch of the proof here and the complete proof is given in Appendix~\ref{apen:compproof:sld}. We will prove the theorem through the chain: i) $\Ra$ ii) $\Ra$ iii) $\Ra$ iv) $\Ra$ i). 
The statements i) $\Ra$ ii) and ii) $\Ra$ iii) hold by definition. Thus, we need to prove iii) $\Ra$ iv) and iv) $\Ra$ i). By equating $\mathcal{C}_{\text{NH}}[\mathbb{I}]$ and $\mathcal{C}_{\text{S}}[\mathbb{I}]$, we show that iii) implies
\begin{equation}
\text{tr}\bigg\{\abs{\sqrt{S}[L_{i}\,,\,L_{j}]\sqrt{S}}\bigg\}=0,\,\forall i,j\;.
\end{equation}
Note that this is equivalent to the partial commutivity condition, identified in Ref.~\cite{yang2019optimal}. When $S$ is full rank, this immediately proves $[L_{i}\,,\,L_{j}]=0,\,\forall i,j$. Thus, we do not need to consider any extension of the Hilbert space. When $S$ is not full rank, iii) implies that the optimisers for the NHCRB can be written as $X_{\text{NH},i}=\sum_{k}(\hat{\theta}_{i}(k)-\theta_{i})\Pi_{k}$, where $\Pi=\{\Pi_{k}\}$ and $\hat{\theta}$ are an optimal POVM and estimator saturating $\mathcal{C}_{\text{MI}}[\mathbb{I}]$. This implies that the SLD operators $L_{i}$ can be written in terms of the POVM elements $\Pi_{k}$. The Naimark extension then implies iv)~\cite{neumark1943spectral}. To show iv) $\Ra$ i), we use the fact that the SLD operators $\tilde{L}_{i}$ for the extended model can be simultaneously diagonalised to find an optimal measurement and estimator function. We then compute the MSE matrix for this measurement and show that it is equal to the inverse of the SLD Fisher information matrix. 

The following corollary presents conditions for when the NHCRB and SLDCRB are equal
\begin{corollary}
\label{theorem:mainSLDNH}
For any $n$-parameter model $S$, which is not necessarily full rank, $\mathcal{C}_{\text{NH}}[W]=\mathcal{C}_{\text{S}}[W]$ if and only if $\text{tr}\big\{\abs{\sqrt{S}[L_{i}\,,\,L_{j}]\sqrt{S}}\big\}=0,\,\forall i,j$.
\end{corollary}
\begin{proof}[\textbf{Proof (Corollary~\ref{theorem:mainSLDNH})}]
    From the uniqueness of the SLD optimiser on the quotient space (lemma~\ref{prop:quotientSLD}), the decomposition $F_\text{NH}[X]=F_\text{S}[X]+A_-^\text{NH}[X]$ (Eq.~\eqref{eq:NHCRBdecomp}), and the fact that $A_-^\text{NH}[X]\geq0$, $\mathcal{C}_{\text{NH}}[W]=\mathcal{C}_{\text{S}}[W]$ implies $A_-^\text{NH}[X_\text{S}]=0$ for equality. From the definition of $A_-^\text{NH}[X]=0$ (Eq.~\eqref{NHfunsecondterm}), this is only possible when $\text{tr}\big\{\abs{\sqrt{S}[X_{\text{S},i}\,,\,X_{\text{S},j}]\sqrt{S}}\big\}=\text{tr}\big\{\abs{\sqrt{S}[L^{i}\,,\,L^{j}]\sqrt{S}}\big\}=0,\,\forall i,j$. As $L^{i}:=\sum_{j} ({J_{\text{S}}}^{-1})_{ji}L_{j}$, and $J_\text{S}\succ0$, this in turn implies $\text{tr}\big\{\abs{\sqrt{S}[L_{i}\,,\,L_{j}]\sqrt{S}}\big\}=0,\,\forall i,j$.

    The converse follows as if $\text{tr}\big\{\abs{\sqrt{S}[L_{i}\,,\,L_{j}]\sqrt{S}}\big\}=0,\,\forall i,j$, then $A_-^\text{NH}[X_\text{S}]=0$, and therefore $F_\text{N}[\mathbf{X}_\text{S}]=F_\text{S}[\mathbf{X}_\text{S}]=\mathcal{C}_\text{S}$
\end{proof}



Although the problem presented in Ref.~\cite{horodecki2022five} was only concerned with the attainability of the SLDCRB when measuring single copies of the quantum state, more generally one will be interested in simultaneously measuring a finite number of copies of the quantum state. To that end, we provide the following theorems
\begin{theorem}
\label{thermSGSLDNH}
A strong gap persistence theorem holds between the SLDCRB and NHCRB.  If $\Delta_{\text{NH}-\text{S}}=0$, then $\Delta_{\text{NH}-\text{S}}(M)=0$ for all finite $M$ and if $\Delta_{\text{NH}-\text{S}}\neq0$, then $\Delta_{\text{NH}-\text{S}}(M)\neq0$ for all finite $M$.
\end{theorem}
\begin{proof}[\textbf{Proof (Theorem~\ref{thermSGSLDNH})}]
 From Corollary~\ref{theorem:mainSLDNH},  $\mathcal{C}_\text{NH}>\mathcal{C}_\text{S}$ implies \mbox{$ \text{tr}\{\abs{\sqrt{S}[X_{\text{S},i}\,,\,X_{\text{S},j}]\sqrt{S}}\}\neq0$} for at least one $i,j$. Denoting $X_{\text{S},i,2}$ as the two-copy SLD optimiser, one can define a matrix
\begin{equation}
\begin{split}
\sqrt{S^{\otimes2}}[X_{\text{S},i,2}\,,\,X_{\text{S},j,2}]\sqrt{S^{\otimes2}}
\end{split}
\end{equation}
If the above matrix is zero, we can take the partial trace over the second-copy to obtain
\begin{equation}
\label{eq:random}
   \frac14\left( \sqrt{S}[X_{\text{S},i}\,,\,X_{\text{S},j}]\sqrt{S}+\text{tr}[\sqrt{S}[X_{\text{S},i}\,,\,X_{\text{S},j}]\sqrt{S}]S\right)=0\;.
\end{equation}
Taking the trace of this equation gives $2\text{tr}[\sqrt{S}[X_{\text{S},i}\,,\,X_{\text{S},j}]\sqrt{S}]=0$, which would in turn imply $\sqrt{S}[X_{\text{S},i}\,,\,X_{\text{S},j}]\sqrt{S}=0$ from Eq.~\eqref{eq:random}. Therefore, if $\sqrt{S}[X_{\text{S},i}\,,\,X_{\text{S},j}]\sqrt{S}\neq0$, it follows that $\sqrt{S^{\otimes2}}[X_{\text{S},i,2}\,,\,X_{\text{S},j,2}]\sqrt{S^{\otimes2}}\neq0$. Therefore, $\mathcal{C}_\text{NH}(2)>\mathcal{C}_\text{S}(2)$. This can be extended to any number of copies in a similar manner. 

This proves weak gap persistence and strong gap persistence follows from Proposition~\ref{prop:additivebounds}.


\end{proof}

Using this we can prove the following theorem, regarding the strong gap persistence theorem between the SLDCRB and the most informative bound

\begin{theorem}
\label{thm:SLD-MI-full-rank-GPT}
For $S\succ0$ a strong gap persistence theorem holds between the SLDCRB and most informative bound.  If $\Delta_{\text{MI}-\text{S}}=0$, then $\Delta_{\text{MI}-\text{S}}(M)=0$ for all finite $M$ and if $\Delta_{\text{MI}-\text{S}}\neq0$, then $\Delta_{\text{MI}-\text{S}}(M)\neq0$ for all finite $M$.
\end{theorem}
This is proven in appendix~\ref{apen:proofMISLD}.

\begin{theorem}
\label{theorem:mainSLDCRBMI}
For general rank deficient states a strong gap persistence theorem does not hold between the most informative bound and the SLDCRB. Specifically, we can find an example, such that $\Delta_{\text{MI}-\text{SLD}}\neq0$, but \mbox{$\Delta_{\text{MI}-\text{SLD}}(2)=0$}.
\end{theorem}
In appendix~\ref{apen:proofMISLD2} we give such an example.

The example in appendix~\ref{apen:proofMISLD2} also gives the following corollary
\begin{corollary}
\label{corol:MINHCRB}
    For general rank deficient states a strong gap persistence theorem does not hold between the most informative bound and the NHCRB. Specifically, we can find an example, such that $\Delta_{\text{MI}-\text{NH}}\neq0$, but \mbox{$\Delta_{\text{MI}-\text{NH}}(2)=0$}.
\end{corollary}
\begin{proof}[\textbf{Proof (Corollary.~\ref{corol:MINHCRB})}]
   The example in appendix~\ref{apen:proofMISLD2} has $\mathcal{C}_\text{MI}>\mathcal{C}_\text{NH}=\mathcal{C}_\text{S}$ with one-copy. With two copies $\mathcal{C}_\text{MI}(2)=\mathcal{C}_\text{NH}(2)=\mathcal{C}_\text{S}(2)$.
\end{proof}

\noindent


\subsection{Gap persistence theorems for the HCRB}
We now turn to the gap persistence theorem for the multi-parameter HCRB. It should first be noted that if $\text{tr}\{S[L^i,L^j]\}=0$ for all $i,j$, then the results in this section immediately hold for the HCRB~\cite{ragy2016compatibility}. In order to prove a gap persistence theorem for the HCRB and NHCRB we require several additional results. We state the main results here, with their proofs deferred to appendices. 
\begin{theorem}\label{thm:lowerB}
  For any $\mathbf{X}$, the second term of the NH function is bounded below by
  \begin{equation}
  A^{\text{NH}}_{-}[\mathbf{X}]\ge \frac{1}{2} \sum_\cP\text{tr}\left\{|\sqrt{S}[X_j,X_k]\sqrt{S}|\right\},
  \end{equation}
  where $\cP$ is arbitrary pairwise partition of $\cI$, an index set $\cI=\{1,2,\ldots,n\}$. 
  This gives a lower bound for the NH bound as
  \begin{equation}
    \mathcal{C}_\text{NH} \ge \min_\mathbf{X}\left\{ \text{Tr}[\Re Z[\mathbf{X}]]+\frac{1}{2}\sum_\cP\text{tr}\left\{|\sqrt{S}[X_j,X_k]\sqrt{S}|\right\} \right\}. 
  \end{equation}  
\end{theorem}
The proof is given in appendix~\ref{apen:proofNHlb}, along with the extension to include a weight matrix. It is important to note that theorem~\ref{thm:lowerB} holds for arbitrary $\cP$.


To proceed we require the following transformation. Given a real antisymmetric matrix $A$ ($A^\intercal=-A$), there exists a real orthogonal matrix $U$ such that $\tilde{A}=UAU^\intercal$ is in normal form: 
  \begin{equation}
  \label{eq:aorthogform}
    \tilde{A}_{jk}=
    \begin{cases}
    \lambda_\ell\ge0&\mbox{ if }j,k\in\cI_\ell\mbox{ and }j<k\\
    -\lambda_\ell\le0&\mbox{ if }j,k\in\cI_\ell\mbox{ and }j>k\\
    0&\mbox{ otherwise}
    \end{cases},
  \end{equation}
  with $\cI_\ell$ defined in appendix~\ref{apen:proofNHlb}. Let $U$ be such an orthogonal matrix for $\Im Z[\mathbf{X}]$. The consider the following transformation:
  \begin{equation}\label{def:xtilde}
    \tilde{X_j}:=\sum_{k=1}^n U_{jk}X_k.
  \end{equation} 
  Denote the vector of matrices under this transformation as $\tilde{\mathbf{X}}$. Then, under this transformation, we have
  \begin{equation}
  \label{eq:pinching}
      F_\text{NH}(\tilde{\mathbf{X}})\geq \text{Tr}[\Re Z[\mathbf{\tilde{X}}]]+\frac{1}{2}\sum_\cP\text{tr}\left\{|\sqrt{S}[\tilde{X}_j,\tilde{X}_k]\sqrt{S}|\right\}\geq F_\text{H}(\tilde{\mathbf{X}})\;.
  \end{equation}

The next lemma uses this transformation to characterize the relationship between the NH and Holevo functions.  
\begin{lemma}\label{lemm:equality_cond2}
For $S>0$, the necessary and sufficient conditions for equality between the Holevo and Nagaoka--Hayashi functions are
\begin{equation}
\FNH(X)=\FH(X)
\quad\Longleftrightarrow\quad
[X_j,X_k]=0\qquad\forall j,k .
\label{eq:functioneq}
\end{equation}
\end{lemma}
The full proof of lemma~\ref{lemm:equality_cond2} is presented in appendix~\ref{apensubsectionanotherproof}.

For full rank $S$, we prove the uniqueness of the HCRB optimiser in proposition~\ref{prop:quotient}. We now use this fact to prove a gap persistence theorem between the HCRB and NHCRB for full rank $S$.
\begin{theorem}
\label{theorem:main2}
For $S>0$, a strong gap persistence theorem holds between the NHCRB and HCRB. If $\Delta_{\text{NH}-\text{H}}=0$, then \mbox{$\Delta_{\text{NH}-\text{H}}(M)=0$} for all finite $M$ and if $\Delta_{\text{NH}-\text{H}}\neq0$, then $\Delta_{\text{NH}-\text{H}}(M)\neq0$ for all finite $M$.
\end{theorem}
\begin{proof}[\textbf{Proof (Theorem~\ref{theorem:main2})}]
(Weak gap persistence theorem) The proof proceeds in a similar manner to the two-parameter case. We prove that $\mathcal{C}_\text{NH}>\mathcal{C}_\text{H}$ for $M=1$ implies $\mathcal{C}_\text{NH}(M)>\mathcal{C}_\text{H}(M)$ for $M\geq2$.
Suppose $\mathcal{C}_\text{NH}>\mathcal{C}_\text{H}$ for $M=1$, Lemma~\ref{lemm:equality_cond2} implies $[X_j,X_k]\neq0$ for at least one $j,k$ pair. As $S>0$ proposition~\ref{prop:quotient} implies that the solution to the HCRB is unique. Denoting $X_{j,M},X_{k,M}$ as the unique $M$-copy extensions of the single-copy estimators satisfying $[X_j,X_k]\neq0$ we find
\begin{equation}
\begin{split}
A(\mathbf{X}_{\text{H},M})
= &\mathrm{i}\sqrt{S^{\otimes M} }[X_{j,M},X_{k,M}]\sqrt{S^{\otimes M}}\\
= &\frac{1}{M^2}\mathrm{i}\sqrt{S}^{\otimes M}\sum_{m,m''}[X_j^{(m)},X_k^{(m')}]\sqrt{S}^{\otimes M}\\
= &\frac{1}{M^2}\mathrm{i}\sqrt{S}^{\otimes M}\sum_{m}[X_j^{(m)},X_k^{(m)}]\sqrt{S}^{\otimes M}\\
= &\frac{1}{M^2}\sum_m \sqrt{S}^{\otimes M} (\mathrm{i}[X_j,X_k])^{(m)}\sqrt{S}^{\otimes M}\neq0\;.
\end{split}
\end{equation}
Therefore, the unique $M$-copy optimisers also satisfy $[X_{j,M},X_{k,M}]\neq0$ for at least one $j,k$ pair. Lemma~\ref{lemm:equality_cond2} then implies that $\mathcal{C}_\text{NH}(M)>\mathcal{C}_\text{H}(M)$. This proves the weak gap persistence theorem. 

(Strong gap persistence theorem) The strong gap persistence theorem then follows from Proposition~\ref{prop:additivebounds} and the additivity of the HCRB, Lemma~\ref{lemma:mcopyhol}.

\end{proof}

Similarly, for the most informative bound we have
\begin{corollary}
\label{cor:multiMIHCRB}
    For $S>0$, a strong gap persistence theorem holds between the most informative bound and HCRB. If $\Delta_{\text{MI}-\text{H}}=0$, then \mbox{$\Delta_{\text{MI}-\text{H}}(M)=0$} for all finite $M$ and if $\Delta_{\text{MI}-\text{H}}\neq0$, then $\Delta_{\text{MI}-\text{H}}(M)\neq0$ for all finite $M$.
\end{corollary}
\begin{proof}[\textbf{Proof (Corollary~\ref{cor:multiMIHCRB})}]
   
    We begin by proving that, for full-rank states, $C_{\rm MI}=C_{\rm H}
\quad\Longleftrightarrow\quad
C_{\rm NH}=C_{\rm H}$.
The forward implication follows immediately from $C_{\rm H}\leq C_{\rm NH}\leq C_{\rm MI}$. Conversely, suppose $C_{\rm NH}=C_{\rm H}$. Since $S>0$, the
HCRB optimiser is unique. Equality of the two bounds therefore implies
that this optimiser also saturates the NHCRB. By Lemma~\ref{lemm:equality_cond2},
its estimator operators commute,
\begin{equation}
    [X_{{\rm H},i},X_{{\rm H},j}]=0
\qquad \forall i,j\;.
\end{equation}
They can therefore be simultaneously diagonalised as
\begin{equation}
    X_{{\rm H},i}
=
\sum_k x_i(k)\Pi_k \;,
\end{equation}
where $\{\Pi_k\}$ is a projective measurement. Using $\Pi_k$ as the
POVM and $x_i(k)$ as the corresponding estimator values gives a
locally unbiased measurement whose MSE is exactly the HCRB. Hence, $C_{\rm MI}=C_{\rm H}$.

Now note that the same equivalence holds for every finite $M$, since
$S^{\otimes M}>0$:
\begin{equation}
    C_{\rm MI}(M)=C_{\rm H}(M)
\quad\Longleftrightarrow\quad
C_{\rm NH}(M)=C_{\rm H}(M)\;.
\end{equation}
 Theorem~\ref{theorem:main2} gives the strong gap persistence theorem between the
NHCRB and HCRB,
\begin{equation}
    C_{\rm NH}(M)=C_{\rm H}(M)
\quad\Longleftrightarrow\quad
C_{\rm NH}=C_{\rm H}\;.
\end{equation}
Combining the above equivalences gives
\begin{equation}
    C_{\rm MI}(M)=C_{\rm H}(M)
\quad\Longleftrightarrow\quad
C_{\rm MI}=C_{\rm H}\;,
\end{equation}
which proves the result.
\end{proof}


For completeness, we state again that for rank-deficient models, the gap persistence theorem does not hold between the HCRB and NHCRB. This is theorem~\ref{theorem:main}.



\section{Finite-copy convergence rate}
\label{sec:convergence}
Above, we have proven a gap persistence theorem in many contexts. However, these theorems alone do not tell us anything about the rate at which the different bounds converge to one another, which is arguably the physically more interesting quantity. As such, in this section, we evaluate the rate of convergence of the HCRB and NHCRB. Note that throughout this section we set the weight matrix $W$ to be the identity matrix for simplicity. However, all of the following results apply for arbitrary positive definite $W$ by a suitable reparameterisation.

We require the following theorem to bound the convergence rate.
\begin{theorem}\label{thm:upperB}
  For any $\mathbf{X}$, the second term of the NH function is bounded above by
  \begin{equation}
  \begin{split}
    \label{eq:upper_NHfn}
        A^{\text{NH}}_{-}[\mathbf{X}]&\le \frac12\sum_{j,k}\text{tr}\{\left|\sqrt{S}[X^j,X^k]\sqrt{S}\right|\}\\
        &=\frac12\sum_{j,k}\text{tr}\{\left|\bbA(\mathbf{X})_{j,k}\right|\}
  \end{split}
  \end{equation}
  This gives an upper bound for the NH bound as
  \begin{equation}
    \mathcal{C}_\text{NH}\le \min_\mathbf{X}\left\{ \text{Tr}[\Re Z[\mathbf{X}]]+\frac12\sum_{j,k}\text{tr}{|\sqrt{S}[X_j,X_k]\sqrt{S}|}\right \}. 
  \end{equation}
\end{theorem}
The proof is presented in Appendix~\ref{apen:asymptotic2}.

We then work with the optimisers defined in Eq.~\eqref{def:xtilde}. In this basis, define the Hermitian commutators
\begin{equation}
\label{eq:K-def}
\begin{split}
     K_{jk}=&i[X_{\mathrm H,j},X_{\mathrm H,k}],\\
 \mu_{jk}=&\tr\{S K_{jk}\},
\\
 v_{jk}=&\tr\!\left\{S(K_{jk}-\mu_{jk}\id)^2\right\}\;.
\end{split}
\end{equation}
Working in this basis implies that $\mu_{jk}=0$ outside the canonical pairs of Eq.~\eqref{def:xtilde} and
\begin{equation}\label{eq:holevo-mean}
 \Tr|\Imm Z[\bm X_{\mathrm H}]|
 =\sum_{j<k}|\mu_{jk}|\;.
\end{equation}
This is the key ingredient in the proof of the following theorem.

\begin{theorem}\label{thm:main-rate}
For every finite-dimensional $n$-parameter model and every integer $M\geq1$,
\begin{equation}\label{eq:raw-rate}
 0\leq \Delta_{\mathrm{NH-H}}(M)
 \leq
 \frac{1}{M^{3/2}}\sum_{j<k}\sqrt{v_{jk}}.
\end{equation}
\end{theorem}
The proof is presented in Appendix~\ref{apen:asymptotic2}. We note three important points about theorem~\ref{thm:main-rate}. First, if the Holevo optimizer is not unique, the right-hand side may be minimized over all Holevo optimizers and over the orthogonal transformations that put $\Imm Z$ in normal form. Second, the normalised gap (which is perhaps more physically relevant) decays at least as fast as $\frac{1}{\sqrt M}$
\begin{equation}\label{eq:main-rate}
 0\leq M\CNH(M)-\CH
 \leq
 \frac{1}{\sqrt M}\sum_{j<k}\sqrt{v_{jk}}\;.
\end{equation}
Finally, note that the gap may decay faster than $\frac{1}{\sqrt M}$. Indeed, in proposition~\ref{prop:exponential}, and section~\ref{sec:eg} below we provide one such example, showing that the gap can decay exponentially in $M$.

We have the following two-parameter corollary that will be used below
\begin{corollary}\label{cor:two-parameter}
For a two-parameter model, $\CNH=\CN$.  If $\bm X_{\mathrm H}=(X_{\mathrm H},Y_{\mathrm H})$, set
\begin{equation}
 K_{\mathrm H}=i[X_{\mathrm H},Y_{\mathrm H}],
 \qquad
 v_{\mathrm H}=\tr\{S K_{\mathrm H}^2\}-\tr\{S K_{\mathrm H}\}^2\;.
\end{equation}
Then
\begin{equation}\label{eq:two-parameter-rate}
 0\leq M\CN(M)-\CH\leq\sqrt{\frac{v_{\mathrm H}}{M}}\;.
\end{equation}
\end{corollary}

\begin{proposition}\label{prop:exponential}
In the setting of Corollary~\ref{cor:two-parameter}, let the spectrum of $K_{\mathrm H}$ be contained in $[k_-,k_+]$, assume $k_-<k_+$, and define
\begin{equation}
 \mu_{\mathrm H}:=\tr\{S K_{\mathrm H}\}.
\end{equation}
If $\mu_{\mathrm H}\neq0$, then
\begin{equation}\label{eq:general-exp}
 0\leq M\CN(M)-\CH
 \leq
 2\norm{K_{\mathrm H}}_{\infty}
 \exp\!\left[-\frac{2M\mu_{\mathrm H}^{2}}{(k_+-k_-)^2}\right].
\end{equation}
If $S\succ0$ and the single-copy Nagaoka--Holevo gap is nonzero, the gap persistence theorem makes the left inequality strict for every finite $M$.
\end{proposition}

\begin{proof}[\textbf{Proof (Proposition~\ref{prop:exponential})}]
Use the same scalar random variable $\xi$ as in the proof of Theorem~\ref{thm:main-rate}.  Suppose first that $\mu_{\mathrm H}>0$.  Since $\mathbb E\overline\xi_M=\mu_{\mathrm H}$,
\begin{equation}
 \mathbb E|\overline\xi_M|-\mu_{\mathrm H}
 =2\mathbb E\!\left[(-\overline\xi_M)_+\right]
 \leq 2\norm{K_{\mathrm H}}_\infty
 \Pr(\overline\xi_M\leq0).
\end{equation}
Hoeffding's inequality gives
\begin{equation}
 \Pr(\overline\xi_M\leq0)
 =\Pr(\overline\xi_M-\mu_{\mathrm H}\leq-\mu_{\mathrm H})
 \leq
 \exp\!\left[-\frac{2M\mu_{\mathrm H}^{2}}{(k_+-k_-)^2}\right].
\end{equation}
The case $\mu_{\mathrm H}<0$ follows by replacing $\xi$ with $-\xi$.  Combining this estimate with the additive-Holevo trial bound proves Eq.~\eqref{eq:general-exp}.
\end{proof}

\subsection{Infinitely many copies of the probe state}
As was done in section~\ref{subsec:inf1} for the two parameter case, here we state a result about the asymptotic equivalence of the NHCRB and HCRB as $M\to\infty$.
\begin{corollary}
\label{theorem:infinite2}
For estimating any number of parameters with any density matrix $S$, we have $\displaystyle\lim_{M\to\infty}M\mathcal{C}_\text{NH}(M)=\mathcal{C}_\text{H}$.
\end{corollary}
This follows trivially by taking the limit $M\to\infty$ in theorem~\ref{thm:main-rate}.

\begin{figure*}[t]
\includegraphics[width=\textwidth]{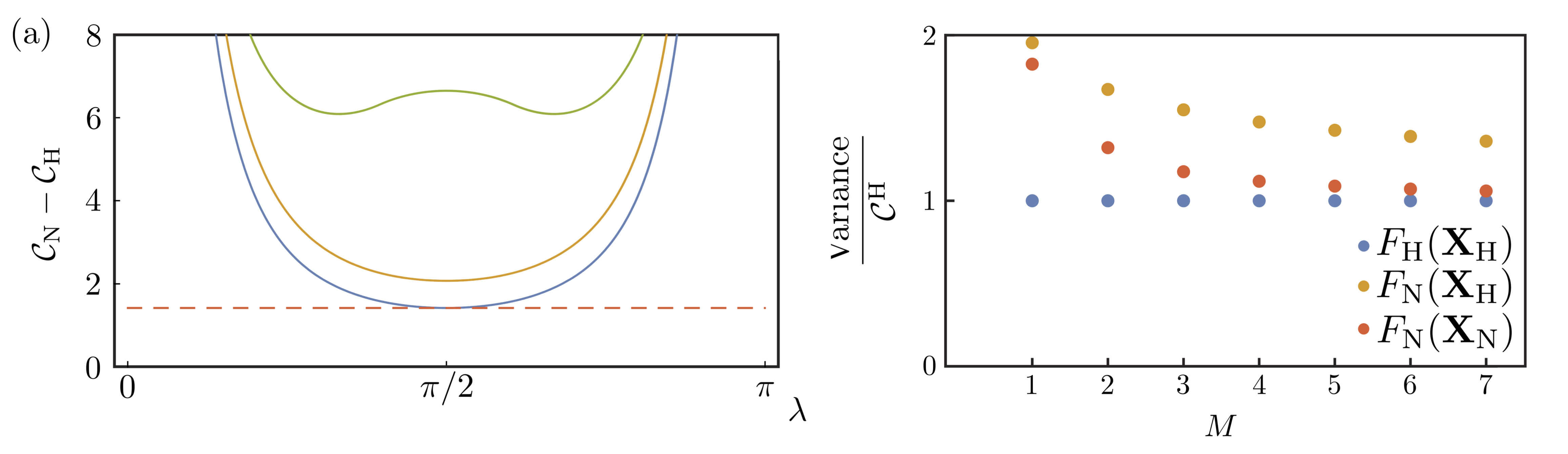}
\caption{\textbf{The unattainability of the HCRB for simultaneously estimating phase and phase diffusion.} (a) The difference between the NCRB and HCRB, $\mathcal{C}_\text{N}-\mathcal{C}_\text{H}$, is plotted as a function of $\lambda$. The blue, orange and green lines correspond to $\delta=0.01,0.5$ and $0.99$ respectively. The red dashed line is at $\sqrt{2}$. (b) Holevo and Nagaoka functions for different input matrices normalised to the HCRB for $\lambda=\phi=\delta=0.3$. The fact that $F_\text{N}(\mathbf{X}_\text{H})$ is never equal to the HCRB is sufficient to show that the HCRB cannot be attained without infinite resources in this case. We also show the NCRB, $\mathcal{C}_\text{N}=F_\text{N}(\mathbf{X}_\text{N})$ for comparison.
}
\label{fig:eg}
\end{figure*}

\section{Examples}
We now illustrate the importance of our results with some physically motivated examples. 
\label{sec:eg}

\subsection{Simultaneous estimation of phase and phase diffusion}
\label{subsecEx1}
We first apply our results to a paradigmatic example of multiparameter estimation; estimating phase and phase diffusion simultaneously~\cite{vidrighin2014joint,altorio2015weak,szczykulska2017reaching,albarelli2022probe}. Following Ref.~\cite{vidrighin2014joint}, the probe state we consider is a single qubit which undergoes a phase shift, $\phi$, and phase diffusion, $\delta$
\begin{equation}
\label{eq:exrho}
S_{\phi,\delta}=\begin{pmatrix}
\text{cos}^2(\frac{\lambda}{2})&\text{cos}(\frac{\lambda}{2})\text{sin}(\frac{\lambda}{2})\mathrm{e}^{-\mathrm{i}\phi-\delta^2}\\
\text{cos}(\frac{\lambda}{2})\text{sin}(\frac{\lambda}{2})\mathrm{e}^{\mathrm{i}\phi-\delta^2}&\text{sin}^2(\frac{\lambda}{2})
\end{pmatrix}\;.
\end{equation}
In this model $\lambda$ is a fixed parameter, which determines the probe state used in the experiment. We restrict the model such that $\lambda\neq0,a\pi$, where $a$ is any integer, to ensure $\phi$ and $\delta$ can be estimated. We assume $\delta\geq0$ and $0\leq\phi<2\pi$. The benefits of collective measurements in this example were examined in Ref.~\cite{vidrighin2014joint}, but only for projective measurements on two copies of the probe state. Hence, the full importance of collective measurements has not yet been understood. Using Refs.~\cite{suzuki2016explicit} and \cite{gill2005state}, it is possible to get analytic expressions for the HCRB and NCRB for this problem. For the HCRB, we define the following function
\begin{equation}
\beta=1-\frac{(-1+\text{e}^{2\delta^2})\abs{\text{cos}(\lambda)}}{2\delta}\;.
\end{equation}
Theorem 1 of Ref.~\cite{suzuki2016explicit} gives the HCRB as (see appendix~\ref{apen:analyticHol} for a detailed calculation)
\begin{equation}
\label{eqn:HCRBdiffformsmain}
\mathcal{C}_\text{H}=
\begin{cases}
\frac{(\text{e}^{2\delta^2}-1)(1+4\delta^2+4\delta\abs{\text{cos}(\lambda)})}{4\delta^2\text{sin}^2(\lambda)}& \text{ if }  \beta\leq0\\
\frac{-1+e^{4\delta^2}+8e^{2\delta^2}\delta^2+\text{cos}(2\lambda)(-1+e^{2\delta^2})^2}{8\delta^2\text{sin}^2(\lambda)}& \text{ if }  \beta\geq0
\end{cases}\;.
\end{equation}
The NCRB is attainable for qubits, meaning $\mathcal{C}_\text{N}=\mathcal{C}_\text{MI}$~\cite{nagaoka2005generalization}. We can therefore use Ref.~\cite{gill2005state} to compute $\mathcal{C}_\text{MI}$, or equivalently for this problem $\mathcal{C}_\text{N}$ (see appendix~\ref{apen:analyticHol} for a detailed calculation).
\begin{equation}
\label{eq:NCRBanalmain}
\mathcal{C}_\text{N}=\frac{\text{e}^{2\delta^2}}{\text{sin}^2(\lambda)}\bigg(1+\frac{1-\text{e}^{-2\delta^2}}{4\delta^2}+\frac{\sqrt{1-\text{e}^{-2\delta^2}}}{\delta}\bigg)\;.
\end{equation}
According to Proposition~\ref{prop:fullrank}, in order to prove that the HCRB is unattainable in this example, it is sufficient to show that the HCRB and NCRB are not equal for a single copy of the probe state. This is proven in Appendix~\ref{apen:strictineqHN}. Equivalently, from Proposition~\ref{prop:fullrank}, one could simply compute the commutator between the HCRB optimisers and prove this quantity is non-zero. Therefore, for this example, the HCRB can \textit{never} be reached.\footnote{Exactly at $\delta=0$ the problem is not well defined, see e.g. Eq.~\eqref{equation:ddelta}.} We plot $\mathcal{C}_\text{N}-\mathcal{C}_\text{H}$ as a function of $\lambda$ for several $\delta$ in Fig.~\ref{fig:eg} (a). In Fig.~\ref{fig:eg} (b) we compare the HCRB to the NCRB, which is the experimentally relevant quantity. For comparison we also plot $F_\text{N}(\mathbf{X_\text{H}})$.

For two parameter estimation with qubit probe states, the NCRB is attainable~\cite{nagaoka2005generalization}, hence $\mathcal{C}_\text{MI}=\mathcal{C}_\text{N}$. Our work, shows that $\mathcal{C}_\text{N}>\mathcal{C}_\text{H}$. For all $\lambda$, except $\lambda=\pi/2$, we have $\mathcal{C}_\text{H}>\mathcal{C}_\text{S}$. Hence, for this example $\mathcal{C}_\text{MI}=\mathcal{C}_\text{N}>\mathcal{C}_\text{H}>\mathcal{C}_\text{S}$. The gap persistence theorem shows that these inequalities will hold for any finite number of copies. 
 

As a final point of interest, we note that as $\delta\to0$ the state becomes a pure state and so we expect $\mathcal{C}_\text{N}\to\mathcal{C}_\text{H}$. Contrary to this, the analytic results presented here and the trade-off relation derived in Ref.~\cite{vidrighin2014joint} appear to show that $\mathcal{C}_\text{N}\to(1+2\sqrt{2}/3)\mathcal{C}_\text{H}$. However, as we discuss in Appendix~\ref{apen:conflict}, this apparent conflict is resolved by noting that solving the optimisation problem in the HCRB and NCRB does not commute with taking the limit $\delta\to0$. (See Appendix~\ref{sec:apen:purestate} for pure states in the general setting.) Similar discontinuities have been observed before for the SLDCRB~\cite{vsafranek2017discontinuities,seveso2019discontinuity,goldberg2021taming}.

 


\subsection{Estimating qubit parameters in the infinite-copy limit}
\begin{figure}[t]
\includegraphics[width=0.47\textwidth]{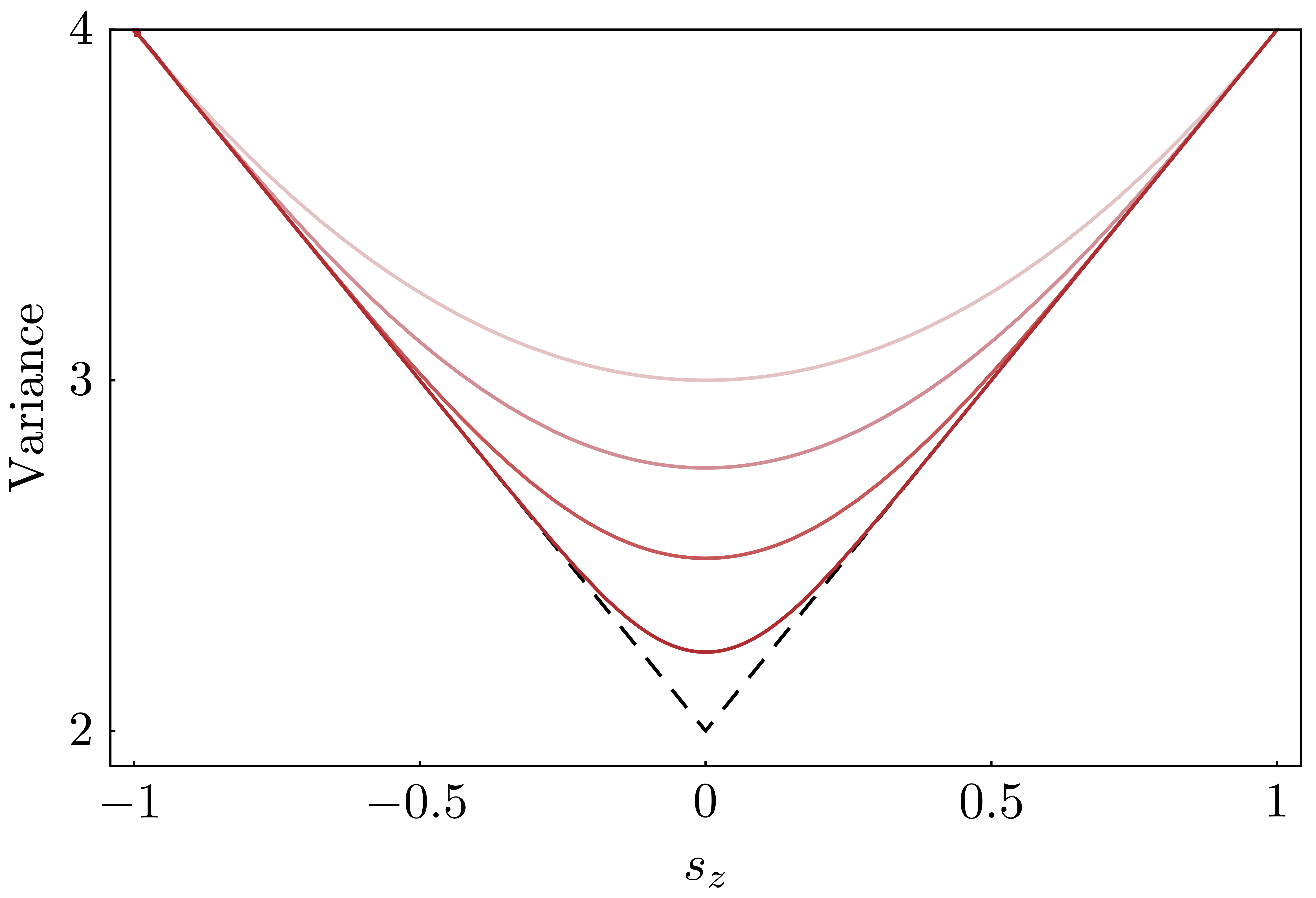}
\caption{\textbf{Asymptotic convergence of the NCRB to the HCRB for estimating qubit parameters.} The dashed black line shows the HCRB. The pink to red lines show $MF_\text{N}(\mathbf{X}_{\text{H},M})$ and are upper bounds on the NCRB. From top to bottom the pink to red lines correspond to $M=2,4,10$ and 60 copies of the probe state.}
\label{fig:convergeg}
\end{figure}
\begin{figure*}[t]
\includegraphics[width=0.9\textwidth]{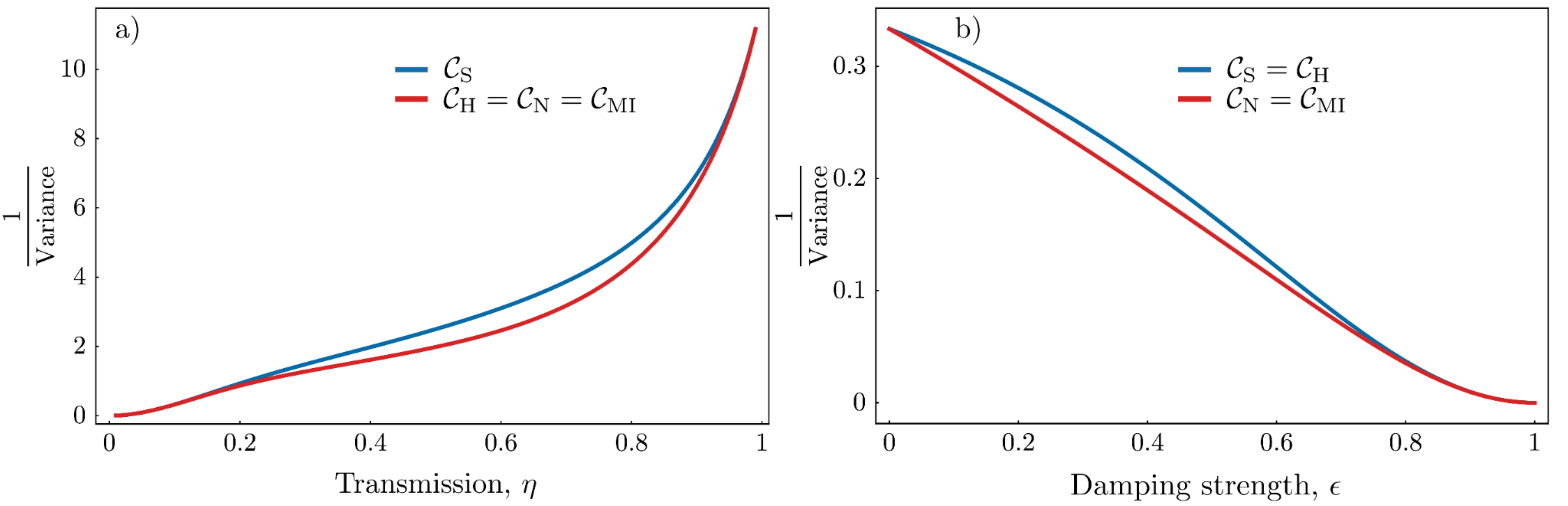}
\caption{\textbf{Hierarchy of various Cram{\'{e}}r-Rao bounds for different examples.} a) Different bounds for the simultaneous estimation of phase and transmissivity in an interferometer with a four photon probe state. In this case collective measurements offer no advantage and the SLDCRB is never attainable. b) Different bounds for estimating qubit rotations subject to the phase damping channel. For this example, as neither the HCRB or SLDCRB can be reached in the single-copy case, they cannot be reached with any measurement on a finite number of copies of the probe state. }
\label{fig:twoegs}
\end{figure*}
Our next example shows the convergence of the NCRB to the HCRB in the infinite-copy setting, as expected from Theorem~\ref{theorem:infinite}. We consider estimating small angles $\theta_x$ and $\theta_y$ in the state $S_\theta=(1+\theta_x\sigma_x+\theta_y\sigma_y+s_z\sigma_z)/2$, where $\sigma_i$ is the $i$th Pauli matrix and $s_z$ is known ($s_z\neq0$) (Note that the three-parameter extension of this model has been studied in Ref.~\cite{PhysRevA.108.032605}). The estimator operators for the HCRB are $\mathbf{X}_\text{H}=(\sigma_x,\sigma_y)$, independent of $s_z$. This allows the HCRB to be computed as
\begin{equation}
\mathcal{C}_\text{H}=2+2\abs{s_z}\;.
\end{equation}
The $M$-copy NCRB, $\mathcal{C}_\text{N}(M)=F_\text{N}(\mathbf{X}_{\text{N},M})$, can be upper bounded by $F_\text{N}(\mathbf{X}_{\text{H},M})$. For this example
\begin{equation}
\frac{\mathcal{C}_\text{H}}{M}\leq F_\text{N}(\mathbf{X}_{\text{H},M})=\frac{2}{M}+\frac{2}{M^2}\text{tr}\bigg\{\abs{\sqrt{S^{\otimes M}}\sum_{i=1}^{M}\sigma_z^i\sqrt{S^{\otimes M}}}\bigg\}\;,
\end{equation}
where $\sigma_z^i$ is the $i$th qubit $\sigma_z$ operator. Hence, to prove convergence, we need to show
\begin{equation}
\lim_{M\to\infty}\frac{1}{M}\text{tr}\bigg\{\abs{\sqrt{S^{\otimes M}}\sum_{i=1}^{M}\sigma_z^i\sqrt{S^{\otimes M}}}\bigg\}=\abs{s_z}\;.
\end{equation}
By calculating the eigenvalues of $S^{\otimes M}\sum_{i=1}^{M}\sigma_z^i$ explicitly, we find
\begin{equation}
\begin{split}
&\text{tr}\bigg\{\abs{\sqrt{S^{\otimes M}}\sum_{i=1}^{M}\sigma_z^i\sqrt{S^{\otimes M}}}\bigg\}=\\
&2^{1-M}\sum_{j=0}^{M}\binom{M}{j}\abs{\frac{M}{2}-j}(1+s_z)^j(1-s_z)^{M-j}\;.
\end{split}
\end{equation}
For even $M$, this evaluates to 
\begin{equation}
\begin{split}
\label{eq:qubitEven}
&Ms_z+\frac{1}{(1+s_z)\Gamma(2+\frac{M}{2})\Gamma(\frac{M}{2})}\bigg[2^{-M}(-1+s_z) \\
&(1-s_z^2)^{M/2}M! \bigg[-(2+M)(1+s_z)+ \\ 
&2Ms_z{}_2F_1\big(1,1-\frac{M}{2};\frac{4+M}{2};\frac{-1+s_z}{1+s_z}\big)\bigg]\bigg]\;,
\end{split}
\end{equation}
where $\Gamma$ is the Gamma function and ${}_2F_1(a,b;c;z)$ is the hypergeometric function. For odd $M$, $\text{tr}\bigg\{\abs{\sqrt{S^{\otimes M}}\sum_{i=1}^{M}\sigma_z^i\sqrt{S^{\otimes M}}}\bigg\}$ evaluates to
\begin{equation}
\begin{split}
\label{eq:qubitOdd}
&Ms_z-\frac{1}{\sqrt{\pi}\Gamma(\frac{3+M}{2})}\bigg[(1-s_z)^{\frac{1+M}{2}}(1+s_z)^{\frac{-1+M}{2}} \Gamma(1+\frac{M}{2})\\
& \bigg[-(1+M)(1+s_z)+\\
&2Ms_z{}_2F_1\big(1,\frac{1-M}{2};\frac{3+M}{2};\frac{-1+s_z}{1+s_z}\big)\bigg]\bigg]\;.
\end{split}
\end{equation}
Taking the limit of this when $M$ goes to infinity, we find
\begin{align}
\lim_{M\to \infty}\frac{1}{M}  \text{tr}\left[\abs{\sqrt{S^{\otimes M}}(\sigma_z^1 + \sigma_z^2 + \ldots +
  \sigma_z^M
  )\sqrt{S^{\otimes M}}}\right]= |s_z|\,,
\end{align}
which again shows that the NCRB converges to the HCRB. A plot of this convergence is shown in Fig.~\ref{fig:convergeg}.

\subsubsection{Rate of convergence of the NCRB to HCRB for this example}
For practical application it is also of interest to study the rate at which the NCRB converges to the HCRB. For this model we can compute
\begin{equation}
\begin{split}
     K_{\mathrm H}&=i[\sigma_x,\sigma_y]=-2\sigma_z \\
 \mu_{\mathrm H}&=-2s_z\\
||K_{\mathrm H}||_\infty&=2\\
 k_+-k_-&=4\;.
\end{split}
\end{equation}
For every fixed $0<|s_z|<1$, Proposition~\ref{prop:exponential} therefore gives
\begin{equation}\label{eq:qubit-exp}
 0<M\CN(M)-\CH
 \leq 4\exp\!\left(-\frac{M s_z^2}{2}\right)\;.
\end{equation}
For all non-zero $s_z$, this gives a convergence rate that is exponential in $M$. 

For completeness we now analyse the convergence rate at $s_z=0$. We know that $\mathcal{C}_\text{H}\leq M\mathcal{C}_\text{N}(M)=MF_\text{N}(\mathbf{X}_{\text{N},M})\leq MF_\text{N}(\mathbf{X}_{\text{H},M})$. Therefore, $M\mathcal{C}_\text{N}(M)-\mathcal{C}_\text{H}\leq MF_\text{N}(\mathbf{X}_{\text{H},M})-\mathcal{C}_\text{H}$. We can then look at this final term to get a lower bound on how quickly the NCRB converges to the HCRB. We are therefore interested in the rate at which 
\begin{equation}
MF_\text{N}(\mathbf{X}_{\text{H},M})-\mathcal{C}_\text{H}=\frac{2}{M}\text{tr}\bigg\{\abs{\sqrt{S^{\otimes M}}\sum_{i=1}^{M}\sigma_z^i\sqrt{S^{\otimes M}}}\bigg\}\;,
\end{equation}
tends to 0 for $S$ being the maximally mixed state. Adding a factor $1/M$ to Eq.~\eqref{eq:qubitEven} gives
\begin{equation}
    2\frac{\Gamma\left(\frac{1+M}{2}\right) }{M\sqrt{\pi}\Gamma\left(\frac{M}{2}\right)}
\end{equation}
when evaluated at $s_z=0$. We now use Stirling's approximation for the gamma function
\begin{equation}
    \Gamma (z)=\sqrt{\frac{2\pi}{z}}\left(\frac{z}{e}\right)^z\left(1+O\left(\frac{1}{z}\right)\right)\approx \sqrt{\frac{2\pi}{z}}\left(\frac{z}{e}\right)^z\;.
\end{equation}
Evaluating 
\begin{equation}
\begin{split}
    2\frac{\Gamma\left(\frac{1+M}{2}\right) }{M\sqrt{\pi}\Gamma\left(\frac{M}{2}\right)}&\approx\frac{2}{\sqrt{\pi}M}\frac{\sqrt{\frac{4\pi}{1+M}\left(\frac{1+M}{2e}\right)^{M/2+1/2}}}{\sqrt{\frac{4\pi}{M}}\left(\frac{M}{2e}\right)^{M/2}}\\
    &=\frac{2}{\sqrt{\pi}M}\frac{\sqrt{\frac{M}{1+M}\left(\frac{1+M}{2e}\right)^{M/2}\left(\frac{1+M}{2e}\right)^{1/2}}}{\left(\frac{M}{2e}\right)^{M/2}}\\
    &=O(\frac{1}{\sqrt{M}})\;.
    \end{split}
\end{equation}
Similarly adding a factor $1/M$ to Eq.~\eqref{eq:qubitOdd} gives
\begin{equation}
    2\frac{\Gamma\left(1+\frac{M}{2}\right) }{M\sqrt{\pi}\Gamma\left(\frac{1+M}{2}\right)}=O(\frac{1}{\sqrt{M}})\;,
\end{equation}
for large $M$. These results demonstrate that the (normalised to $M$) gap between the NCRB and the HCRB shrinks at a rate of at least $O(\frac{1}{\sqrt{M}})$.

\subsection{Simultaneous estimation of phase and loss in interferometry}
For some rank deficient probe states it is possible to have equality between the NCRB and HCRB. This is trivial for pure states (see Appendix~\ref{sec:apen:purestate}), but is also true for other, more interesting problems. One such problem, is that of simultaneously estimating a phase shift and the transmissivity of one arm of a two mode interferometer. We consider using Holland-Burnett states, obtained by interfering two equal photon number Fock states on a balanced beam splitter, to estimate the phase shift and transmissivity~\cite{holland1993interferometric}. It has been shown that, for this problem, collective measurements offer no advantage over separable measurements, see Example 2 of Ref.~\cite{conlon2021efficient}. Interestingly, it is known for this problem that the SLDCRB is not attainable~\cite{crowley2014tradeoff}, hence we have $\mathcal{C}_\text{MI}=\mathcal{C}_\text{N}=\mathcal{C}_\text{H}>\mathcal{C}_\text{S}$. This hierarchy is shown in \mbox{Fig.~\ref{fig:twoegs} a)} for a four photon state.

\subsection{Estimating qubit rotations}
The estimation of qubit rotations about the $x,y$ and $z$ axes of the Bloch sphere with a two qubit probe was considered in Ref.~\cite{conlon2021efficient}. We consider estimating the qubit rotations using the maximally entangled two-qubit state, $(\ket{01}+\ket{10})/\sqrt{2}$, which is subject to the phase damping channel after experiencing the rotations. For this example the HCRB and NHCRB have particularly simple forms,
\begin{equation}
\mathcal{C}_\text{H}=2+\frac{1}{(1-\epsilon)^2}
\end{equation}
and
\begin{equation}
\mathcal{C}_\text{NH}=\frac{4}{2-\epsilon}+\frac{1}{(1-\epsilon)^2}
\end{equation}
where $\epsilon$ is the damping strength. A measurement saturating the NHCRB was presented in Ref.~\cite{conlon2021efficient} (see Example 1 and Supplementary Note 5) and for this problem the HCRB and SLDCRB coincide, hence $\mathcal{C}_\text{MI}=\mathcal{C}_\text{NH}>\mathcal{C}_\text{H}=\mathcal{C}_\text{S}$. By the gap persistence theorem we then know that neither the HCRB or SLDCRB is attainable for this problem. In Fig.~\ref{fig:twoegs} b) we show the different bounds as a function of the damping strength.

\subsection{Random examples}
\begin{figure}[b]
\includegraphics[width=0.45\textwidth]{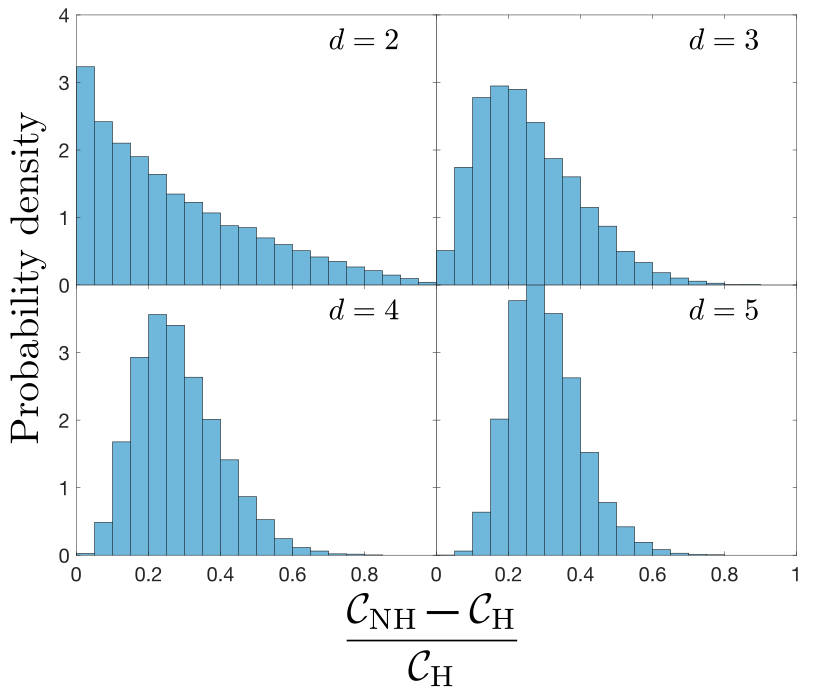}
\caption{\textbf{Gap between the HCRB and NCRB for two parameter estimation with random full rank probe states.} We plot $(\mathcal{C}_\text{N}-\mathcal{C}_\text{H})/\mathcal{C}_\text{H}$ for random full rank probe states of different dimensions, up to and including dimensions of five. For each dimension 10,000 random probe states are shown. For each plot we use 20 equally spaced bins between 0 and 1.}
\label{fig:egrandom}
\end{figure}

Proposition~\ref{prop:fullrank} shows that for full rank states and finite dimensional systems $\mathcal{C}_\text{N}(M)>\mathcal{C}_\text{H}(M)$ for all $M$. However, this says nothing about how close the two bounds can be. To investigate this, we plot $(\mathcal{C}_\text{N}-\mathcal{C}_\text{H})/\mathcal{C}_\text{H}$ for 40,000 random full rank probe states in Fig.~\ref{fig:egrandom}. As expected, for all probe states tested, we found $(\mathcal{C}_\text{N}-\mathcal{C}_\text{H})/\mathcal{C}_\text{H}$ to be greater than zero. The gap between the two bounds shows some dependence on the dimension of the system. For estimating two parameters, the quantity $(\mathcal{C}_\text{N}-\mathcal{C}_\text{H})/\mathcal{C}_\text{H}$ is upper bounded by 1. 

\section{Open problems and conjectures}
\label{sec:OP}
Before concluding, we present some important questions that our work opens up. 

\subsection{Does the gap persistence theorem hold in other areas of quantum information?}
Next, we look more broadly and ask whether we can expect similar results to hold in other areas of quantum information where collective measurements are important. For example, in quantum state discrimination, collective measurements can help discriminate between mixed states~\cite{calsamiglia2010local,conlon2023discriminating,conlon2025attainability}. For this task, the error probability for distinguishing two quantum states is given by the Helstrom bound~\cite{helstrom1969quantum}, and the asymptotic error rate when allowing for infinitely many copies of the quantum state is given by the quantum Chernoff bound~\cite{nussbaum2009chernoff,audenaert2007discriminating}, or the quantum Hoeffding bound~\cite{nagaoka2006converse,hayashi2007error} or quantum Stein's lemma in the asymmetric case~\cite{hiai1991proper,ogawa2000strong}. In this setting there has been significant work done on examining the differences between the asymptotic and finite copy scenarios~\cite{pereira2023analytical,audenaert2012quantum,rouze2017finite,li2014second}, something akin to a gap persistence theorem. Similarly, for quantum channel capacities, there is a gap between asymptotic and finite copy results~\cite{tomamichel2016quantum}.

Even within quantum metrology, the gap persistence theorem may find use in slightly different settings. Recently, it was shown that many entangled states obtain their maximal metrological advantage over separable states when infinitely many copies of the state are available~\cite{trenyi2022multicopy}. In this scenario it may also be possible that such states cannot attain the maximal possible advantage with any finite number of copies of the probe state. 

\begin{table*}[t]
\begin{center}
\begin{tabular}{ |p{1cm}|p{5cm}|p{5cm}|p{5cm}| } 
\hline
& $\mathcal{C}_\text{NH}$ & $\mathcal{C}_\text{H}$ & $\mathcal{C}_\text{S}$ \\
\hline
{$\mathcal{C}_\text{MI}$} & GPT does not hold in general (Corollary~\ref{corol:MINHCRB}). $S\succ0$ remains open& GPT holds when $S\succ0$ (Corollary~\ref{cor:multiMIHCRB}). GPT does not hold in the general case when $S\succeq0$ (Theorem~\ref{theorem:mainHCRBMI})&GPT holds when $S\succ0$ (Theorem~\ref{thm:SLD-MI-full-rank-GPT}). GPT does not hold in general when $S\succeq0$ (Theorem~\ref{theorem:mainSLDCRBMI})\\ \hline
$\mathcal{C}_\text{NH}$& N/A & GPT holds when $S\succ0$ (Theorem~\ref{theorem:main2}). GPT does not hold in general when $S\succeq0$ (Theorem~\ref{theorem:main})&GPT is true (Theorem~\ref{thermSGSLDNH}) \\ \hline
$\mathcal{C}_\text{H}$& See above row& N/A&GPT is true from additivity of the two bounds \\ 
\hline
\end{tabular}
\caption{Summary of the conditions under which a gap persistence theorem (GPT) has been established. }
\label{tab:results}
\end{center}
\end{table*}

\section{Discussions and conclusion}
\label{sec:disc}
The SLDCRB and HCRB both play a fundamental role in quantum metrology. Given that these bounds differ only by a factor of 2~\cite{carollo2019quantumness}, the common motivation for computing the computationally more demanding HCRB~\cite{albarelli2019evaluating, sidhu2021tight}, is that the additional complexity is worthwhile as the HCRB is in some sense attainable~\cite{kahn2009local, yamagata2013quantum, yang2019attaining}. Our results challenge this viewpoint and suggest that for practical applications of quantum metrology the NCRB and NHCRB may be the more relevant quantities. For full-rank states (which are practically relevant), we have proven that the gap persistence theorem holds for the NHCRB and HCRB/SLDCRB and also for the most informative bound and the HCRB/SLDCRB. Therefore, if the HCRB/SLDCRB cannot be saturated with a single copy of the probe state, it cannot be saturated with measurements on any finite number of copies of the probe state. These results immediately imply that the gap persistence theorem always holds for qubit states, and are summarised in Table~\ref{tab:results}. The difference between full rank and rank deficient states is somewhat striking.



We applied our results to show that for simultaneously estimating phase and phase diffusion, the HCRB can \textit{never} be attained for any finite number of probe states. This result can be generalised to many other important problems which can be mapped to qubit multiparameter estimation problems, such as quantum superresolution~\cite{chrostowski2017super,vrehavcek2017multiparameter} or qubit tomography~\cite{hou2016achieving,razavian2020quantumness}. Furthermore, this result can be extended to any scenario where the probe is full rank and the NHCRB does not equal the HCRB for a single copy of the probe state~\cite{conlon2021efficient,friel2020attainability}. We anticipate that our results will have an important role to play in understanding the attainable precisions in quantum multiparameter estimation. While the benefits of the gap persistence theorem in quantum metrology are immediately obvious, the concept may prove beneficial for other areas of quantum information. Beyond quantum metrology, collective measurements are important in quantum illumination~\cite{zhuang2017optimum,bradshaw2021optimal}, quantum communications~\cite{tserkis2020teleportation}, entanglement distillation~\cite{bennett1996purification} and for better Bell inequality violations~\cite{liang2006better}. Future work will involve extending our results to these scenarios.


\section*{Acknowledgements}
This research was funded by the Australian Research Council Centre of Excellence CE170100012, Laureate Fellowship FL150100019 and the Australian Government Research Training Program Scholarship. JS is partially supported by JSPS KAKENHI Grant Numbers JP21K04919, JP21K11749, JP24K14816. JS acknowledges useful discussions with Prof. Fujiwara.

 ChatGPT 5.6 sol was used to generate the counter examples to various rank deficient gap persistence theorems.

 \section*{Note Added}
During revision of this manuscript we became aware of Ref.~\cite{li2026closing}. This work proves finite-copy gap persistence between the most informative bound and the HCRB for full-rank models, and constructs an explicit three-parameter rank-two model where the gap persistence theorem does not hold. In this example there is a gap at one copy but no gap at two copies. The rank-deficient counterexamples presented here show that such finite-copy gap closure can occur already for two-parameter models.

\appendix
\renewcommand{\thetheorem}{\arabic{theorem}}

\section{The most informative bound}
In this appendix we discuss several important properties of the most informative bound. In sections~\ref{apen:CMI} and \ref{apen:CMI2} we discuss the additivity and attainability of the most informative bound respectively.

\subsection{Additivity of most informative bound under separable measurements}
\label{apen:CMI}
Here we present the proof of proposition~\ref{prop1}. This demonstrates that the most informative bound is additive under separable measurements when working in the local estimation framework.

Recall that a separable measurement is of the form $\Pi_\text{sep}=\{\Pi_{k,1}\}\otimes\{\Pi_{k,2}\}\hdots\otimes\{\Pi_{k,N}\}$. In general $\Pi_{k,i}$ depends on the previous $i-1$ measurement results $(x_1,...x_{i-1})$. By the Cram\'er-Rao inequality for locally unbiased estimators, we have
\begin{equation}
        \label{eq:apenA1}
\begin{split}
V^{(N)}_{\theta}[\Pi_\text{sep},\hat{\theta}]&\geq (J_\theta[\Pi_\text{sep};S^{\otimes N}])^{-1}\\
&=\left(\sum_{m=1}^N \sum_{h_{m-1}}p_\theta(h_{m-1})J_\theta[\Pi^{(m)}_{h_{m-1}};S]\right)^{-1}
\end{split}
\end{equation}
where we use $V^{(N)}_{\theta}[\Pi_\text{sep},\hat{\theta}]$ to denote the MSE matrix at $\theta$, given a specified separable POVM and estimator, $J_\theta[\Pi_\text{sep};S^{\otimes N}]$ to denote the classical Fisher information given a certain POVM and density matrix. The second line follows from the chain rule for the classical Fisher information matrix of a sequential experiment, where we use $\Pi_{h_{m-1}}^{(m)}$ to denote the POVM for the $m$th probe state which may depend on the $m-1$ previous measurement results. Past measurement results are denoted through a history $h_{m-1}=(x_1,x_2,...x_{m-1})$, such that $p_\theta(h_{m-1})$ is the $\theta$-dependent probability of a certain set of past measurement results. As we are interested in local estimation we shall evaluate this quantity only at $\theta=\theta_0$ with $\theta_0$ known.

To prove $ \mathcal{C}_{\text{MI;separable}}(S^{\otimes N},W)=\frac{1}{N}\mathcal{C}_{\text{MI}}(S,W)$, we prove both $ \mathcal{C}_{\text{MI;separable}}(S^{\otimes N},W)\leq\frac{1}{N}\mathcal{C}_{\text{MI}}(S,W)$ and $ \mathcal{C}_{\text{MI;separable}}(S^{\otimes N},W)\geq\frac{1}{N}\mathcal{C}_{\text{MI}}(S,W)$. 

$ \mathcal{C}_{\text{MI;separable}}(S^{\otimes N},W)\leq\frac{1}{N}\mathcal{C}_{\text{MI}}(S,W)$ follows almost immediately by considering a separable POVM. All single-copy POVMs must satisfy: 
\begin{equation}
\text{Tr}[W \,J_{\theta_0}[\Pi;S]^{-1}]\geq \mathcal{C}_{\text{MI}}(S,W)=\text{Tr}[W\,\left( J_{\theta_0}[\Pi^*;S]\right)^{-1}]
\end{equation}
where $\Pi^*$ is the optimal POVM. We therefore have that
\begin{equation}
    \begin{split}
        \mathcal{C}_{\text{MI}}(S,W)&=\text{Tr}[W\,\left( J_{\theta_0}[\Pi^*;S]\right)^{-1}]\\
        &=N\text{Tr}[W\,\left( J_{\theta_0}[(\Pi^*)^{\otimes N};S^{\otimes N}]\right)^{-1}]\\
        &\geq N\mathcal{C}_{\text{MI;separable}}(S^{\otimes N},W)
    \end{split}
\end{equation}
In the second line we have used the additivity of the Fisher information. This demonstrates the $ \mathcal{C}_{\text{MI;separable}}(S^{\otimes N},W)\leq\frac{1}{N}\mathcal{C}_{\text{MI}}(S,W)$ part of the argument. 



We next prove $ \mathcal{C}_{\text{MI;separable}}(S^{\otimes N},W)\geq\frac{1}{N}\mathcal{C}_{\text{MI}}(S,W)$. Define the average CFI as
\begin{equation}
  \bar{J}_{\theta_0}= \frac{1}{N}\sum_{m=1}^N\sum_{h_{m-1}}p_{\theta_0}(h_{m-1}) J_{\theta_0}[\Pi_{h_{m-1}}^{(m)};S] \;.
\end{equation}
Consider the following single-copy POVM
\begin{equation}
\label{eq:POVMadaptive}
    \bar{\Pi}_{m,k,h_{m-1}}=\frac{1}{N}p_{\theta_0}(h_{m-1})\Pi_
    {k,h_{m-1}}^{(m)}\;.
\end{equation}
This is a valid POVM, satisfying $\sum_{m,k,h_{m-1}} \bar{\Pi}_{m,k,h_{m-1}}=\mathbb{I}_d$, and positivity $\bar{\Pi}_{m,k,h_{m-1}}\geq0$. This POVM satisfies
\begin{equation}
  \bar{J}_{\theta_0}=J_{\theta_0}[\bar{\Pi};S]= \frac{1}{N}\sum_{m=1}^N \sum_{h_{m-1}}p_{\theta_0}(h_{m-1})J_{\theta_0}[\Pi_{h_{m-1}}^{(m)};S] \;.
\end{equation}
This can be verified by explicitly computing the probabilities of the different POVM outcomes. Therefore, any separable POVM over $S^{\otimes N}$ achieves a CFI that differs by a factor of $N$ compared to a POVM acting on $S$, i.e. taking the weighted trace of Eq.~\eqref{eq:apenA1} we get:
\begin{equation}
\begin{split}
    &\;\;\;\;\;\text{Tr}[W\,V^{(N)}_{\theta_0}[\Pi_\text{sep},\hat{\theta}]]\\
    &\geq\text{Tr}\left[W\,\left(\sum_{m=1}^N\sum_{h_{m-1}}p_{\theta_0}(h_{m-1}) J_{\theta_0}[\Pi_{h_{m-1}}^{(m)};S]\right)^{-1}\right]\\
    &=\frac{1}{N}\text{Tr}\left[W\,\left(\bar{J}_{\theta_0}[\bar{\Pi}
    ;S]\right)^{-1}\right]\\
    &\geq \frac{1}{N} \mathcal{C}_{\text{MI}}(S,W)\;.
\end{split}
\end{equation}
As this holds for all separable POVMs and estimators, this proves $ \mathcal{C}_{\text{MI;separable}}(S^{\otimes N},W)\geq\frac{1}{N}\mathcal{C}_{\text{MI}}(S,W)$. Taken together with $ \mathcal{C}_{\text{MI;separable}}(S^{\otimes N},W)\leq\frac{1}{N}\mathcal{C}_{\text{MI}}(S,W)$, this proves $ \mathcal{C}_{\text{MI;separable}}(S^{\otimes N},W)=\frac{1}{N}\mathcal{C}_{\text{MI}}(S,W)$. It is simple to extend this proof to $ \mathcal{C}_{\text{MI;separable}}(S^{\otimes NM},W)=\frac{1}{N}\mathcal{C}_{\text{MI}}(S^{\otimes M},W)$.

We stress the importance of working in the local estimation regime for this result. The POVM in Eq.~\eqref{eq:POVMadaptive} explicitly allows for dependence on $\theta_0$. Thus it reproduces the relevant averaged Fisher information at $\theta_0$, and is not optimal for all $\theta$ simultaneously. In global estimation, when the true parameter is not known locally, a preliminary estimation stage or more general adaptive scheme may still be
required to attain the same local asymptotic precision, as discussed in the next section, appendix~\ref{apen:CMI2}.

It is worth remarking that the above proof can be considerably simplified if one makes the additional assumption that the optimal separable POVM is of the form $\Pi_\text{sep}^*=\{\Pi_{k}^*\}\otimes\{\Pi_{k}^*\}...\otimes\{\Pi_{k}^*\}$, i.e. repeatedly implementing the optimal single-copy POVM. The above argument shows that this assumption is not required for the proof.



\subsection{Attainability of the most informative bound}
\label{apen:CMI2}

Above, we have proven that when working in the local estimation setting, the bound $ \mathcal{C}_{\text{MI;separable}}(S^{\otimes NM},W)=\frac{1}{N}\mathcal{C}_{\text{MI}}(S^{\otimes M},W)$ is attainable by $M$-copy measurements. Here we discuss briefly a simple adaptive protocol, which asymptotically saturates this bound even for global estimation. Given $S_M^{\otimes N}=S^{\otimes MN}$, we wish to show the bound 
$\mathcal{C}_{\text{MI}}(S^{\otimes M},W)$ is attained in the large $N$ limit by performing suitable separable measurements. This requires that we prove the existence of a sequence of POVMs and estimators such that 
\begin{equation} \label{eq:adaptive}
\lim_{N\to\infty}N\times \mathcal{C}_{\text{MI}}(S^{\otimes NM},W)=\mathcal{C}_{\text{MI}}(S^{\otimes M},W)
\end{equation}
holds under the condition that the estimate is asymptotically unbiased. Here the crucial point that each POVM is separable with respect to the partition $\mathcal{H}^{\otimes M}\otimes\cdots\otimes\mathcal{H}^{\otimes M}$. Let us denote the POVM as
\begin{equation}
\Pi_{\text{sep},\mathbf{k}}^N(N)=\Pi_{k_1}(1)\otimes \Pi_{k_2}(2)\otimes\cdots\otimes\Pi_{k_N}(N)
\end{equation}
A sequence of separable POVMs is not sufficient to attain the bound. Instead, we need an adaptive estimation strategy. There are many variants known in the theory of optimal design of experiments in statistics. A common one in quantum estimation is the two-stage estimation method. We split a sufficiently large sample $N$ into $N_1$ and the rest $N_2=N-N_1$. 
For $N_1$ samples, we perform a (tomographically complete) separable measurement. Based on the estimate from $N_1$ samples, we next implement the optimal POVM, which depends on the unknown parameter. It is well known that this scheme gives the result \eqref{eq:adaptive}~\cite{barndorff2000fisher,hayashi2005statistical}. A more general approach is to use an adaptive approach with each subsequent measurement depending on all previous results.

\section{$M$-copy Nagaoka--Hayashi Cram{\'{e}}r-Rao bound}
\label{apen:McopyNHCRB}
For clarity and completeness we present a discussion of the NHCRB here and how it is extended to the $M$-copy setting. We present a fundamentally different derivation compared to Ref.~\cite{conlon2021efficient}.
\subsection{Alternative expression for the weighted mean squared error}

Given a POVM with the outcome set $\cX$ and an estimator, introduce the following Hermitian matrices on $\mathcal{H}$
\begin{equation}
\begin{split}
\label{eq:apenLdef1}
  \hat{\mathbb{L}}_{jk}[\Pi,\hat{\theta}]
&=\sum_{x\in\mathcal{X}}\hat{\theta}_j(x)\Pi_x\hat{\theta}_k(x)\qquad (j,k = 1,2,...n)\\
\hat{X}_j[\Pi,\hat{\theta}]&=\sum_{x\in\mathcal{X}}\hat{\theta}_j(x)\Pi_x \qquad (j = 1,2,...n)
  \end{split}
\end{equation}
By definition, $\mathbb{L}_{jk}[\Pi,\hat{\theta}]$ and $X_j[\Pi,\hat{\theta}]$ satisfy the conditions
\begin{equation}
\begin{split}
\label{eq:Ldef2}
  \hat{\mathbb{L}}_{jk}[\Pi,\hat{\theta}]
&=\hat{\mathbb{L}}_{kj}[\Pi,\hat{\theta}]\qquad (j,k = 1,2,...n)\\
  \hat{\mathbb{L}}_{jk}[\Pi,\hat{\theta}]
&: \text{Hermitian}\qquad (j,k = 1,2,...n)\\
\hat{X}_j[\Pi,\hat{\theta}]&: \text{Hermitian} \qquad (j = 1,2,...n)
  \end{split}
\end{equation}
We express the $(jk)$ element of the MSE matrix $V(\theta)$ as
\begin{widetext}
\begin{equation}
    \begin{split}
V(\theta)_{jk}&=\sum_{x\in\mathcal{X}}
 \left(\hat{\theta}_{j}(x)-\theta_j \right)
    \left(\hat{\theta}_{k}(x)-\theta_k \right)\mathrm{tr}\big\{ S \Pi_x \big\}\\
    &=\sum_{x\in\mathcal{X}}\hat{\theta}_{j}(x)\hat{\theta}_{k}(x)\mathrm{tr}\big\{ S \Pi_x \big\}-\sum_{x\in\mathcal{X}}\theta_{j}\hat{\theta}_{k}(x)\mathrm{tr}\big\{ S \Pi_x \big\}-\sum_{x\in\mathcal{X}}\hat{\theta}_{j}(x)\theta_{k}\mathrm{tr}\big\{ S \Pi_x \big\}+\theta_j\theta_k\\
    &=\mathrm{tr}\big\{ S(\hat{\mathbb{L}}_{jk}[\Pi,\hat{\theta}\}-\theta_j\hat{X}_k[\Pi,\hat{\theta}] -\theta_k\hat{X}_j[\Pi,\hat{\theta}]  +\theta_j\theta_k)\big]
    \end{split}
\end{equation}

We next define a matrix and a vector which on an extended Hilbert space. Consider
the Hilbert space $\mathbb{H}=\mathbb{C}^n\otimes\mathcal{H}$, and define $\hat{\mathbb{L}}$ on $\mathbb{H}$ whose $(j, k)$ block-matrix component is given by $\hat{\mathbb{L}}_{jk}$. We also define a column vector $\hat{X}$ whose $j$th component is $X_j$: $\hat{X} = [\hat{X}_1 , \hat{X}_2 , . . . , \hat{X}_n ]^\intercal$ . Here, we denote transpose of matrices and vectors with respect to the first Hilbert space =$\mathbb{C}^n$ by $(\cdot)^\intercal$ . Let $W = [W_{jk} ]$ be a weight matrix, and define
\begin{equation}
    \begin{split}
        \mathbb{S}&=[\mathbb{S}_{jk}]\quad\text{with}\quad\mathbb{S}_{jk}:=W_{jk}S\\
        D&=[D_{j}]\quad\text{with}\quad D_{j}:=\sum_{k=1}^{n}W_{jk}\theta_kS\\
        \mathrm{w}&=\sum_{j,k=1}^{n}\theta_jW_{jk}\theta_k
    \end{split}
\end{equation}

By definition, $\mathbb{S} = W\otimes S$ is an operator on the extended Hilbert space. $D = [D_1, D_2,..., D_n]^\intercal$
is a column vector with Hermitian matrix elements. 
With these notations, the weighted trace of the MSE matrix is expressed as follows.
\begin{equation}
    \begin{split}
        \mathrm{Tr}[WV(\theta)]&=\sum_{j,k=1}^{n}W_{jk}\text{tr}\big\{ S(\hat{\mathbb{L}}_{jk}[\Pi,\hat{\theta}]-\theta_j\hat{X}_k[\Pi,\hat{\theta}] -\theta_k\hat{X}_j[\Pi,\hat{\theta}]  +\theta_j\theta_k\big\}\\
        &=\sum_{j,k=1}^{n}\mathrm{tr}\big\{ \mathbb{S}_{jk}\hat{\mathbb{L}}_{jk}[\Pi,\hat{\theta}]-D_j\hat{X}_k[\Pi,\hat{\theta}] -D_k\hat{X}_j[\Pi,\hat{\theta}] \big\} +\theta_jW_{jk}\theta_k\\
        &=\mathbb{T}\mathrm{r}\big[ \mathbb{S}\hat{\mathbb{L}}[\Pi,\hat{\theta}]\big] -\mathbb{T}\mathrm{r}\big[D\hat{X}[\Pi,\hat{\theta}]^\intercal\big]  -\mathbb{T}\mathrm{r}\big[\hat{X}[\Pi,\hat{\theta}]D^\intercal\big] +\mathrm{w}\\
    \end{split}
\end{equation}
where $\mathbb{T}\mathrm{r}[\cdot]$ denotes the trace on the extended Hilbert space $\mathbb{H}$.

\subsection{Fundamental lemmas}

The fundamental inequality is stated in the following lemma, which is a variant of Holevo's lemma~\cite{holevo2011probabilistic}.
  \begin{lemma}
   \label{lemmaNH1}
 For all POVMs and estimators, $\hat{\mathbb{L}}$ and $\hat{X}$ obey the matrix inequality:
 \begin{equation}
     \hat{\mathbb{L}}[\Pi,\hat{\theta}]\geq \hat{X}[\Pi,\hat{\theta}]\left(\hat{X}[\Pi,\hat{\theta}]\right)^\intercal
 \end{equation}
  \end{lemma}
  \begin{proof}[\textbf{Proof (Lemma~\ref{lemmaNH1})}]
For this we first recall the following result from Holevo. 
For a given POVM $\Pi=\{\Pi_x\}$ and a function $f$ from $\mathcal{X}$ to $\mathbb{C}$,  
define a matrix $F=\sum_{x\in\mathcal{X}}f(x)\Pi_x$. 
The lemma (Proposition II.9.1 of Ref.~\cite{holevo2011probabilistic}) proves the inequality.
\begin{equation}
\sum_{x\in\mathcal{X}}f(x)\Pi_x f(x)^*\geq FF^\dagger.
\end{equation}

We generalise this result as follows. 
Consider a set of functions $f_j$ from $\mathcal{X}$ to $\mathbb{C}$ for $j=1,2,\ldots,n$. 
Define matrices associated with them by $F_j=\sum_{x\in\mathcal{X}}f_j(x)\Pi_x$ ($j=1,2,\ldots,n$). 
Introduce column vectors of matrices on the extended Hilbert space $\mathbb{H}$ by
\begin{equation}
F(x)=[F_1(x),F_2(x),\ldots,F_n(x)]^\intercal\mbox{ with } F_{j}(x):=F_j-f_j(x) I, 
\end{equation}
for $x\in\mathcal{X}$. 
Since $\Pi_x$ is positive semi-definite, we have
\begin{equation}
F(x)\Pi_x F(x)^\dagger\geq0\mbox{ for all }x\in\mathcal{X}. 
\end{equation} 
Taking the summation over $x$ yields
\begin{align*}
&\sum_{x\in\mathcal{X}}F(x)\Pi_x F(x)^\dagger\geq0\\
\Leftrightarrow&\sum_{x\in\mathcal{X}} 
   \begin{pmatrix}
   F_{1}(x)\Pi_x F_{1}(x)^\dagger & F_{1}(x)\Pi_x F_{2}(x)^\dagger & \cdots &F_{1}(x)\Pi_x F_{n}(x)^\dagger\\
   F_{2}(x)\Pi_x F_{1}(x)^\dagger & F_{2}(x)\Pi_x F_{2}(x)^\dagger & \cdots &F_{2}(x)\Pi_x F_{n}(x)^\dagger\\
   \vdots &  &\ddots  &\vdots\\
   F_{n}(x)\Pi_x F_{1}(x)^\dagger & F_{n}(x)\Pi_x F_{2}(x)^\dagger & \cdots &F_{n}(x)\Pi_x F_{n}(x)^\dagger
    \end{pmatrix} \geq 0
\end{align*}
For each $(jk)$ element, we have
\begin{align*}
&\sum_{x\in\mathcal{X}}F_{j}(x)\Pi_x F_{k}(x)^\dagger\\
&=\sum_{x\in\mathcal{X}}(F_j-f_j(x) I)\Pi_x (F_k-f_k(x) I)^\dagger\\
&=F_j\sum_{x\in\mathcal{X}}\Pi_x F_k^\dagger-F_j\sum_{x\in\mathcal{X}}\Pi_xf_k(x)^*-\sum_{x\in\mathcal{X}}f_j(x)\Pi_x F_k^\dagger+\sum_{x\in\mathcal{X}}f_j(x)\Pi f_k(x)^*\\
&=F_j F_k^\dagger-F_jF_k^\dagger-F_j F_k^\dagger+\sum_{x\in\mathcal{X}}f_j(x)\Pi f_k(x)^*\\
&=\sum_{x\in\mathcal{X}}f_j(x)\Pi_x f_k(x)^*-F_j F_k^\dagger. 
\end{align*}
The above matrix inequality is then expressed as
\begin{align*}
&   \begin{pmatrix}
   \sum_{x\in\mathcal{X}}f_1(x)\Pi_x f_1(x)^*-F_1 F_1^\dagger & \sum_{x\in\mathcal{X}}f_1(x)\Pi_x f_2(x)^*-F_1 F_2^\dagger & \cdots &\sum_{x\in\mathcal{X}}f_1(x)\Pi_x f_n(x)^*-F_1 F_n^\dagger\\
   \sum_{x\in\mathcal{X}}f_2(x)\Pi_x f_1(x)^*-F_2 F_1^\dagger & \sum_{x\in\mathcal{X}}f_2(x)\Pi_x f_2(x)^*-F_2 F_2^\dagger & \cdots &\sum_{x\in\mathcal{X}}f_2(x)\Pi_x f_n(x)^*-F_2 F_n^\dagger\\ 
     \vdots &  &\ddots  &\vdots\\
   \sum_{x\in\mathcal{X}}f_n(x)\Pi_x f_1(x)^*-F_n F_1^\dagger & \sum_{x\in\mathcal{X}}f_n(x)\Pi_x f_2(x)^*-F_n F_2^\dagger & \cdots &\sum_{x\in\mathcal{X}}f_n(x)\Pi_x f_n(x)^*-F_n F_n^\dagger
    \end{pmatrix} \geq 0\\
\Leftrightarrow&\sum_{x\in\mathcal{X}} 
\begin{pmatrix}
f_1(x)\Pi_x f_1(x)^* & f_1(x)\Pi_x f_2(x)^* & \cdots &f_1(x)\Pi_x f_n(x)^*\\
   f_2(x)\Pi_x f_1(x)^* & f_2(x)\Pi_x f_2(x)^* & \cdots &f_2(x)\Pi_x f_n(x)^*\\ 
     \vdots &\ddots  &  &\vdots\\
   f_n(x)\Pi_x f_1(x)^* & f_n(x)\Pi_x f_2(x)^* & \cdots &f_n(x)\Pi_x f_n(x)^*
\end{pmatrix}\geq
\begin{pmatrix}
 F_1 F_1^\dagger &F_1 F_2^\dagger & \cdots &F_1 F_n^\dagger\\
   F_2 F_1^\dagger & F_2 F_2^\dagger & \cdots &F_2 F_n^\dagger\\ 
     \vdots &  &\ddots  &\vdots\\
   F_n F_1^\dagger & F_n F_2^\dagger & \cdots &F_n F_n^\dagger
    \end{pmatrix}\\
\Leftrightarrow&\sum_{x\in\mathcal{X}} 
\begin{pmatrix}
f_1(x)\Pi_x f_1(x)^* & f_1(x)\Pi_x f_2(x)^* & \cdots &f_1(x)\Pi_x f_n(x)^*\\
   f_2(x)\Pi_x f_1(x)^* & f_2(x)\Pi_x f_2(x)^* & \cdots &f_2(x)\Pi_x f_n(x)^*\\ 
     \vdots &\ddots  &  &\vdots\\
   f_n(x)\Pi_x f_1(x)^* & f_n(x)\Pi_x f_2(x)^* & \cdots &f_n(x)\Pi_x f_n(x)^*
\end{pmatrix}\geq
\begin{pmatrix}
F_1\\F_2\\ \vdots\\ F_n
\end{pmatrix}
(F_1^\dagger, F_2^\dagger,\ldots,F_n^\dagger).
\end{align*}

Setting $f_j(x) = \hat{\theta}_j(x)$ gives $F_j= \hat{X}_j[\Pi,\hat{\theta}]$ 
and $\sum_{x\in\mathcal{X}}f_j(x)\Pi_x f_k(x)^*=\hat{\mathbb{L}}_{jk}[\Pi,\hat{\theta}]$. 
Therefore, we prove Lemma \ref{lemmaNH1}. 
\end{proof}

 \begin{lemma}
 \label{lemmaNH2}For all POVMs and estimators, which are locally unbiased at $\theta$, $\hat{X}[\Pi,\hat{\theta}]$ satisfy the following conditions:
 \begin{equation}
     \begin{split}
         \mathrm{tr}\big\{\hat{X}_j[\Pi,\hat{\theta}]S\big\}&=\theta_j\\
         \mathrm{tr}\big\{\hat{X}_j[\Pi,\hat{\theta}]\partial_kS\big\}&=\delta_{jk}\\
     \end{split}
 \end{equation}

\end{lemma}
\begin{proof}[\textbf{Proof (Lemma~\ref{lemmaNH2})}]
    For a POVM and estimator to be globally unbiased requires
     \begin{equation}
     \begin{split}
\sum_{x\in\mathcal{X}}\hat{\theta}_j(x)\mathrm{tr}\big\{\Pi_x.S\big\}&=\theta_j\quad\forall j\;,
     \end{split}
 \end{equation}
 for all possible values of $\theta$. By Taylor expanding around the true value of $\theta$ we obtain the locally unbiased conditions
\begin{equation}
     \begin{split}
\sum_{x\in\mathcal{X}}\hat{\theta}_j(x)\mathrm{tr}\big\{\Pi_x.S\big\}&=\theta_j\\
\sum_{x\in\mathcal{X}}\hat{\theta}_j(x)\mathrm{tr}\big\{\Pi_x.\partial_kS\big\}&=\delta_{jk}\;.
     \end{split}
 \end{equation}
From the definition of $\hat{X}_j$ (Eq.~\eqref{eq:apenLdef1}) this lemma immediately follows.
 \end{proof}

\subsection{The Nagaoka-Hayashi bound}

\begin{theorem}[\textbf{The Nagaoka--Hayashi bound}]
\label{theoremNH1}
For any POVM $\Pi$ and estimator $\hat{\theta}$ that are locally unbiased, the following lower bound holds for the weighted trace of the MSE matrix.
\begin{equation}
\begin{split}    
    &\mathrm{Tr}[WV(\theta)]\geq\mathcal{C}_\text{NH}\\
    &\mathcal{C}_\text{NH}\coloneq \min_{\mathbb{L},X}\mathbb{T}\mathrm{r}\big[ \mathbb{S}\mathbb{L}\big] -\mathbb{T}\mathrm{r}\big[DX\big]  -\mathbb{T}\mathrm{r}\big[X D\big] +\mathrm{w}
\end{split}
\end{equation}
Here optimization is subject to the constraints: $\forall j,k \quad\mathbb{L}_{jk}=\mathbb{L}_{kj}$ , $\mathbb{L}_{jk}$ is Hermitian, $X_j$ is Hermitian, and $\mathbb{L}\geq XX^\intercal$. Additionally, $X_j$ satisfies the locally unbiased condition: $\mathrm{tr}\big\{X_jS\big\}=\theta_j$ and $\mathrm{tr}\big\{X_j\partial_kS\big\}=\delta_{jk}$.
\end{theorem}

\begin{proof}[\textbf{Proof (Theorem~\ref{theoremNH1})}]
The proof is based on applying Lemma~\ref{lemmaNH1} and Lemma~\ref{lemmaNH2} to the alternative expression for the weighted trace of the MSE matrix together with the conditions in Eq.~\eqref{eq:Ldef2}. We have the following chains of inequalities
\begin{equation}
    \begin{split}
&\operatorname{Tr}\!\left[ W \, V(\theta) \right]
    \;\;\ge\;\;
    \min_{\Pi,\hat{\theta} }\left\{
    \operatorname{Tr}\!\left[ W \, V(\theta) \right]\bigg| \Pi,\hat{\theta}:\text{l.u.~at } \theta\right\}\\
    &=\min_{\Pi,\hat{\theta} }\left\{\mathbb{T}\mathrm{r}\big[ \mathbb{S}\hat{\mathbb{L}}[\Pi,\hat{\theta}]\big] -\mathbb{T}\mathrm{r}\big[D\hat{X}[\Pi,\hat{\theta}]^\intercal\big]  -\mathbb{T}\mathrm{r}\big[\hat{X}[\Pi,\hat{\theta}]D^\intercal\big] +\mathrm{w}\bigg|\Pi,\hat{\theta}:\text{l.u.~at } \theta\right\}\\
    &=\min_{\Pi,\hat{\theta} }\bigg\{\mathbb{T}\mathrm{r}\big[ \mathbb{S}\hat{\mathbb{L}}[\Pi,\hat{\theta}]\big] -\mathbb{T}\mathrm{r}\big[D\hat{X}[\Pi,\hat{\theta}]^\intercal\big]  -\mathbb{T}\mathrm{r}\big[\hat{X}[\Pi,\hat{\theta}]D^\intercal\big] +\mathrm{w}  \bigg|\Pi,\hat{\theta}:\text{l.u.~at } \theta, \hat{\mathbb{L}}[\Pi,\hat{\theta}]\geq \hat{X}[\Pi,\hat{\theta}]\left(\hat{X}[\Pi,\hat{\theta}]\right)^\intercal\bigg\}\\
    &=\min_{\mathbb{L},X}\bigg\{\mathbb{T}\mathrm{r}\big[ \mathbb{S}\mathbb{L}\big] -\mathbb{T}\mathrm{r}\big[DX^\intercal\big]  -\mathbb{T}\mathrm{r}\big[XD^\intercal\big] +\mathrm{w} \bigg|\exists \Pi,\hat{\theta} ,\mathbb{L}=\hat{\mathbb{L}}[\Pi,\hat{\theta}],X=\hat{X}[\Pi,\hat{\theta}],\mathrm{tr}\{X_jS\}=\theta_j,\text{tr}\{X_j\partial_kS\}=\delta_{jk}, \mathbb{L}\geq XX^\intercal\bigg\}\\
    &\geq\min_{\mathbb{L},X}\bigg\{\mathbb{T}\mathrm{r}\big[ \mathbb{S}\mathbb{L}\big] -\mathbb{T}\mathrm{r}\big[DX^\intercal\big]  -\mathbb{T}\mathrm{r}\big[XD^\intercal\big] +\mathrm{w} \bigg|\mathbb{L}_{jk}=\mathbb{L}_{kj}:\text{Hermitian},X_j:\text{Hermitian},\mathrm{tr}\{X_jS\}=\theta_j,\text{tr}\{X_j\partial_kS\}=\delta_{jk}, \mathbb{L}\geq XX^\intercal\bigg\}\\
   & =\mathcal{C}_\text{NH}
    \end{split}
\end{equation}
\end{proof}

\subsection{Alternative form for the Nagaoka-Hayashi bound}

By introducing different variables, we can also express the Nagaoka-Hayashi bound in the following form.

\begin{theorem}[\textbf{The Nagaoka--Hayashi bound}]
\label{theoremapenNH2} For any POVM $\Pi$ and estimator $\hat{\theta}$ that are locally unbiased the following lower bound
holds for the weighted trace of the MSE matrix:
\begin{equation}
    \begin{split}
        &\mathrm{Tr}[WV(\theta)]\geq\mathcal{C}_\mathrm{NH}\\
        &\mathcal{C}_\mathrm{NH}=\min_{\mathbb{L},X}\mathbb{T}\mathrm{r}[\mathbb{S}\mathbb{L}]
    \end{split}
\end{equation}

Here optimization is subject to the constraints: $\forall j,k \quad\mathbb{L}_{jk}=\mathbb{L}_{kj}$ , $\mathbb{L}_{jk}$ is Hermitian, $X_j$ is Hermitian, and $\mathbb{L}\geq XX^\intercal$. Additionally, $X_j$ satisfies the locally unbiased condition: $\mathrm{tr}\big\{X_jS\big\}=0$ and $\mathrm{tr}\big\{X_j\partial_kS\big\}=\delta_{jk}$.
\end{theorem}
Remark: The differences from Theorem~\ref{theoremNH1} are (i) the objective function and (ii) the locally unbiasedness
condition on $X$.

    \begin{proof}[\textbf{Proof (Theorem~\ref{theoremapenNH2})}]
Define the new Hermitian variables by
\begin{equation}
\begin{split}    
    \tilde{X}&=X_j-\theta_jI\quad(j=1,2,... n)\\
    \tilde{\mathbb{L}}_{jk}&=\mathbb{L}_{jk}-\theta_jX_k-X_j\theta_k+\theta_j\theta_kI\quad(j,k=1,2,... n)
    \end{split}
\end{equation}
We see that the three conditions required from Eq.~\eqref{eq:Ldef2} are invariant:
\begin{equation}
\begin{split}
\label{eq:Ldef3}
  &\tilde{\mathbb{L}}_{jk}
=\tilde{\mathbb{L}}_{kj}\qquad (j,k = 1,2,...n)\\&
  \tilde{\mathbb{L}}_{jk}
: \text{Hermitian}\qquad (j,k = 1,2,...n)\\&\tilde{X}_j: \text{Hermitian} \qquad (j = 1,2,...n)
  \end{split}
\end{equation}
Next, the two conditions required by the locally unbiasedness condition of Lemma 2 are changed to
 \begin{equation}
     \begin{split}
\mathrm{tr}\big\{\tilde{X}_jS\big\}&=0\quad (j = 1,2,...n)\\
\mathrm{tr}\big\{\tilde{X}_j\partial_kS\big\}&=\delta_{jk}\quad (j,k = 1,2,...n)\\
     \end{split}
 \end{equation}
The first condition follows directly from the definition of $\tilde{X}_j$. The second condition follows from $\text{tr}\{(\tilde{X}_j+\theta_jI)\partial_kS\}=\delta_{jk}$ and the fact that
\begin{equation}
    \text{tr}\{\theta_jI\partial_kS\}=\theta_j\text{tr}\{\partial_kS\}=\theta_j\partial_k\text{tr}\{S\}=\theta_j\partial_k(1)=0\;.
\end{equation}
Next, the condition $\mathbb{L}\geq XX^\intercal$ is also invariant due to the identity:
\begin{equation}
    \mathbb{L}-XX^\intercal=\tilde{\mathbb{L}}-\tilde{X}\tilde{X}^\intercal\;.
\end{equation}
This is shown as follows. Let us define a column vector of matrices by $X_\theta=(\theta_1I,\theta_2I,...,\theta_nI)^\intercal$ , then we
have
\begin{equation}
\begin{split}
    &\tilde{X}=X-X_\theta\\
    &\tilde{\mathbb{L}}=\mathbb{L}-X_\theta X^\intercal-XX_\theta^\intercal+X_\theta X_\theta^\intercal\;.
\end{split}
\end{equation}
Substitution gives
\begin{equation}
    \begin{split}
        \tilde{\mathbb{L}}-\tilde{X}\tilde{X}^\intercal&=[\mathbb{L}-X_\theta X^\intercal-XX_\theta^\intercal+X_\theta X_\theta^\intercal]-(X-X_\theta)(X-X_\theta)^\intercal\\
        &=\mathbb{L}-XX^\intercal \;.
    \end{split}
\end{equation}
Lastly, the objective function is expressed in terms of new variables as
\begin{equation}
    \begin{split}
        \mathbb{T}\mathrm{r}\left[\mathbb{S}\tilde{\mathbb{L}}\right]&=\sum_{j,k=1}^{n}\text{tr}\left\{\mathbb{S}_{jk}\tilde{\mathbb{L}}_{jk}\right\}\\
        &=\sum_{j,k=1}^{n}\text{tr}\left\{W_{jk}S\left\{\mathbb{L}_{jk}-\theta_jX_k-X_j\theta_k+\theta_j\theta_kI\right\}\right\}\\
        &=\sum_{j,k=1}^{n}\text{tr}\left\{W_{jk}S\mathbb{L}_{jk}\right\}
        -\sum_{k=1}^{n}\text{tr}\{D_kX_k\}-\sum_{j=1}^{n}\text{tr}\{X_jD_j\}+\mathrm{w}\\
        &=\mathbb{T}\mathrm{r}[\mathbb{S}\mathbb{L}]-\mathbb{T}\mathrm{r}[DX^\intercal]-\mathbb{T}\mathrm{r}[XD^\intercal]+\mathrm{w}
    \end{split}
\end{equation}
Combining all these elements, we prove theorem~\ref{theoremapenNH2}.
\end{proof}
\end{widetext}

By carefully comparing theorems~\ref{theoremNH1} and~\ref{theoremapenNH2}, we see that the form of the NHCRB presented in Ref.~\cite{conlon2021efficient} corresponds to the latter theorem. As such, the unbiased conditions presented in Eq.~(4) of Ref.~\cite{conlon2021efficient} is incorrect, and should read $\mathrm{tr}\big\{X_jS\big\}=0$ (as opposed to $\mathrm{tr}\big\{X_jS\big\}=\theta_j$).

For clarity, we now illustrate how the NHCRB can be applied to determine a bound on $M$-copy entangling measurements. Observe that there is nothing special about the Hilbert space $H$ on which $S$ exists. Therefore, one can compute the original NHCRB using $\bbS=I\otimes S^{\otimes M}$, and change the unbiased conditions to incorporate $S_i=\partial S^{\otimes M}/\partial\theta_i$. From the above derivation, it follows that the computed quantity $\mathcal{C}_{\text{NH}}(M)$ is a bound on the MSE attainable by any POVM acting on $\mathcal{H}^{\otimes M}$.

\section{Proof of basic results}
In this appendix we prove some basic results, necessary for the main results. In Sec.~\ref{apen:hol2copy} we discuss the optimisers for the two-copy HCRB. In Sec.~\ref{apen:subadditivity}, we discuss the subadditivity of the NCRB. In Secs.~\ref{apen:strictconvex} and \ref{subsec:MultipleSols} we discuss the multiplicity of solutions to the HCRB.

\subsection{Optimal $X$ matrices for two-copy HCRB}
\label{apen:hol2copy}
For simplicity we will denote the optimal estimator operators in the single-copy case as $X$ and $Y$, and in the two-copy case as $X_2$ and $Y_2$, given in Eq.~\eqref{eq:twocopyest}. We now demonstrate that the two-copy estimator operators do in fact attain a HCRB which is half that of the single-copy case. First note that
\begin{equation}
\text{tr}\left\{ (S\otimes S)(X_2^2+Y_2^2)\right\}=\text{tr}\left\{ S(X^2+Y^2)\right\}/2\;,
\end{equation}
i.e. the first term in the HCRB is two times smaller as expected. The second term in the HCRB is the sum of the eigenvalues of $\mathrm{i}(\sqrt{S}\otimes\sqrt{S})([X_2,Y_2])(\sqrt{S}\otimes\sqrt{S})$. We can write
\begin{equation}
\begin{split}
&\mathrm{i}(\sqrt{S}\otimes\sqrt{S})([X_2,Y_2])(\sqrt{S}\otimes\sqrt{S})\\
=&\frac{\mathrm{i}}{4}(S\otimes(\sqrt{S}([X,Y])\sqrt{S})+(\sqrt{S}([X,Y])\sqrt{S})\otimes S)\;.
\end{split}
\end{equation}
The sum of the eigenvalues of the above matrix is given by the trace of this matrix. We then see that
\begin{equation}
\begin{split}
&\text{tr}\{\mathrm{i}(\sqrt{S}\otimes\sqrt{S})([X_2,Y_2])(\sqrt{S}\otimes\sqrt{S})\}\\
=&\frac{\mathrm{i}}{4}\text{tr}\{S\otimes(S[X,Y])+(S[X,Y])\otimes S\}\\
=&\frac{\mathrm{i}}{4}\text{tr}\left\{S[X,Y]\right\}+\frac{\mathrm{i}}{4}\text{tr}\left\{S[X,Y]\right\}\\
=&\frac{\mathrm{i}}{2}\text{tr}\left\{S[X,Y]\right\}\;.
\end{split}
\end{equation}
Therefore, the second term in the HCRB is also a factor of two smaller confirming that the new matrices $X_2$ and $Y_2$ achieve the value $\mathcal{C}_\text{H}/2$ for the two-copy case. Similarly, by direct substitution it can be verified that the matrices $X_2$ and $Y_2$ satisfy the unbiased conditions in the two-copy case:
\begin{equation}
\begin{split}
&\text{tr}\{S\otimes S X_2\}\\
=&\frac{1}{2}\text{tr}\{S\otimes S (X\otimes \mathbb{I}+\mathbb{I}\otimes X)\}\\
=&\text{tr}\{SX\}=0\;,
\end{split}
\end{equation}
and similarly $\text{tr}\{(S\otimes S )Y_2\}=0$. Second, we have
\begin{equation}
\begin{split}
&\text{tr}\{(S\otimes S_i+S_i\otimes S) X_2\}\\
=&\frac{1}{2}\text{tr}\{(S\otimes S_i+S_i\otimes S) (X\otimes \mathbb{I}+\mathbb{I}\otimes X)\}\\
=&\frac{1}{2}\text{tr}\{(S\otimes S_i)(X\otimes \mathbb{I})+(S_i\otimes S)(X\otimes \mathbb{I})\\
&\qquad+(S\otimes S_i)( \mathbb{I}\otimes X)+(S_i\otimes S)( \mathbb{I}\otimes X)\}\\
=&\frac{1}{2}\text{tr}\{SX\}\text{tr}\{ S_i\}+\frac{1}{2}\text{tr}\{S_iX\}\text{tr}\{ S\}\\
&+\frac{1}{2}\text{tr}\{ S\}\text{tr}\{S_iX\}+\frac{1}{2}\text{tr}\{ S_i\}\text{tr}\{SX\}\\
=&\text{tr}\{S_iX\}=\delta_{x,i}\;,
\end{split}
\end{equation}
using the facts that $\text{tr}\{S\}=1$, $\text{tr}\{S_i\}=0$, $\text{tr}\{SX\}=0$, and $\text{tr}\{S_iX\}=\delta_{x,i}$. $\text{tr}\{(S\otimes S_i+S_i\otimes S) Y_2\}=\delta_{y,i}$ can be verified in a similar manner. The extension to estimating $M$ copies is handled in a similar manner, substituting in $S^{\otimes M}$, $X_M$ and $Y_M$. Note that this only proves that $\mathcal{C}_\text{H}(2)\leq\mathcal{C}_\text{H}/2$, as there could be some other matrices $X_2'$ and $Y_2'$, which give a lower bound. For the full proof of the additivity of the HCRB, we refer the reader to Lemma 4 of Ref.~\cite{hayashi2008asymptotic}.

\subsection{Subadditivity of the NCRB}
\label{apen:subadditivity}
Lemma~\ref{lemma:mcopyhol} relates to the additivity of the HCRB, $\mathcal{C}_\text{H}(M)=\mathcal{C}_\text{H}/M$. However, for the NCRB, we have $\mathcal{C}_\text{N}(M)\leq\mathcal{C}_\text{N}/M$. We now show this to be true by evaluating $F_\text{N}(\mathbf{X}_{\text{N},2})$. From Appendix~\ref{apen:hol2copy} we know that the first term in the Nagaoka function evaluated with $\mathbf{X}_{\text{N},2}$ is exactly half of the corresponding term evaluated with $\mathbf{X}_{\text{N}}$. The second term in the Nagaoka function is 
\begin{equation}
\begin{split}
&\text{tr}\left\{ \abs{\sqrt{S\otimes S} [X_{2},Y_{2}]\sqrt{S\otimes S}}\right\}\\
=&\frac{1}{4}\text{tr}\left\{\abs{ \sqrt{S}[X,Y]\sqrt{S}\otimes S+ S\otimes \sqrt{S}[X,Y]\sqrt{S}}\right\}\\
\leq&\frac{1}{4}[\text{tr}\left\{\abs{ S\otimes\sqrt{ S}[X,Y]\sqrt{ S}}\right\}\\
&+\text{tr}\left\{\abs{\sqrt{ S}[X,Y]\sqrt{ S}\otimes S}\right\}]\\
=&\frac{1}{2}\text{tr}\left\{\abs{\sqrt{ S}[X,Y]\sqrt{ S}}\right\}\;,
\end{split}
\end{equation}
where the inequality follows from the sub-additive property of matrix norms, $\norm{A+B}\leq\norm{A}+\norm{B}$~\cite{horn2012matrix} (Here we use $\norm{A}=\text{tr}\{\abs{A}\}$). We therefore arrive at $\mathcal{C}_\text{N}(2)\leq F_\text{N}(\mathbf{X}_{\text{N},2})\leq\mathcal{C}_\text{N}/2$. This argument can then be repeated many times to get the result for any $M$.

\subsection{Multiplicity of solutions to the HCRB and SLDCRB}
\label{apen:strictconvex}
In this Appendix we discuss the uniqueness of the solutions to the HCRB and SLDCRB. Specifically, we demonstrate that while the solution to the SLDCRB is unique, the solution to the HCRB need not be.

We first introduce an equivalence relation for linear operators
\begin{equation}
\label{eqequiv}
\begin{split}
X&\sim_SX'\iff \\
S(X-X')=0\quad&\text{and}\quad(X-X')S=0\;.
\end{split}
\end{equation}
If $\mathcal{L}(\mathcal{H})$ denotes the set of all matrices, this allows us to define a quotient space of $\mathcal{L}(\mathcal{H})$ as $\mathcal{L}^{\sim_S}(\mathcal{H})=\mathcal{L}(\mathcal{H})/\sim_S$. For rank deficient states, the SLD operators are not uniquely determined. However, the SLD operators are unique on the quotient space defined here~\cite{liu2014quantum,fujiwara1995quantum}.

\begin{proof}
Let $X$ and $Y$ be two vectors of Hermitian estimator operators and
define
\begin{equation}
D=X-Y.
\end{equation}
For $0<t<1$, set $X_t=tX+(1-t)Y$. Since $Z[X]$ is quadratic
in $X$,
\begin{equation}
Z[X_t]
=
tZ[X]+(1-t)Z[Y]-t(1-t)Z[D].
\label{eq:Zconvex}
\end{equation}
Using the triangle inequality for the trace norm in the imaginary
part gives
\begin{align}
F_{\rm H}(X_t)
&\leq
tF_{\rm H}(X)+(1-t)F_{\rm H}(Y)
\nonumber\\
&\quad
-t(1-t)
\left\{
\Tr[\operatorname{Re}Z[D]]
-
\Tr\!\left[|\operatorname{Im}Z[D]|\right]
\right\}.
\label{eq:FHconvex}
\end{align}
To prove the convexity of $F_{\rm H}$, it therefore remains to show that
\begin{equation}
\Tr[\operatorname{Re}Z[D]]
\geq
\Tr\!\left[|\operatorname{Im}Z[D]|\right]\;.
\label{eq:keyconvexineq}
\end{equation}

To see this, make a real orthogonal transformation of the parameter
indices such that $\operatorname{Im}Z[D]$ is in its canonical
antisymmetric form,
\begin{equation}
\operatorname{Im}Z[D]
=
\bigoplus_{\ell}
\begin{pmatrix}
0 & \lambda_\ell\\
-\lambda_\ell & 0
\end{pmatrix}
\oplus 0,
\qquad
\lambda_\ell\geq0.
\end{equation}
Denote the transformed estimator operators by $\widetilde D_i$.
For each canonical pair, set
\begin{equation}
A=\widetilde D_{2\ell-1},
\qquad
B=\widetilde D_{2\ell}.
\end{equation}
Then
\begin{equation}
\lambda_\ell
=
\left|
\operatorname{Im}\tr\{SAB\}
\right|.
\end{equation}
Since $(A\pm iB)^\dagger(A\pm iB)\geq0$,
\begin{equation}
0
\leq
\tr\!\left\{
S(A\pm iB)^\dagger(A\pm iB)
\right\}
=
\tr\{S(A^2+B^2)\}
\mp
2\operatorname{Im}\tr\{SAB\}.
\end{equation}
Hence
\begin{equation}
\tr\{S(A^2+B^2)\}
\geq
2\lambda_\ell .
\label{eq:pairbound}
\end{equation}
Summing over all canonical pairs, together with any unpaired
operators, gives
\begin{equation}
\Tr[\operatorname{Re}Z[D]]
=
\sum_i\tr\{S\widetilde D_i^2\}
\geq
2\sum_\ell\lambda_\ell
=
\Tr\!\left[|\operatorname{Im}Z[D]|\right],
\end{equation}
which proves Eq.~\eqref{eq:keyconvexineq}. Substitution into
Eq.~\eqref{eq:FHconvex} proves convexity.

Now suppose $S>0$. Equality in Eq.~\eqref{eq:pairbound} for a
canonical pair implies, for one choice of sign,
\begin{equation}
\tr\!\left\{
S(A\pm iB)^\dagger(A\pm iB)
\right\}=0.
\end{equation}
Since $S>0$, this is possible only if
\begin{equation}
A\pm iB=0.
\end{equation}
As $A$ and $B$ are Hermitian, this implies $A=B=0$.
Likewise, for any unpaired nonzero Hermitian operator
$\widetilde D_i$,
\begin{equation}
\tr\{S\widetilde D_i^2\}>0.
\end{equation}
Therefore, if $S>0$ and $D\neq0$,
\begin{equation}
\Tr[\operatorname{Re}Z[D]]
>
\Tr\!\left[|\operatorname{Im}Z[D]|\right].
\end{equation}
Equation~\eqref{eq:FHconvex} then gives
\begin{equation}
F_{\rm H}(tX+(1-t)Y)
<
tF_{\rm H}(X)+(1-t)F_{\rm H}(Y)
\end{equation}
for all $X\neq Y$ and $0<t<1$. Thus $F_{\rm H}$ is strictly
convex when $S>0$.

Finally, the locally unbiased conditions define an affine, and hence
convex, feasible set. A strictly convex function has at most one
minimiser on a convex set. Therefore the HCRB optimiser is unique for
a full-rank probe state.
\end{proof}

\noindent


Similar arguments can be used to prove Lemma~\ref{prop:quotientSLD}, regarding the strict convexity of the SLDCRB. From Eq.~\eqref{equation:slddefinition} and the arguments above, the minimisation problem involved in computing the SLDCRB is minimising a sum of strictly convex functions, which is itself strictly convex.

\subsection{Example with multiple solutions to the HCRB}
\label{subsec:MultipleSols}
Here we present a simple example, demonstrating that the HCRB can have multiple solutions and is therefore not a strictly convex function. We consider the following probe state
\begin{equation}
S=\begin{pmatrix}
1-\epsilon&0&0\\
0&\epsilon&0\\
0&0&0
\end{pmatrix}\;,
\end{equation}
with derivatives
\begin{equation}
S_x=\begin{pmatrix}
0&0&\mathrm{i}\\
0&0&0\\
-\mathrm{i}&0&0
\end{pmatrix}\;,
\end{equation}
and
\begin{equation}
S_y=\begin{pmatrix}
0&0&1\\
0&0&0\\
1&0&0
\end{pmatrix}\;.
\end{equation}

Then the following matrices form a family of solutions to the HCRB, which are not equivalent
\begin{equation}
\label{eq:HCRBXnonconvex}
X=\frac{1}{2}\begin{pmatrix}
0&0&\mathrm{i}\\
0&0&\delta\\
-\mathrm{i}&\delta&0
\end{pmatrix}\;,
\end{equation}
and
\begin{equation}
\label{eq:HCRBYnonconvex}
Y=\frac{1}{2}\begin{pmatrix}
0&0&1\\
0&0&\delta\mathrm{i}\\
1&-\delta\mathrm{i}&0
\end{pmatrix}\;,
\end{equation}
for $0<\delta<\sqrt{(1-\epsilon)/\epsilon}$. This example gives a HCRB of $\mathcal{C}_\text{H}=1-\epsilon$. We can use this example to illustrate why the HCRB is not strictly convex. Consider two solutions to the HCRB $Z_1=X_1+\mathrm{i}Y_1$ and $Z_2=X_2+\mathrm{i}Y_2$ given by Eqs.~\eqref{eq:HCRBXnonconvex} and \eqref{eq:HCRBYnonconvex}, with $\delta=\delta_1$ and $\delta=\delta_2$ respectively. Then $\text{tr}\{S(Z_1-Z_2)(Z_1-Z_2)^\dagger\}=0$ for \mbox{$0<\delta_1<\delta_2<\sqrt{(1-\epsilon)/\epsilon}$}.

\section{Proof of two parameter estimation results}
In this appendix we prove the main two parameter results. We begin by proving theorem~\ref{theorem:main} (that a gap persistence theorem does not hold between the HCRB and NCRB in general for rank deficient states) in Sec.~\ref{apen:SGPmiHCRB}. In Sec.~\ref{apen:asymptotic}, we prove the asymptotic equivalence of the HCRB and NCRB.

\subsection{Example proving Theorem~\ref{theorem:main}.}
\label{apen:SGPmiHCRB}
We now present an example proving that the gap persistence theorem does not hold for the NCRB and the HCRB. We do this by presenting a specific example where $\mathcal{C}_\text{N}>\mathcal{C}_\text{H}$  but $\mathcal{C}_\text{N}(2)=\mathcal{C}_\text{H}(2)$ .

We consider the following density matrix
\begin{equation}
    S=\frac12\begin{pmatrix}
        1&0&0&0\\
        0&1&0&0\\
        0&0&0&0\\
        0&0&0&0\\
    \end{pmatrix}\;,
\end{equation}
such that the projector onto the support of $S$ is given by $P=2S$. Next define
\begin{equation}
   Z= \begin{pmatrix}
     A&B\\C&0   
    \end{pmatrix}\;,
\end{equation}
with
\begin{equation}
    \begin{split}
        A=&\begin{pmatrix}
            0&0\\1&0
        \end{pmatrix}\\
        B=&\frac45\begin{pmatrix}
            1&0\\0&1
        \end{pmatrix}\\
        C=&\begin{pmatrix}
            1&0\\0&1
        \end{pmatrix}\;.
    \end{split}
\end{equation}
The derivatives for our problem are given by
\begin{equation}
\begin{split}
        S_x&=\frac{10}{23}G_x\\
            S_y&=\frac{-10i}{7}G_y\;,
\end{split}
\end{equation}
with $G_x=(ZS+SZ^\dagger)$ and $G_y=(ZS-SZ^\dagger)$. Note that although we have simply stated the derivatives at a particular point, it is simple to construct a statistical model based on this.

\subsubsection{Single-copy HCRB}
The optimisers for the HCRB are
\begin{equation}
        \label{counterexampleoptimisers}X_\text{H}=\frac{Z+Z^\dagger}{2}\;,\;Y_\text{H}=\frac{Z-Z^\dagger}{2i}\;.
\end{equation}
One can verify by direct calculation that $X_\text{H}$ and $Y_\text{H}$ satisfy the unbiased conditions, and that for this particular solution
\begin{equation}
    F_\text{H}(\mathbf{X})=\frac{3}{2}\;.
\end{equation}

We now prove that this is the minimum possible value of $F_\text{H}(\mathbf{X})$, proving that this solution is optimal. In order to do so, we note that for $Z=X+\mathrm{i}Y$ the two-parameter HCRB can be written as ($Z$ here should not be confused with $Z[\mathbf{X}]$ in Eq.~\eqref{eq_zmat})
 \begin{align}
 \label{eq:HCRBconvexform}
\mathcal{C}_\text{H}= \min_{Z}\max\Big\{ q_+(Z),q_-(Z) \Big\}\,,
 \end{align}
with
\begin{equation}
    \begin{split}
        q_+&=\tr{S Z^\dagger Z}\\q_-&=\tr{S Z Z^\dagger}\;.
    \end{split}
\end{equation}
Note that for the optimisers in Eq.~\eqref{counterexampleoptimisers} we have
\begin{equation}
    \begin{split}
        q_+(Z)=\frac12\text{Tr}[A^\dagger A+C^\dagger C]=\frac32\\q_-(Z)=\frac12\text{Tr}[A^\dagger A+B^\dagger B]<\frac32\;.
    \end{split}
\end{equation}

Next let $Z'=Z+\Delta$ be any other locally unbiased estimator with $\Delta=\Delta X+i\Delta Y$ where ($\Delta X,\Delta Y$) satisfy the locally unbiased conditions. We have
\begin{equation}
\begin{split}
        q_+(Z+\Delta)=&q_+(Z)+q_+(\Delta)+\text{Tr}[SZ^\dagger \Delta]+\text{Tr}[S\Delta^\dagger Z]\\
        =&q_+(Z)+q_+(\Delta)+\text{Tr}[SZ^\dagger (\Delta X+i\Delta Y)]\\
        &+\text{Tr}[S(\Delta X-i\Delta Y) Z]\\
        =&q_+(Z)+q_+(\Delta)+\text{Tr}[(SZ^\dagger +ZS)\Delta X]\\
        &+i\text{Tr}[(SZ^\dagger-ZS)\Delta Y]\\
        =&q_+(Z)+q_+(\Delta)+\text{Tr}[G_x\Delta X]-i\text{Tr}[G_y\Delta Y]
\end{split}
\end{equation}
As $S_i$ is proportional to $G_i$, to ensure the unbiased conditions remain satisfied we require $\text{Tr}[G_x\Delta X]=0$ and $\text{Tr}[G_y\Delta Y) ]=0$. Therefore, we have $q_+(Z+\Delta)\geq q_+(Z)$, proving optimality of our solution.

We can further characterise the class of optimisers for the HCRB. We require $q_+(\Delta)=0$, which is equivalent to $\Delta P=0$, meaning all single-copy HCRB optimisers have the following form
\begin{equation}
    \label{Zgeneral}Z_\text{H}+\Delta=\begin{pmatrix}
        A&B+E\\C&F
    \end{pmatrix}\;,
\end{equation}
with $\text{Tr}[E]=0$.

\subsubsection{Single-copy NCRB}
We next prove that the single-copy NCRB is strictly larger than the HCRB for this example. Recall Lemma~\ref{lemm:equality_cond} that gives the conditions for $F_\text{H}(\mathbf{X})=F_\text{N}(\mathbf{X})$. In terms of $Z=X+iY$, the equivalent condition is that
\begin{equation}
\label{lemma5equiv}
    P(Z^\dagger Z-ZZ^\dagger)P
\end{equation}
is positive semidefinite (or negative semidefinite). For $Z$ in Eq.~\eqref{Zgeneral}, this condition becomes that
\begin{equation}
    A^\dagger A+C^\dagger C-A A^\dagger -(B+E)(B+E)^\dagger
\end{equation}
must be positive semidefinite on the support space. Note that on this space
\begin{equation}
    D=A^\dagger A+C^\dagger C-A A^\dagger=\begin{pmatrix}
       2 &0\\0&0
    \end{pmatrix}\;.
\end{equation}
Defining $M=B+E$ we then require $D-MM^\dagger\succeq0$. Observe that the 2,2 component of this matrix is
\begin{equation}
    (D-MM^\dagger)_{2,2}=-(|M_{21}|^2+|M_{22}|^2)\;.
\end{equation}
For this to be non-negative we require $M_{21}=M_{22}=0$ which forces $E_{21}=0$ and $E_{22}=-4/5$. $\text{Tr}[E]=0$ then imposes $E_{11}=4/5$. We then have 
\begin{equation}
\begin{split}
       (D-MM^\dagger)_{1,1}&=2-(|M_{11}|^2+|M_{12}|^2)\\&\leq 2-\left(\frac85\right)^2\\
       &=-\frac{14}{25}<0\;.
\end{split}
\end{equation}
Therefore, Eq.~\eqref{lemma5equiv} cannot be positive semidefinite. As $q_+\geq q_-$, Eq.~\eqref{lemma5equiv} cannot be negative semidefinite either. Therefore, the condition for equality between the HCRB and NCRB cannot be satisfied for this example, and so we have
\begin{equation}
    \mathcal{C}_\text{N}>\mathcal{C}_\text{H}=\frac32\;.
\end{equation}
Interestingly, we find numerically that $\mathcal{C}_\text{N}\approx 1.50008$ for this example (numerical precision is better than $10^{-12}$), and so the two bounds are close.

\subsubsection{Two-copy bounds}
The two-copy HCRB follows trivially from the additivity of the HCRB, $\mathcal{C}_\text{H}(2)=3/4$. We now present an optimiser $Z_2$ such that $F_\text{N}(Z_2)=F_\text{H}(Z_2)=\mathcal{C}_\text{H}(2)$. For simplicity, we work in the reordered basis $(P\otimes P)\oplus(Q\otimes P)\oplus (P\otimes Q)\oplus(Q\otimes Q)$ where $Q=I-P$.
\begin{equation}
\alpha=\frac{8(\sqrt{2}-1)}{5},
\qquad
\beta=\frac{4(2-\sqrt{2})}{5}.
\end{equation}
\begin{widetext}
\begin{equation}
Z_2=
\begin{pmatrix}
0&0&0&0&\alpha&0&0&0&\alpha&0&0&0&0&0&0&0\\
\frac12&0&0&0&0&\beta&0&0&0&\beta&0&0&0&0&0&0\\
\frac12&0&0&0&0&0&\beta&0&0&0&\beta&0&0&0&0&0\\
0&\frac12&\frac12&0&0&0&0&0&0&0&0&0&0&0&0&0\\

\frac12&0&0&0&0&0&0&0&0&0&0&0&\frac25&0&0&0\\
0&\frac12&0&0&\frac12&0&0&0&0&0&0&0&0&\frac25&0&0\\
0&0&\frac12&0&0&0&0&0&0&0&0&0&0&0&\frac25&0\\
0&0&0&\frac12&0&0&\frac12&0&0&0&0&0&0&0&0&\frac25\\

\frac12&0&0&0&0&0&0&0&0&0&0&0&\frac25&0&0&0\\
0&\frac12&0&0&0&0&0&0&0&0&0&0&0&\frac25&0&0\\
0&0&\frac12&0&0&0&0&0&\frac12&0&0&0&0&0&\frac25&0\\
0&0&0&\frac12&0&0&0&0&0&\frac12&0&0&0&0&0&\frac25\\

0&0&0&0&\frac12&0&0&0&\frac12&0&0&0&0&0&0&0\\
0&0&0&0&0&\frac12&0&0&0&\frac12&0&0&0&0&0&0\\
0&0&0&0&0&0&\frac12&0&0&0&\frac12&0&0&0&0&0\\
0&0&0&0&0&0&0&\frac12&0&0&0&\frac12&0&0&0&0
\end{pmatrix}.
\end{equation}
\end{widetext}
One can verify by direct computation that this satisfies both the unbiased conditions and $F_\text{N}(Z_2)=F_\text{H}(Z_2)=\mathcal{C}_\text{H}(2)$.

\subsection{Example proving theorem~\ref{theorem:mainHCRBMI}}
\label{apen2parMIHCRB}
Building on the above example, we note that if we could find an optimal POVM saturating the two-copy NCRB, this would immediately prove that a strong gap persistence theorem does not hold between the most informative bound and the HCRB. This is because we have $\mathcal{C}_\text{MI}\geq\mathcal{C}_\text{N}>\mathcal{C}_\text{H}$, but a matching optimal 2-copy POVM would prove $\mathcal{C}_\text{MI}(2)=\mathcal{C}_\text{N}(2)=\mathcal{C}_\text{H}(2)$. In this section we provide such a POVM. More precisely, we provide an alternative solution $Z_\text{P}$, such that $Z_\text{P}^\dagger Z_\text{P}=Z_\text{P}Z_\text{P}^\dagger$, implying that the corresponding $X$ and $Y$ commute. Therefore, $X$ and $Y$ can be simultaneously decomposed into the same POVM basis and can be jointly measured optimally.

Let
\begin{equation}
\lvert ij\rangle_{\pm}:=
\frac{\lvert ij\rangle\pm\lvert ji\rangle}{\sqrt2}
\qquad (i\neq j).
\end{equation}
Use the orthonormal basis
\begin{equation}
\begin{aligned}
\mathcal B_+=\bigg(&
\lvert00\rangle,\,
\lvert01\rangle_+,\,
\lvert11\rangle,\,
\lvert20\rangle_+,\\
&\frac{\lvert21\rangle_++\lvert30\rangle_+}{\sqrt2},\,
\frac{\lvert21\rangle_+-\lvert30\rangle_+}{\sqrt2},\,\\&
\lvert31\rangle_+,
\lvert22\rangle,\,
\lvert23\rangle_+,\,
\lvert33\rangle
\bigg),
\end{aligned}
\end{equation}
for the swap-symmetric sector, and
\begin{equation}
\begin{split}
    \mathcal B_-=\bigg(&
\lvert01\rangle_-,\,
\frac{\lvert21\rangle_- -\lvert30\rangle_-}{\sqrt2},\,
\lvert20\rangle_-,\\&\,
\frac{\lvert21\rangle_-+\lvert30\rangle_-}{\sqrt2},\,
\lvert31\rangle_-,\,
\lvert23\rangle_-
\bigg)
\end{split}
\end{equation}
for the swap-antisymmetric sector.  Let
\begin{equation}
a=\frac1{\sqrt2},\qquad
b=\frac{3\sqrt2}{5},\qquad
r=\frac{\sqrt7}{5},\qquad
x=-\frac{\sqrt{2177}}{50}.
\end{equation}

Define
\begin{equation}
Z_{\mathrm P}=Z_+\oplus Z_-
\end{equation}
in the basis \(\mathcal B_+\oplus\mathcal B_-\), where
\begin{equation}
Z_+=
\begin{pmatrix}
A&B\\
C&F
\end{pmatrix},
\end{equation}
with
\begin{equation}
A=
\begin{pmatrix}
0&0&0\\
a&0&0\\
0&a&0
\end{pmatrix},
\qquad
B=
\begin{pmatrix}
b&0&0&0&r&0&0\\
0&a&0&0&0&0&0\\
0&0&0&0&0&0&0
\end{pmatrix},
\end{equation}
\begin{equation}
C=
\begin{pmatrix}
a&0&0\\
0&a&0\\
0&0&0\\
0&0&a\\
0&0&0\\
0&0&0\\
0&0&0
\end{pmatrix},
\qquad
F=F_5\oplus 0_2,
\end{equation}
and
\begin{equation}
F_5=
\begin{pmatrix}
x&-\frac a2&b&0&-\frac{x}{\sqrt{14}}\\[1mm]
\frac a2&\frac{15x}{14}&0&a&-\frac1{4\sqrt7}\\[1mm]
a&0&0&-a&0\\
0&a&0&0&0\\[1mm]
\frac{x}{\sqrt{14}}&-\frac1{4\sqrt7}&-\frac{\sqrt7}{5}&0&-\frac{11x}{14}
\end{pmatrix}.
\end{equation}
On the antisymmetric sector,
\begin{equation}
Z_-=
\begin{pmatrix}
0&a\\
a&0
\end{pmatrix}
\oplus 0_4.
\end{equation}

For arbitrary real \(x\), direct multiplication gives
\begin{equation}
\begin{split}
    Z_+^\dagger Z_+-
    &Z_+Z_+^\dagger
=\\&
-\frac{\sqrt{14}\left(2500x^2-2177\right)}{9800}
\left(
\lvert p_0\rangle\langle q_0\rvert+
\lvert q_0\rangle\langle p_0\rvert
\right),
\end{split}
\end{equation}
where \(\lvert p_0\rangle=\lvert20\rangle_+\) and
\(\lvert q_0\rangle=\lvert22\rangle\).
Thus \(x=-\sqrt{2177}/50\) makes \(Z_+\), and hence \(Z_{\mathrm P}\),
normal:
\begin{equation}
Z_{\mathrm P}^\dagger Z_{\mathrm P}
=
Z_{\mathrm P}Z_{\mathrm P}^\dagger.
\end{equation}

For the rank-two counterexample, let
\begin{equation}
S^{(2)}=S\otimes S,\qquad
S_j^{(2)}=S_j\otimes S+S\otimes S_j.
\end{equation}
Exact substitution gives
\begin{equation}
\begin{split}
    \operatorname{tr}\!\left(S^{(2)}Z_{\mathrm P}\right)=0,\\
\qquad
\operatorname{tr}\!\left(S_1^{(2)}Z_{\mathrm P}\right)=1,\\
\qquad
\operatorname{tr}\!\left(S_2^{(2)}Z_{\mathrm P}\right)=i,
\end{split}
\end{equation}
and
\begin{equation}
\operatorname{tr}\!\left(
S^{(2)}Z_{\mathrm P}^\dagger Z_{\mathrm P}
\right)=\frac34.
\end{equation}

Let
\begin{equation}
Z_{\mathrm P}=\sum_{\lambda\in\sigma(Z_{\mathrm P})}
\lambda\,\Pi_\lambda
\end{equation}
be its spectral resolution.  Since \(Z_{\mathrm P}\) is normal,
the operators \(\{\Pi_\lambda\}\) are mutually orthogonal projectors
summing to the identity.  Use the PVM \(\{\Pi_\lambda\}\), assigning
the two real estimates
\begin{equation}
\widehat\theta_1(\lambda)=\operatorname{Re}\lambda,\qquad
\widehat\theta_2(\lambda)=\operatorname{Im}\lambda.
\end{equation}
The three trace identities above prove local unbiasedness.  Its total
MSE is
\begin{equation}
\sum_\lambda
|\lambda|^2\operatorname{tr}\!\left(S^{(2)}\Pi_\lambda\right)
=
\operatorname{tr}\!\left(
S^{(2)}Z_{\mathrm P}^\dagger Z_{\mathrm P}
\right)
=
\frac34.
\end{equation}
Therefore
\begin{equation}
C_{\mathrm{MI}}(2)=C_{\mathrm N}(2)=C_{\mathrm H}(2)=\frac34.
\end{equation}
Since the one-copy model has
\(C_{\mathrm{MI}}\geq C_{\mathrm N}>C_{\mathrm H}=3/2\),
this is also a counterexample to gap persistence between the
most-informative bound and the HCRB for rank-deficient models.

\subsection{Asymptotic equivalence of HCRB and NCRB}
\label{apen:asymptotic}
We now present the full proof of  $\displaystyle\lim_{M\to\infty}M\mathcal{C}_\text{N}(M)=\mathcal{C}_\text{H}$. 
As $M\mathcal{C}_\text{N}(M)\geq M\mathcal{C}_\text{H}(M)=\mathcal{C}_\text{H}$, it is sufficient to prove
\begin{equation}
\label{eq:asproof1}
\lim_{M\to\infty}M\mathcal{C}_\text{N}(M)\leq\mathcal{C}_\text{H}\;.
\end{equation}
For this we require the following lemma
\begin{lemma}[Hayashi~\cite{hayashi1999}]
\label{lemmahayashi}
Given a Hermitian matrix $X$ and a quantum state $S$, the following relations hold.
\begin{equation}
\begin{split}
\label{eqs:hayashilemma}
&\lim_{M\to\infty}\frac{1}{M}\text{tr}\{S^{\otimes M}\abs{X(M)}\}=\lim_{M\to\infty}\frac{1}{M}\text{trAbs}\{S^{\otimes M}X(M)\}\\
=&\lim_{M\to\infty}\frac{1}{M}\abs{\text{tr}\{S^{\otimes M}X(M)\}}=\abs{\text{tr}\{SX\}}\;,
\end{split}
\end{equation}
where $X(M)=\sum_{k=1}^{M}X^{(k)}$ with $X^{(k)}=\mathbb{I}\otimes\mathbb{I}\hdots\otimes X\otimes\hdots\otimes\mathbb{I}$, with $X$ in the $k$th position, and $\text{trAbs}\{A\}$ denotes the sum of the absolute values of the eigenvalues of $A$.
\end{lemma}
\begin{proof}[\textbf{Proof (Lemma~\ref{lemmahayashi})}]
Although this proof is presented in Ref.~\cite{hayashi1999}, we recreate it here for self-consistency. Let $X=\sum_{a=1}^{d}x_a\ket{a}\bra{a}$ be an eigenvalue decomposition of $X$. By definition, we have
\begin{equation}
X^{(k)}=\sum_{a_1,a_2,\hdots a_M}x_{a_k}\ket{a_1,a_2,\hdots a_M}\bra{a_1,a_2,\hdots a_M}\;,
\end{equation}
where $\sum\ket{a_k}\bra{a_k}=\mathbb{I}$. Using this we can write
\begin{align}
X(M)&=\sum_{a_1,a_2,\hdots a_M}\sum_{k=1}^{M}x_{a_k}\ket{a_1,a_2,\hdots a_M}\bra{a_1,a_2,\hdots a_M}\\
\abs{X(M)}&=\sum_{a_1,a_2,\hdots a_M}\abs{\sum_{k=1}^{M}x_{a_k}}\ket{a_1,a_2,\hdots a_M}\bra{a_1,a_2,\hdots a_M}\;.
\end{align}

We next define $p(a):=\text{tr}\{S\ket{a}\bra{a}\}=\bra{a}S\ket{a}$, so that $p(a)$ forms a probability distribution. The expectation value of $S^{\otimes M}$ with respect to the state $\ket{a_1,a_2,\hdots a_M}$ leads to the i.i.d. distribution of $p(a)$.
\begin{equation}
\begin{split}
&\bra{a_1,a_2,\hdots a_M}S^{\otimes M}\ket{a_1,a_2,\hdots a_M}\\
&=\prod_{k=1}^{M}p(a_k)=:p_M(a_1,a_2,\hdots,a_M)\;.
\end{split}
\end{equation}
We can then get the following result
\begin{equation}
\begin{split}
\frac{1}{M}\text{tr}\{S^{\otimes M}\abs{X(M)}\}&=\frac{1}{M}\sum_{a_1,a_2,\hdots a_M}\abs{\sum_{k=1}^{M}x_{a_k}}\prod_{k=1}^{M}p(a_k)\\
&=\frac{1}{M}E_{p_M}\bigg[\abs{\sum_{k=1}^{M}X_{a_k}}\bigg]\\
&\to \abs{E_p[X]} \quad(M\to\infty)\\
&= \abs{\text{tr}\{SX\}}\;,
\end{split}
\end{equation}
where the third line follows from the law of large numbers and the dominated convergence theorem. We can apply the dominated convergence theorem as $\abs{\sum_{k=1}^{M}X_{a_k}}/M$ is bounded. This shows $\lim\limits_{M\to\infty}\frac{1}{M}\text{tr}\{S^{\otimes M}\abs{X(M)}\}=\abs{\text{tr}\{SX\}}$. To prove the remaining relations we use the following inequalities
\begin{equation}
\text{tr}\{\sqrt{S}\abs{X}\sqrt{S}\}\geq\text{tr}\left\{\abs{\sqrt{S}X\sqrt{S}}\right\}\geq\abs{\text{tr}\{\sqrt{S}X\sqrt{S}\}}\;.
\end{equation}
Using the identity $\text{trAbs}\{SX\}=\text{tr}\left\{\abs{\sqrt{S}X\sqrt{S}}\right\}$, Eq.~\eqref{eqs:hayashilemma} is proven. 
\end{proof}

Using this lemma, we now prove
\begin{equation}
\label{eq:aseq1}
\lim_{M\to\infty}MF_\text{N}(\mathbf{X}_M)=F_\text{H}(\mathbf{X})\;,
\end{equation}
for any $\mathbf{X}\in\mathcal{L}(\mathcal{H})^2$. We have 
\begin{equation}
\begin{split}
MF_\text{N}[\mathbf{X}_M]&=MF_\text{S}[\mathbf{X}_M]+M\text{tr}\bigg\{\abs{\sqrt{S^{\otimes M}}\mathrm{i}[X_M,Y_M]\sqrt{S^{\otimes M}}}\bigg\}\\
&=MF_\text{S}[\mathbf{X}_M]+M\text{trAbs}\{S^{\otimes M}\mathrm{i}[X_M,Y_M]\}\\
&=F_\text{S}[\mathbf{X}]+\frac{1}{M}\text{trAbs}\{S^{\otimes M}\mathrm{i}[X(M),Y(M)]\}\;.
\end{split}
\end{equation}
By Lemma~\ref{lemmahayashi}, the second term converges to $\abs{\text{tr}\{S\mathrm{i}[X,Y]\}}$, in the limit $M\to\infty$. Therefore Eq.~\eqref{eq:aseq1} is proven.  

Finally, we prove Eq.~\eqref{eq:asproof1}. 
\begin{equation}
M\mathcal{C}_\text{N}(M)=MF_\text{N}(\mathbf{X}_{N,M})\leq MF_\text{N}(\mathbf{X}_{H,M})\;.
\end{equation}
This final term tends to $F_\text{H}(\mathbf{X}_H)=\mathcal{C}_\text{H}$ as $M\to\infty$. Hence, when performing a collective measurement on infinitely many copies of the probe state the HCRB and NCRB are equal as expected. 

\subsection{Equivalence of the HCRB and NCRB for pure states}
\label{sec:apen:purestate}
Matsumoto proved that for pure states, $\mathcal{C}_{\text{H}}=\mathcal{C}_{\text{MI}}$~\cite{matsumoto2002new}, which implies $\mathcal{C}_{\text{H}}=\mathcal{C}_{\text{N}}$. We now prove that $\mathcal{C}_{\text{H}}=\mathcal{C}_{\text{N}}$, consistent with Matsumoto's result. For a pure state we can write $ S=\ket{\psi}\bra{\psi}$. Then $ A(\mathbf{X})=\mathrm{i}\sqrt{S}[X,Y]\sqrt{S}$ has at most one non-zero eigenvalue for all $\mathbf{X}$. Hence, either $A_+(\mathbf{X})=0$ or $A_-(\mathbf{X})=0$. Lemma~\ref{lemm:equality_cond} then implies $\mathcal{C}_{\text{H}}=\mathcal{C}_{\text{N}}$. However, this does not prove $\mathcal{C}_{\text{H}}=\mathcal{C}_{\text{MI}}$, only the weaker statement that $\mathcal{C}_{\text{H}}=\mathcal{C}_{\text{N}}$.

\section{Proof of multiparameter estimation results}
Here we prove our main multiparameter estimation results. We begin in Sec.~\ref{apen:NHSLD} by rewriting the NH function. Then in Sec.~\ref{apen:compproof:sld}, we prove theorem~\ref{theorem:mainSLD}, providing necessary and sufficient conditions for the attainability of the SLDCRB. In Sec.~\ref{apen:proofNHlb} we prove theorem~\ref{thm:lowerB}, proving a lower bound on the NHCRB. In Sec.~\ref{apensubsectionanotherproof} we present the proof of Lemma~\ref{lemm:equality_cond2}. Finally, in section~\ref{apen:asymptotic2}, we prove the asymptotic equivalence of HCRB and NHCRB (Corollary~\ref{theorem:infinite2}).

\subsection{Re-writing the Nagaoka--Hayashi function}
\label{apen:NHSLD}
\begin{lemma}[Hayashi~\cite{hayashi1999}] 
\label{lemmaHayashi}
The NH function can be written as 
\begin{equation}
F_\text{NH}[\mathbf{X}|W]= \text{Tr}[W\Re Z[\mathbf{X}]] + A^{\text{NH}}_{-}[\mathbf{X}|W],
\end{equation}
where $Z[\mathbf{X}]$ and $A^{\text{NH}}_{-}[\mathbf{X}|W]$ are defined in Eqs.~\ref{eq_zmat} and \ref{NHfunsecondterm} respectively. 
\end{lemma}
\begin{proof}[\textbf{Proof (Lemma~\ref{lemmaHayashi})}]
Define the symmetrized matrix of $\bbX$ by $s(\bbX)$. 
Explicitly, for $\bbX=[\bbX_{ij}]$ with $\bbX_{ij}\in\lofh$, $s(\bbX)=\frac12(\bbX_{ij}+\bbX_{ji})$. 
Then, the identity $\bbX=s(\bbX)+a(\bbX)$ holds. 
We next rewrite the NH function as follows. 
\begin{align}
\begin{split}
F_\text{NH}[\mathbf{X}|W]=& \min_{\bbL'}\{\mathbb{T}\mathrm{r}\{\bbW\,\bbS\, s(\mathbf{X}\mathbf{X}^{\intercal})\}+\mathbb{T}\mathrm{r}\{\bbW\bbS \bbL'\}\,|\\ 
&\qquad\bbL'\ge a(\mathbf{X}\mathbf{X}^{\intercal}),\,\bbL'_{ij}=\bbL'_{ji}\mbox{: Hermitian} \}
\end{split}\\
\begin{split}
=& \text{Tr}[W\Re Z[\mathbf{X}]]+ \min_{\bbL'}\{\mathbb{T}\mathrm{r}\{\bbW\bbS \bbL'\}\,|\\
&\qquad \bbL'\ge a(\mathbf{X}\mathbf{X}^{\intercal}),\,\bbL'_{ij}=\bbL'_{ji}\mbox{: Hermitian} \}
\end{split}\\
\begin{split}
=&\text{Tr}[W\Re Z[\mathbf{X}]]+\\
&\min_{\bbV}\{\mathbb{T}\mathrm{r}\{\bbV\}\,|\, \bbV\ge (\bbW\bbS)^{1/2}a(\mathbf{X}\mathbf{X}^{\intercal})(\bbW\bbS)^{1/2},\\
&\qquad\qquad\bbV_{ij}=\bbV_{ji}\mbox{: Hermitian} \}
\end{split}\\
\begin{split}
=&\text{Tr}[W\Re Z[\mathbf{X}]]+\\
&\min_{\bbV'}\{\mathbb{T}\mathrm{r}\{\bbV'\}\,|\, \bbV'\ge 0,\\
&\quad a(\bbV')+(\bbW\bbS)^{1/2}a(\mathbf{X}\mathbf{X}^{\intercal})(\bbW\bbS)^{1/2} =0\}\;.
\end{split}
\end{align}

The first line follows by setting $\bbL=s(\mathbf{X}\mathbf{X}^{\intercal})+\bbL'$. 
The second line holds by the definition of the $Z[\mathbf{X}]$ matrix. 
The third line is obtained by introducing a new variable $\bbV:=(\bbW\bbS)^{1/2}\bbL'(\bbW\bbS)^{1/2}$. 
To get the final line, we set $\bbV'=\bbV- (\bbW\bbS)^{1/2}a(\mathbf{X}\mathbf{X}^{\intercal})(\bbW\bbS)^{1/2}$, 
and use the fact that $(\bbW\bbS)^{1/2}a(\mathbf{X}\mathbf{X}^{\intercal})(\bbW\bbS)^{1/2}$ is Hermitian and 
the antisymmetrized matrix of $\bbV'$ is equal to 's negation. 
We remark that this new variable $\bbV'$ does not necessarily satisfy the condition $\bbV'_{ij}=\bbV'_{ji}$. 

Note that when obtaining the third line, there is minor caveat when $S$ is rank deficient: the change of variables (from $\bbL'$ to $\bbV$) is
understood on the support of $R:=(\mathbb W\mathbb S)^{1/2}$. As we now show, this doesn't affect the conclusion. Let $P$ be the projector onto $\operatorname{supp}R$ and let
$Q=\mathbb I-P$. Every feasible $\bbL'$ gives a feasible
$\bbV=R\bbL'R$ with the same objective value. Conversely, given a feasible
$\bbV$, we may replace it by $P\bbV P$, which remains feasible and cannot
increase its trace. Defining $K:=R_P^{-1}\bbV R_P^{-1}$, where $R_P$ is the restriction of $R$ to its support, gives
\begin{equation}
    K\succeq P\,a(\mathbf X\mathbf X^{\mathsf T})\,P\;.
\end{equation}
For every $\epsilon>0$, define the operator
\begin{equation}
    \bbL'_{\epsilon,t}:=K+\epsilon P+tQ\;.
\end{equation}
Since the $P$ block of
$\mathbb L'_{\epsilon,t}
-a(\mathbf X\mathbf X^{\mathsf T})$
is strictly positive, the Schur-complement condition shows that
$\mathbb L'_{\epsilon,t}\succeq
a(\mathbf X\mathbf X^{\mathsf T})$
for sufficiently large $t$.

 Moreover,
\begin{equation}
    \operatorname{Tr}(\mathbb W\mathbb S \bbL'_{\epsilon,t})
=
\operatorname{Tr}\bbV+
\epsilon\operatorname{Tr}(\mathbb W\mathbb S)\;.
\end{equation}
Letting $\epsilon\downarrow0$ proves that the two programmes have the
same optimal value even when $S$ is rank deficient.

\end{proof}

\subsection{Proof of Theorem~\ref{theorem:mainSLD}}
\label{apen:compproof:sld}
\begin{proof}[\textbf{Proof (Theorem~\ref{theorem:mainSLD})}]
As mentioned in the main text, we will prove the theorem by the chain: i) $\Ra$ ii) $\Ra$ iii) $\Ra$ iv) $\Ra$ i). We prove iii) $\Ra$ iv) and iv) $\Ra$ i) below. 

\noindent
\textit{Proof that iii) $\Ra$ iv):}\\
We assume condition iii). $\mathcal{C}_{\text{NH}}[\mathbb{I}]=\mathcal{C}_{\text{S}}[\mathbb{I}]$ implies 
\begin{equation}
\begin{split}
\min_{\mathbf{X}}\{ F_\text{NH}&[\mathbf{X}]\,|\, X_i:\,\mathrm{l.u.}\text{ } \mathrm{and}\text{ } \mathrm{Hermitian}\}= \mathcal{C}_\text{S}=\text{Tr}[{J_S}^{-1}]\\
\Lra\min_{\mathbf{X}}\{ &\text{Tr}[\Re Z[\mathbf{X}]] + A^{\text{NH}}_{-}[\mathbf{X}]\,|\, \\
& X_i:\,\mathrm{l.u.}\text{ } \mathrm{and}\text{ } \mathrm{Hermitian}\}=\text{Tr}[\Re Z[\mathbf{X}_\text{S}]]\;,
\end{split}
\end{equation}
where $\mathbf{X}_\text{S}$ denotes the optimal $\mathbf{X}$ which minimises the first term of the NH function. Explicitly, $\mathbf{X}_\text{S}=(X_{\text{S},i})=(L^{i})$ with $L^{i}:=\sum_{j} ({J_{S}}^{-1})_{ji}L_{j}$. 

Lemma~\ref{prop:quotientSLD} shows that the function $\text{Tr}[\Re Z[\mathbf{X}]]$ is strictly convex about $\mathbf{X}$ 
(on the quotient space when $S$ is not full rank) and the optimiser exists uniquely. 
By definition, it holds that the second term $A^{\text{NH}}_{-}[\mathbf{X}]$ is non-negative function of $\mathbf{X}$. We can apply Lemma~\ref{lemmasupportSLD} (Appendix~\ref{apen:lemmasupportSLD}) to show that this is possible if and only if the following condition holds. 
\begin{equation}
A^{\text{NH}}_{-}[\mathbf{X}_\text{S}]=0\;.
\end{equation}
The uniqueness of the optimiser for the SLDCRB is proven in Appendix~\ref{apen:strictconvex}. $\mathbb{T}\mathrm{r}\{\bbV\}=0$ holds if and only if $\bbV=0$ for a positive matrix $\bbV$. 
This is possible if and only if $\bbS^{1/2}a(\mathbf{X}_\text{S}\mathbf{X}_\text{S}^{\intercal})\bbS^{1/2}=0$, since this matrix is Hermitian. 
Therefore, it is equivalent to the condition such that all elements are zero. 
In other words, we have
\begin{equation}
\forall i,j,\, \sqrt{S}[L^{i}\,,\,L^{j}]\sqrt{S}=0\ \Lra\  
\forall i,j,\, \sqrt{S}[L_{i}\,,\,L_{j}]\sqrt{S}=0\;. 
\end{equation}
This is also equivalent to 
\begin{equation}
\forall i,j,\, \text{tr}\bigg\{\abs{\sqrt{S}[L_{i}\,,\,L_{j}]\sqrt{S}}\bigg\}=0\;,
\end{equation}
or using our alternative notation $\forall i,j,\, \text{tr}\{\abs{\sqrt{S}[X_{\text{S},i}\,,\,X_{\text{S},j}]\sqrt{S}}\}=0$. When $S$ is full rank, this immediately proves $[L_{i}\,,\,L_{j}]=0,\,\forall i,j$. Thus, we do not need to consider any extension of the Hilbert space.

If $S$ is rank deficient, note that the optimizer $X_{\text{NH}}$ also needs to satisfy 
the following condition from $\mathcal{C}_{\text{NH}}[\mathbb{I}]=\mathcal{C}_{\text{S}}[\mathbb{I}]$
\begin{equation}
X_{\text{NH},i}= L^{i} \mbox{ up to equivalence }\sim_{S},\,\forall i. 
\end{equation}
Let $\Pi=\{\Pi_{k}\}$ and $\hat{\theta}$ be an optimal POVM and estimator, then $\mathcal{C}_{\text{MI}}[\mathbb{I}]=\mathcal{C}_{\text{NH}}[\mathbb{I}]$ implies that the optimiser for the NHCRB can be decomposed into some POVM (see Ref.~\cite{conlon2021efficient}, Eqs. (18-21))
\begin{equation}
X_{\text{NH},i}=\sum_{k}(\hat{\theta}_{i}(k)-\theta_{i})\Pi_{k}\;. 
\end{equation}
 Combining the above two relations, after some algebra we get 
\begin{equation}
L_{i}=\sum_{k} \ell_{i}(k)\Pi_{k}\mbox{ up to equivalence }\sim_{S},\,\forall i, 
\end{equation}
where $\ell_i(k):=\sum_j J_{\text{S},ji}(\hat{\theta}_j(k)-\theta_j)$. 
Under the assumption $\mathcal{C}_\text{MI}=\mathcal{C}_\text{S}$, Appendix~B.1 of
Ref.~\cite{suzuki2020quantum} then implies that we can apply the Naimark extension to find a larger Hilbert space 
in which the POVM elements are expressed as mutually orthogonal projectors~\cite{neumark1943spectral} (See Ref.~\cite{pozza2017effective} for an iterative method of constructing the Naimark extension and Ref.~\cite{dalla2019naimark} for an explicit example of this). 
After this extension, the SLDs $\tilde{L}_{i}$ commute with each other. 

Note that a decomposition of either $L_i$, or
$L^i$, in terms of a POVM as \begin{equation}
L^i=\sum_{k}(\hat{\theta}'_{i}(k)-\theta_{i})\Pi_{k}\;,
\end{equation}
does not by itself imply that the
corresponding diagonal operators in an arbitrary Naimark dilation are
SLD operators for the extended model. The saturation hypothesis in condition
iii) is essential; it is this additional equality condition that
allows the commuting Naimark lifts to be chosen as genuine SLD
representatives~\cite{suzuki2020quantum}.

\noindent
\underline{iv) $\Rightarrow$ i)}:\\
Suppose we find such an extension where the SLDs $\tilde{L}_{i}$ for the extended model $\tilde{S}_{\theta}=U(S_\theta\otimes S_{0})U^{\dagger}$ are commutative. 
We can diagonalize all $\tilde{L}_{i}$ simultaneously as 
\begin{equation}
\tilde{L}_{i}=\sum_{k} \ell_{i}(k) E_{k}\;,
\end{equation}
where $\{E_{k}\}$ are mutually orthogonal projectors. 
When $S$ is not full rank, we can always extend $\{E_{k}\}$ to form a projection measurement. 
By substituting into the SLD equation, we explicitly obtain $\ell_{i}(k)$ as 
\begin{equation}
\ell_{i}(k)=\frac{\text{tr}\{\tilde{S}_i E_{k}\}}{\text{tr}\{\tilde{S} E_{k}\}}\;,
\end{equation}
which is the score function. We define $ \ell_{i}(k)=0$ when $\text{tr}\{\tilde{S} E_{k}\}=0$.
Thus, $\sum_{k} \ell_{i}(k) \ell_{j}(k) \text{tr}\{\tilde{S} E_{k}\}=J_{S,ij}$ holds. Note that the SLD Fisher information remains invariant under the extension (see e.g. Proposition 2.1 of Ref.~\cite{liu2019quantum}). Under this circumstance, an optimal measurement is given by $\{E_{k}\}$ 
and a locally unbiased estimator is constructed by 
\begin{equation}
\hat{\theta}_{i}(k)=\theta_{i}+\sum_{j}(J_{S}^{-1})_{ji}\ell_{j}(k)\;.
\end{equation}

Finally, we calculate the MSE matrix to show that this is equal to the inverse of the SLD Fisher information matrix. From Eq.~\eqref{eqMSEmatrix}, we have 
\begin{equation}
\begin{split}
V_{ij}&=\sum_{k}\big(\sum_{m}(J_{S}^{-1})_{mi}\ell_{m}(k)\big)\big(\sum_{n}(J_{S}^{-1})_{nj}\ell_{n}(k)\big)\text{tr}\{\tilde{S} E_{k}\}\\
&=\sum_{m}(J_{S}^{-1})_{mi}\sum_{n}(J_{S}^{-1})_{nj}(J_{S})_{mn}\\
&=(J_{S}^{-1})_{ji}\;,
\end{split}
\end{equation}
where we use the fact that $\sum_{n}(J_{S}^{-1})_{nj}(J_{S})_{mn}=\delta_{jm}$, which follows from $(J_S)(J_{S})^{-1}=\mathbb{I}$.
\end{proof}

%
%

\subsection{Lemma to support Theorem~\ref{theorem:mainSLD}}
\label{apen:lemmasupportSLD}
\begin{lemma}
\label{lemmasupportSLD}
Consider the sum of two non-negative functions $f(x)=f_{1}(x)+f_{2}(x)$ on the domain $\cX$. 
Suppose the optimization of $f_{1}$ under a constraint $C$ exists and the optimiser is unique, 
then the following statement holds. 
\begin{equation}
\begin{split}
&\min_{x\in\cX:\, C}f(x)=\min_{x\in\cX:\, C}f_{1}(x)\ \\
\Lra\ & f_{2}(x_{*})=0\mbox{ with }x_{*}:=\arg\min_{x\in\cX:\, C}f_{1}(x)
\end{split}
\end{equation}
\end{lemma}

\subsection{Proof of Theorem~\ref{thm:SLD-MI-full-rank-GPT}}
\label{apen:proofMISLD}
We now prove that for $S\succ0$ a strong gap persistence theorem holds between the most informative bound and the SLDCRB.
\begin{proof}[\textbf{Proof (Theorem~\ref{thm:SLD-MI-full-rank-GPT})}]
Let $L_1,\ldots,L_n$ be the SLD operators of the single-copy model,
defined by
\begin{equation}
 S_i=\frac{1}{2}(SL_i+L_iS),
 \qquad
 (J_{\rm S})_{ij}
 =
 \frac{1}{2}\operatorname{tr}\!\left[
 S(L_iL_j+L_jL_i)
 \right].
 \label{eq:SLD-definition-GPT}
\end{equation}
We first prove the following full-rank attainability criterion:
\begin{equation}
 C_{\rm MI}(W)=C_{\rm S}(W)
 \quad\Longleftrightarrow\quad
 [L_i,L_j]=0
 \quad\text{for all }i,j.
 \label{eq:full-rank-SLD-attainability}
\end{equation}

Suppose first that the SLDs commute. They can then be simultaneously
diagonalised,
\begin{equation}
 L_i=\sum_x \ell_i(x)E_x,
 \label{eq:joint-SLD-PVM}
\end{equation}
where $\{E_x\}$ is a projective measurement. For
$p_x=\operatorname{tr}(SE_x)$, the corresponding classical score is
\begin{equation}
 \partial_i\log p_x=\ell_i(x).
\end{equation}
Consequently, the classical Fisher information matrix of this
measurement is exactly $J_{\rm S}$. The locally unbiased estimator
\begin{equation}
 \widehat{\theta}_i(x)-\theta_i
 =
 \sum_j (J_{\rm S}^{-1})_{ij}\ell_j(x)
 \label{eq:SLD-optimal-estimator}
\end{equation}
has covariance matrix $J_{\rm S}^{-1}$. Hence
\begin{equation}
 C_{\rm MI}(W)=
 \operatorname{Tr}[WJ_{\rm S}^{-1}]
 =
 C_{\rm S}(W).
\end{equation}

Conversely, suppose that $C_{\rm MI}(W)=C_{\rm S}(W)$. Let
$\{E_x\}$ be an attaining POVM, and let $J_{\rm C}$ be its classical
Fisher information matrix. The classical and quantum
Cram\'er--Rao inequalities give
\begin{equation}
 V\succeq J_{\rm C}^{-1}\succeq J_{\rm S}^{-1}.
 \label{eq:matrix-CR-chain}
\end{equation}
Since $W>0$ and
\begin{equation}
 \operatorname{Tr}[WV]
 =
 \operatorname{Tr}[WJ_{\rm S}^{-1}],
\end{equation}
both positive-semidefinite differences in
Eq.~\eqref{eq:matrix-CR-chain} must vanish. In particular,
\begin{equation}
 J_{\rm C}=J_{\rm S}.
 \label{eq:classical-equals-quantum-FI}
\end{equation}

Equality in the SLD Fisher-information inequality implies that, for
every real linear combination
\begin{equation}
 L_a=\sum_i a_iL_i
\end{equation}
and every nonzero POVM element $E_x$, there exists a real number
$\ell_a(x)$ such that
\begin{equation}
 E_x^{1/2}L_aS^{1/2}
 =
 \ell_a(x)E_x^{1/2}S^{1/2}.
 \label{eq:BC-equality}
\end{equation}
Since $S>0$, we may multiply Eq.~\eqref{eq:BC-equality} on the right
by $S^{-1/2}$, obtaining
\begin{equation}
 E_x^{1/2}L_a
 =
 \ell_a(x)E_x^{1/2}.
\end{equation}
Taking the adjoint gives
\begin{equation}
 L_aE_x^{1/2}
 =
 \ell_a(x)E_x^{1/2}.
\end{equation}
Applying this separately to $L_i$ and $L_j$ therefore gives
\begin{equation}
 [L_i,L_j]E_x^{1/2}=0.
\end{equation}
Because $\sum_xE_x=I$,
\begin{equation}
 [L_i,L_j]
 =
 \sum_x [L_i,L_j]E_x^{1/2}E_x^{1/2}
 =
 0.
\end{equation}
This proves Eq.~\eqref{eq:full-rank-SLD-attainability}.

We now apply this criterion to the $M$-copy model. Its SLD operators
are
\begin{equation}
 L_{i,M}
 =
 \sum_{r=1}^{M}L_i^{(r)},
 \label{eq:additive-M-copy-SLD}
\end{equation}
where $L_i^{(r)}$ denotes $L_i$ acting on copy $r$. Its SLD Fisher
information matrix and SLDCRB are
\begin{equation}
 J_{\rm S}(M)=MJ_{\rm S},
 \qquad
 C_{\rm S}(M;W)
 =
 \frac{1}{M}C_{\rm S}(W).
 \label{eq:SLD-additivity}
\end{equation}
Since $S^{\otimes M}>0$, the full-rank attainability criterion also
applies to the $M$-copy model:
\begin{equation}
 C_{\rm MI}(M;W)=C_{\rm S}(M;W)
 \quad\Longleftrightarrow\quad
 [L_{i,M},L_{j,M}]=0
 \quad\text{for all }i,j.
 \label{eq:M-copy-attainability}
\end{equation}

Operators acting on distinct copies commute, and hence
\begin{equation}
 [L_{i,M},L_{j,M}]
 =
 \sum_{r=1}^{M}
 \bigl([L_i,L_j]\bigr)^{(r)}.
 \label{eq:additive-commutator}
\end{equation}
It is immediate that $[L_i,L_j]=0$ implies
$[L_{i,M},L_{j,M}]=0$.

Conversely, suppose that $[L_{i,M},L_{j,M}]=0$. Taking the partial
trace over copies $2,\ldots,M$ in
Eq.~\eqref{eq:additive-commutator} gives
\begin{align}
 0
 &=
 d^{M-1}[L_i,L_j]
 +(M-1)d^{M-2}
   \operatorname{tr}([L_i,L_j])I
 \nonumber\\
 &=
 d^{M-1}[L_i,L_j],
 \label{eq:partial-trace-commutator}
\end{align}
where $d=\dim\mathcal H$ and we used
$\operatorname{tr}([L_i,L_j])=0$. Therefore
\begin{equation}
 [L_i,L_j]=0.
\end{equation}
We have proved
\begin{equation}
 [L_{i,M},L_{j,M}]=0\ \forall i,j
 \quad\Longleftrightarrow\quad
 [L_i,L_j]=0\ \forall i,j.
\end{equation}
Combining this equivalence with
Eqs.~\eqref{eq:full-rank-SLD-attainability} and
\eqref{eq:M-copy-attainability} proves
Theorem~\ref{thm:SLD-MI-full-rank-GPT}.
\end{proof}

\subsection{Example proving there is no general gap persistence theorem between the most informative bound and the SLDCRB}
\label{apen:proofMISLD2}

This note gives a three-parameter rank-three model for which
\begin{equation}
   \mathcal{C}_\text{MI}>\mathcal{C}_\text{S}=\frac{54}{25},
   \qquad
   \mathcal{C}_\text{MI}(2)=\mathcal{C}_\text{S}(2)=\frac{27}{25}.
\end{equation}
Thus weak, and hence strong, gap persistence between the SLDCRB and the
most informative bound fails for general rank-deficient models.

\subsubsection*{1. A Gram criterion}
Let the state be maximally mixed on an $r$-dimensional support and zero on
its kernel.  Suppose the fixed support blocks of the SLDs vanish, and write
chosen SLD representatives as
\begin{equation}
 L_i=\begin{pmatrix}0&B_i\\B_i^\dagger&0\end{pmatrix}.
\end{equation}
Set
\begin{equation}
 H_{ij}:=B_iB_j^\dagger,
 \qquad
 H:=\sum_{i,j=1}^p |i\rangle\langle j|\otimes H_{ij}.
\end{equation}
If the SLDCRB is attainable, a Naimark dilation gives commuting Hermitian
SLD representatives.  Simultaneous diagonalisation then yields vectors
$x_a\in\mathbb R^p$ and $u_a\in\mathbb C^r$ such that
\begin{equation}
 H=\sum_a x_ax_a^T\otimes u_au_a^\dagger.
\end{equation}
Therefore separability of $H$ across the parameter--support cut is a
necessary condition for attainability. This idea was used in Ref.~\cite{conlon2026deciding}.

Conversely, when the support blocks vanish, a decomposition of $H$
of the form above, with $x_a\in\mathbb R^p$, is also sufficient.  Indeed, split every positive support factor into rank-one
terms and write
\begin{equation}
 H_{ij}=\sum_a x_{a,i}x_{a,j}u_au_a^\dagger.
\end{equation}
Choose $t>0$ so that
$t^{-2}\sum_a u_au_a^\dagger\preceq I_r$.  For every $a$, introduce the
paired atoms
\begin{equation}
 x_{a,\pm}=\pm t x_a,
 \qquad
 v_{a,\pm}=\frac{u_a}{\sqrt2\,t}.
\end{equation}
They have zero first moment and second moment $H$.  Complete their zeroth
moment to $I_r$ with atoms having $x=0$.  The resulting finite moment
representation is the compression of diagonal, hence commuting, Hermitian
operators.  If $\widehat B_i$ are the support--kernel blocks of this
compression, then
$\widehat B_i\widehat B_j^\dagger=B_iB_j^\dagger$ for all $i,j$.
Stacking the blocks gives
$\widehat{\mathcal B}\widehat{\mathcal B}^\dagger=
\mathcal B\mathcal B^\dagger$.  Polar decomposition therefore supplies a
common partial isometry $U$ with
$\widehat{\mathcal B}=\mathcal B U$.  After zero padding, $U$ extends to a
unitary on the kernel.  Conjugating all commuting operators by the same
block unitary makes their support--kernel blocks equal to the prescribed
$B_i$.  Their joint spectral measurement therefore supplies a POVM
attaining the SLD Fisher matrix.

\subsubsection*{2. The one-copy model}
Let
\begin{equation}
 S=\frac13 I_3\oplus0_7
\end{equation}
on $\mathbb C^3\oplus\mathbb C^7$.  Define the real $3\times7$ matrices
\begin{equation}
B_0=\begin{pmatrix}
1&0&0&0&0&0&0\\
0&1/\sqrt2&0&0&0&0&0\\
0&0&0&0&0&\sqrt2&\sqrt{2/3}
\end{pmatrix},
\end{equation}
\begin{equation}
B_1=\begin{pmatrix}
0&\sqrt2&\sqrt{2/3}&0&0&0&0\\
1&0&0&0&0&0&0\\
0&0&0&1/\sqrt2&0&0&0
\end{pmatrix},
\end{equation}
\begin{equation}
B_2=\begin{pmatrix}
0&0&0&0&0&1/\sqrt2&0\\
0&0&0&\sqrt2&\sqrt{2/3}&0&0\\
1&0&0&0&0&0&0
\end{pmatrix}.
\end{equation}
Take
\begin{equation}
 L_i=\begin{pmatrix}0_3&B_i\\ B_i^T&0_7\end{pmatrix},
 \qquad i=0,1,2,
\end{equation}
and define the tangent matrices by
\begin{equation}
 \dot S_i=\frac12(SL_i+L_iS).
\end{equation}
These are realised by the smooth positive model
\begin{equation}
 S_\theta=
 \frac{T(\theta)S T(\theta)}{\tr[T(\theta)S T(\theta)]},
 \qquad
 T(\theta)=I+\frac12\sum_{i=0}^2\theta_iL_i.
\end{equation}

Put $\mathcal B=(B_0^T,B_1^T,B_2^T)^T$ and $H=\mathcal B\mathcal B^T$.
In the basis
\begin{equation}
 (|00\rangle,|01\rangle,|02\rangle,
   |10\rangle,|11\rangle,|12\rangle,
   |20\rangle,|21\rangle,|22\rangle)
\end{equation}
one obtains
\begin{equation}
H=\begin{pmatrix}
1&0&0&0&1&0&0&0&1\\
0&\frac12&0&1&0&0&0&0&0\\
0&0&\frac83&0&0&0&1&0&0\\
0&1&0&\frac83&0&0&0&0&0\\
1&0&0&0&1&0&0&0&1\\
0&0&0&0&0&\frac12&0&1&0\\
0&0&1&0&0&0&\frac12&0&0\\
0&0&0&0&0&1&0&\frac83&0\\
1&0&0&0&1&0&0&0&1
\end{pmatrix}.
\end{equation}
The displayed factorisation proves $H\succeq0$.  Its blocks satisfy
$H_{ij}=H_{ji}=H_{ij}^T$, so the sandwiched SLD commutators vanish and
$C_{\rm NH}=C_{\rm S}$.

The QFI matrix is
\begin{equation}
 (J_{\rm S})_{ij}=\frac13\tr(H_{ij})
 =\frac{25}{18}\delta_{ij},
\end{equation}
and hence
\begin{equation}
 \mathcal{C}_\text{S}=\Tr(J_{\rm S}^{-1})=\frac{54}{25}.
\end{equation}

Nevertheless, $H$ is entangled.  Let $R(H)$ denote its realignment,
$R(H)_{ij,ab}=H_{ia,jb}$.  Its singular values are
\begin{equation}
 \frac{25}{6},\quad
 2,2,2,\quad
 \frac{\sqrt{139}}6,\frac{\sqrt{139}}6,\quad
 0,0,0.
\end{equation}
Consequently,
\begin{equation}
 \|R(H)\|_1-\tr H
 =\frac{\sqrt{139}-7}{3}>0.
\end{equation}
The realignment criterion therefore proves that $H$ is not separable.
By the Gram criterion above, no single-copy POVM attains the SLDCRB, and
thus
\begin{equation}
\mathcal{C}_\text{MI}>\mathcal{C}_\text{S}=\frac{54}{25}.
\end{equation}

\subsubsection*{3. Two-copy activation}
For two copies the support dimension is $9$.  The additive SLDs have zero
support blocks and support--kernel Gram blocks
\begin{equation}
 K_{ij}=H_{ij}\otimes I_3+I_3\otimes H_{ij}.
\end{equation}
Let $K=\sum_{i,j=0}^2|i\rangle\langle j|\otimes K_{ij}$.

Define the signed cyclic group
\begin{equation}
 \mathcal G=
 \{C^k\operatorname{diag}(1,\epsilon_1,\epsilon_2):
 k=0,1,2,\ \epsilon_1,\epsilon_2\in\{\pm1\}\},
\end{equation}
where $Ce_j=e_{j+1\ ({\rm mod}\ 3)}$, and take the ten integer seeds
\begin{equation}
\begin{split}
&t_1=(1,0,0)^T,\quad t_2=(1,1,0)^T,\quad t_3=(1,1,1)^T,\\
&t_4=(1,2,0)^T,\quad t_5=(1,2,1)^T,\quad t_6=(1,2,2)^T,\\
&t_7=(1,3,0)^T,\quad t_8=(1,3,1)^T,\quad t_9=(1,3,2)^T,\\
&t_{10}=(1,2,3)^T.
\end{split}
\end{equation}
The accompanying exact certificate gives ten rational positive-definite
$9\times9$ matrices $Q_s$ satisfying
\begin{equation}
K=\sum_{s=1}^{10}\sum_{g\in\mathcal G}
 (gt_s)(gt_s)^T\otimes
 (g\otimes g)Q_s(g\otimes g)^T.
\end{equation}
Every summand is separable across the parameter--support cut, so $K$ is
separable.  The Gram criterion therefore yields commuting two-copy SLD
representatives and a POVM whose covariance matrix is
$J_{\rm S}(2)^{-1}=(2J_{\rm S})^{-1}$.  Hence
\begin{equation}
\mathcal{C}_\text{MI}(2)=\mathcal{C}_\text{S}(2)=\frac12\mathcal{C}_\text{S}=\frac{27}{25}\;.
\end{equation}
Combining the one- and two-copy statements gives
\begin{equation} \mathcal{C}_\text{MI}>\mathcal{C}_\text{S},
 \qquad
 \mathcal{C}_\text{MI}(2)=\mathcal{C}_\text{S}(2),
\end{equation}
which disproves weak and strong SLD--most-informative gap persistence for
general rank-deficient models.

\paragraph{Exact certificate.}
The files \texttt{sld\_mi\_counterexample\_Q.txt} and
\texttt{verify\_sld\_mi\_counterexample.ipynb} found in Ref.~\cite{ConlonGapPersistenceGitHub} contain the ten rational
matrices and check, over exact rational arithmetic, that every $Q_s$ is
positive definite and that the orbit sum is exactly $K$.

\subsection{Proof of Theorem~\ref{thm:lowerB}}
\label{apen:proofNHlb}
We first describe the index set and partition in more detail. Given an index set $\cI=\{1,2,\ldots,n\}$, we consider a pairwise partition of $\cI$ as follows. 
For even $n=2m$, $\cI=\bigcup_{\ell=1}^m \cI_\ell$ such that $|\cI_\ell|=2$ for all $\ell$ 
and $\cI_\ell\cap \cI_{\ell'}=\emptyset$. For example, $\cI_1=\{1,2\}, \cI_2=\{3,4\}$ for $n=4$. 
For odd $n=2m+1$, a pairwise partition of $\cI$ is given by an additional set $\cI_0$ with a single element 
as $\cI=\bigcup_{\ell=0}^m \cI_\ell$. 
For example, $\cI_0=\{5\},\cI_1=\{1,2\}, \cI_2=\{3,4\}$ for $n=5$. 
Another example is $\cI_0=\{1\},\cI_1=\{2,3\}$ for $n=3$. 
The pairwise partition of $\cI=\bigcup_\ell\cI_\ell$ is denoted by $\cP$. 
Finally, we say a partition $\cP_{\rm st}$ is standard when each pair is chosen numerical order. 
The first two examples above are standard partition, whereas the last one is not. 

Given a pairwise partition, define the following summation with respect to it. 
For a double-index sequence of real numbers $\{a_{jk}\}$, 
\begin{equation}
  \sum_\cP a_{jk} := \sum_\ell\sum_{j\neq k\in\cI_\ell}a_{jk}.
\end{equation} 
As an example of $\cI_1=\{1,2\}, \cI_2=\{3,4\}$ for $n=4$, 
$\sum_\cP a_{jk}=a_{12}+a_{21}+a_{34}+a_{43}$. 
When $n$ is odd, we will omit the set $\cI_0$ from the summation. This is the summation used in theorem~\ref{thm:lowerB}. 

We now move onto the proof of theorem~\ref{thm:lowerB}. We first define $\bbA(\mathbf{X})$ as the matrix with $\bbA_{i,j}=\sqrt{S}[X_i,X_j]\sqrt{S}$ (note that we drop the $i$ compared to $A(\mathbf{X})$. Also denote the joint classical-quantum Hilbert space as $\bbH=\mathcal{H}_\text{Cl}\otimes\mathcal{H}$. Given an arbitrary pairwise partition $\cP$, we define a pinching map on $\bbA(\mathbf{X})\in\mathcal{L}(\bbH)$: 
  \begin{equation}
   \cE_\cP(\bbA):= \sum_\ell \bbP_{\cI_\ell}\bbA \bbP_{\cI_\ell}, 
  \end{equation} 
  where $\bbP_{\cI_\ell}$ is the projector on the subspace specified by $\cI_\ell$ as 
  \begin{equation}
    \bbP_{\cI_\ell}:=\sum_{p\in\cI_\ell}\ket{p}\bra{p}\otimes \mathbb{I}_d.   
  \end{equation}
  Since the pinching map is a trace-preserving positive map, we have the following lower bound by applying it:
\begin{multline}\label{eq:Loptimization}
    A^{\text{NH}}_{-}[\mathbf{X}]\ge \min_{\bbV}\{\bbTr{\bbV}\,|\,\bbV\in\mathcal{L}_{+,\mathrm{sym}}(\bbH),\\
    \cE_\cP(\bbV)\ge \cE_\cP(a(\sqrt{\bbS}\mathbf{X} \mathbf{X}^\intercal\sqrt{\bbS})) \}.
  \end{multline}
  The effect of the pinching map produces 
  $\cE_\cP(a(\sqrt{\bbS}\mathbf{X} \mathbf{X}^\intercal\sqrt{\bbS}))
  = \sum_\cP|j)(k|\otimes \frac12\sqrt{S}[X_j,X_k]\sqrt{S}$, 
  the optimization from the antisymmetrized term can now be carried out as follows.  
  Note that $\sqrt{\bbS} a(\cE_\cP(\mathbf{X} \mathbf{X}^\intercal))\sqrt{\bbS}$ is expressed as direct sum of a $2\times2$ block matrices in an anti-diagonal form:
  \begin{align}
    \begin{bmatrix} 
      0&\bbA_{jk}\\
      \bbA_{kj}&0
    \end{bmatrix}=
    \begin{bmatrix} 
      0&\bbA_{jk}\\
      -\bbA_{jk}&0
    \end{bmatrix}.
  \end{align}
  Within each $2\times 2$ block-matrix subspace, we apply the following optimization for $\bbV$. 
  \begin{align*}
   \hspace{-3mm} \min_{\bbV}\left\{\bbTr{\bbV}\,\Big|\,\bbV\in\mathcal{L}_{\rm sym}(\mathbb{C}^2\otimes\cH), 
   \bbV\ge\frac12{\scriptsize\begin{bmatrix} 
      0&\bbA_{jk}\\
      -\bbA_{jk}&0
    \end{bmatrix}} \right\}\\
    =\text{tr}\{|\bbA_{jk}|\},
  \end{align*}
  where the optimizer $\bbV_*$ is given by 
  $\bbV_*={\frac12\scriptsize \begin{bmatrix} 
    |\bbA_{jk}|&0\\
    0&|\bbA_{jk}|
  \end{bmatrix}}$. Each block therefore contributes $\text{tr}\{|\bbA_{jk}|\}=\text{tr}\{|\sqrt{S}[X^j,X^k]\sqrt{S}|\}$. As $\sum_\cP$ sums over all pairs twice, by including a factor $\frac12$, we arrive at theorem~\ref{thm:lowerB}.

To extend the above to the case with a non-zero weight matrix $W>0$ define $\widetilde X_i=\sum_r(\sqrt W)_{ir}X_r$. Then $((WS)^{1/2}a(XX^\top)(WS)^{1/2})_{ij} =\frac12\sqrt S[\widetilde X_i,\widetilde X_j]\sqrt S$ and the same pinching argument applies verbatim with $X_i\mapsto \widetilde X_i$.

\subsection{Proof of Lemma~\ref{lemm:equality_cond2}}
\label{apensubsectionanotherproof}
 
 \begin{proof}[\textbf{Proof (Lemma~\ref{lemm:equality_cond2})}]
Suppose $\FNH(X)=\FH(X)$. Then both inequalities in
\eqref{eq:pinching} are equalities. Hence, for every canonical pair
$(j,k)$,
\begin{equation}
\tr\!\left|
\sqrt S\,i[\widetilde X_j,\widetilde X_k]\sqrt S
\right|
=
\left|
\tr\!\left(
\sqrt S\,i[\widetilde X_j,\widetilde X_k]\sqrt S
\right)
\right|.
\end{equation}
Thus $\sqrt S\,i[\widetilde X_j,\widetilde X_k]\sqrt S$
has a definite sign. Since $S>0$,
$i[\widetilde X_j,\widetilde X_k]$ also has a definite sign.
However,
\begin{equation}
\tr\{i[\widetilde X_j,\widetilde X_k]\}=0,
\end{equation}
so
\begin{equation}
[\widetilde X_j,\widetilde X_k]=0
\end{equation}
for every canonical pair. It follows that every canonical singular
value of $\operatorname{Im}Z[\widetilde X]$ vanishes, and hence $\operatorname{Im}Z[\widetilde X]=0.$

Now
\begin{equation}
\FH(X)=\Tr[\operatorname{Re}Z[X]] .
\end{equation}
Writing the NH function as
\begin{equation}
\FNH(X)=\Tr[\operatorname{Re}Z[X]]+A_-^{\mathrm{NH}}[X],
\end{equation}
equality implies $A_-^{\mathrm{NH}}[\tilde{X}]=0$.  In the SDP defining
$A_-^{\mathrm{NH}}$, the optimisation variable is positive
semidefinite, so zero objective value forces the optimiser to be zero.
The constraint then gives
\begin{equation}
(I_n\otimes\sqrt S)\,
a(\tilde{X}\tilde{X}^{\mathsf T})\,
(I_n\otimes\sqrt S)=0 .
\end{equation}
Its $(j,k)$ block is proportional to
$\sqrt S[\tilde{X}_j,\tilde{X}_k]\sqrt S$.  Since $S>0$,
\begin{equation}
[\tilde{X}_j,\tilde{X}_k]=0\qquad\forall j,k \;.
\end{equation}
The converse is immediate, since all antisymmetric terms then vanish. Finally as $U$ in Eq.~\eqref{def:xtilde} is invertible, this implies
\begin{equation}
[X_j,X_k]=0\qquad\forall j,k \;.
\end{equation}
 \end{proof}

\subsection{Convergence rate of HCRB and NHCRB}
\label{apen:asymptotic2}
Here we present the full proof of  $\displaystyle\lim_{M\to\infty}M\mathcal{C}_\text{NH}(M)=\mathcal{C}_\text{H}$. First, we show that for any $\mathbf{X}$, $\lim_{M\to\infty}M F_{\mathrm{NH}}(\mathbf{X}_{M})=F_{\rm H}(\mathbf{X})$. 
  Since the first term of the NH function is identical to the Holevo function, 
  it suffices to show $\lim_{M\to\infty}M A^{\text{NH}}_{-}[\mathbf{X}_{M}|W]=\text{Tr}[|\Im Z[\mathbf{X}]|]$. In order to prove this, we need to introduce an upper bound for the NHCRB. For the upper bound we require the following simple lemma:


 We first prove the upper bound on the NHCRB (theorem~\ref{thm:upperB}).
\begin{proof}[\textbf{Proof (Theorem~\ref{thm:upperB})}]
  To simplify notation, we omit dependence on $\mathbf{X}$. 
  We define a block diagonal operator-valued matrix by 
  \begin{equation*}
    \bbV_D:=\frac12\mathrm{diag}\Big(\sum_{j=1}^n|\bbA_{j1}|, \sum_{j=1}^n|\bbA_{j2}|,\cdots,\sum_{j=1}^n|\bbA_{jn}| \Big).
  \end{equation*}
  From $\bbV_D$ we can construct a solution that satisfies the constraints in the function $A^{\text{NH}}_{-}$.  
  Note that $\bbV_D-\bbA/2$ consists of the sum of block matrices of the form: 
  {\scriptsize$\frac12\begin{bmatrix}|\bbA_{jk}|&-\bbA_{jk}\\-\bbA_{kj}& |\bbA_{jk}|\end{bmatrix}$}. 
  Using the fact that each block matrix is positive semidefinite, $\bbV_D-\bbA/2\ge0$. 
  Moreover, $a(\bbV_D-\bbA/2)=-\bbA/2$ and
$\operatorname{Tr}(\bbV_D-\bbA/2)=\operatorname{Tr}\bbV_D$, so this feasible choice gives the stated upper bound. Thus, the trace of $\bbV_D$ gives an upper bound as in the theorem. 
\end{proof}  

We now prove theorem~\ref{thm:main-rate} about the rate at which the NHCRB converges to the HCRB.
\begin{proof}[\textbf{Proof (Theorem~\ref{thm:main-rate})}]
Use the additive Holevo operators
\begin{equation}
 X_{\mathrm H,j,M}=\frac{1}{M}\sum_{a=1}^{M}X_{\mathrm H,j}^{(a)}.
\end{equation}
They are locally unbiased for $S^{\otimes M}$ and attain $\FH[\bm X_{\mathrm H,M}]=\CH/M$.  Theorem~\ref{thm:upperB}, together with
\begin{equation}
 i[X_{\mathrm H,j,M},X_{\mathrm H,k,M}]
 =\frac{1}{M^2}K_{jk}^{(M)},
 \qquad
 K_{jk}^{(M)}:=\sum_{a=1}^{M}K_{jk}^{(a)},
\end{equation}
gives
\begin{align}
 M\CNH(M)
 &\leq M\FNH[\bm X_{\mathrm H,M}]\\
 &\leq \Tr\!\left[\Ree Z[\bm X_{\mathrm H}]\right]
 +\sum_{j<k}\frac{1}{M}
 \norm{\sqrt{S^{\otimes M}}K_{jk}^{(M)}\sqrt{S^{\otimes M}}}_1.
\end{align}
Subtracting the Holevo value and using Eq.~\eqref{eq:holevo-mean},
\begin{equation}\label{eq:pair-decomp}
 M\CNH(M)-\CH
 \leq
 \sum_{j<k}
 \left[
 \frac{1}{M}
 \norm{\sqrt{S^{\otimes M}}K_{jk}^{(M)}\sqrt{S^{\otimes M}}}_1
 -|\mu_{jk}|
 \right].
\end{equation}

We now bound each term.  For any density operator $\rho$ and Hermitian $B$, write $B=B_+-B_-$ with $B_\pm\geq0$.  The triangle inequality gives
\begin{equation}\label{eq:sandwich-bound}
 \norm{\sqrt\rho B\sqrt\rho}_1
 \leq \tr(\rho B_+)+\tr(\rho B_-)
 =\tr(\rho|B|).
\end{equation}
Let $K_{jk}=\sum_a\kappa_a|a\rangle\langle a|$ and set
$p_a=\langle a|S|a\rangle$.  If $\xi_1,\ldots,\xi_M$ are independent random variables taking the value $\kappa_a$ with probability $p_a$, then
\begin{equation}
\begin{split}
     \tr\!\left\{S^{\otimes M}|K_{jk}^{(M)}|\right\}
 &=\mathbb E\left|\sum_{a=1}^{M}\xi_a\right|,
\\
 \mathbb E\xi_a&=\mu_{jk},
\\
 \operatorname{Var}(\xi_a)&=v_{jk}\;.
\end{split}
\end{equation}
Writing $\overline\xi_M=M^{-1}\sum_a\xi_a$, Eqs.~\eqref{eq:sandwich-bound} and the Cauchy--Schwarz inequality yield
\begin{align}
 \frac{1}{M}
 \norm{\sqrt{S^{\otimes M}}K_{jk}^{(M)}\sqrt{S^{\otimes M}}}&_1
 -|\mu_{jk}|\\
 &\leq \mathbb E|\overline\xi_M|-|\mu_{jk}|\\
 &\leq \mathbb E|\overline\xi_M-\mu_{jk}|\\
 &\leq \sqrt{\operatorname{Var}(\overline\xi_M)}
 =\sqrt{\frac{v_{jk}}{M}}.
\end{align}
Substitution into Eq.~\eqref{eq:pair-decomp} proves the upper bound.  The lower bound follows from $\CNH(M)\geq\CH(M)=\CH/M$.
\end{proof}

\section{Simultaneous estimation of phase and phase diffusion}
\subsection{Analytic solution to the HCRB and NCRB for simultaneously estimating phase and phase diffusion}
\label{apen:analyticHol}

We first note that the possible solutions to the HCRB and NCRB are greatly restricted by the unbiased conditions for this problem. This makes an explicit optimisation of the matrices $\mathbf{X}$ feasible for this problem. However, it is considerably simpler to use the results of Ref.~\cite{suzuki2016explicit} and Refs.~\cite{gill2005state,nagaoka2005generalization} to obtain analytic expressions for the HCRB and NCRB respectively. Ref.~\cite{suzuki2016explicit} provides a simple method for obtaining an analytic form for the HCRB, which only requires that we evaluate the SLD and right logarithmic derivative (RLD) operators. In what follows we set $\phi=0$ as the HCRB has no $\phi$ dependence. The SLD operators are given by
\begin{equation}
L_\phi=\mathrm{i}\text{e}^{-\delta^2}\text{sin}(\lambda)\begin{pmatrix}
0&-1\\
1&0
\end{pmatrix}\;,
\end{equation}
and
\begin{equation}
L_\delta=\begin{pmatrix}
L_{\delta,11}&L_{\delta,12}\\
L_{\delta,21}&L_{\delta,22}
\end{pmatrix}\;,
\end{equation}
where
\begin{equation}
\begin{split}
L_{\delta,11}&=\delta(1-\text{cos}(\lambda))(-1+\text{coth}(\delta^2))\\
L_{\delta,12}&=\mathcal{L}_{\delta,21}=-\delta\text{cosech}(\delta^2)\text{sin}(\lambda)\\
L_{\delta,22}&=\delta(1+\text{cos}(\lambda))(-1+\text{coth}(\delta^2))\;.
\end{split}
\end{equation}
The RLD operator for estimating $\phi$ is given by
\begin{equation}
\tilde{\mathcal{L}}_\phi=\begin{pmatrix}
\tilde{\mathcal{L}}_{\phi,11}&\tilde{\mathcal{L}}_{\phi,12}\\
\tilde{\mathcal{L}}_{\phi,21}&\tilde{\mathcal{L}}_{\phi,22}
\end{pmatrix}\;,
\end{equation}
where 
\begin{equation}
\begin{split}
\tilde{\mathcal{L}}_{\phi,11}&=-\frac{\mathrm{i}}{2}(-1+\text{coth}(\delta^2))\\
\tilde{\mathcal{L}}_{\phi,12}&=-\frac{\mathrm{i}}{2}\text{cosech}(\delta^2)\text{tan}(\lambda/2)\\
\tilde{\mathcal{L}}_{\phi,21}&=\frac{\mathrm{i}}{2}\text{cosech}(\delta^2)\text{cot}(\lambda/2)\\
\tilde{\mathcal{L}}_{\phi,22}&=\frac{\mathrm{i}}{2}(-1+\text{coth}(\delta^2))\;.
\end{split}
\end{equation}
The RLD operator for estimating $\delta$ is given by
\begin{equation}
\tilde{\mathcal{L}}_\delta=\begin{pmatrix}
\tilde{\mathcal{L}}_{\delta,11}&\tilde{\mathcal{L}}_{\delta,12}\\
\tilde{\mathcal{L}}_{\delta,21}&\tilde{\mathcal{L}}_{\delta,22}
\end{pmatrix}\;,
\end{equation}
where 
\begin{equation}
\begin{split}
\tilde{\mathcal{L}}_{\delta,11}&=\delta(-1+\text{coth}(\delta^2))\\
\tilde{\mathcal{L}}_{\delta,12}&-\delta\text{cosech}(\delta^2)\text{tan}(\lambda/2)\\
\tilde{\mathcal{L}}_{\delta,21}&-\delta\text{cosech}(\delta^2)\text{cot}(\lambda/2)\\
\tilde{\mathcal{L}}_{\delta,22}&=\delta(-1+\text{coth}(\delta^2))\;.
\end{split}
\end{equation}

Theorem 1 of Ref.~\cite{suzuki2016explicit} gives two analytic forms of the HCRB. One of these forms depends on the RLD Cram{\'{e}}r-Rao bound, given by~\cite{yuen1973}
\begin{equation}
\mathcal{C}_\text{R}=\frac{(\text{e}^{2\delta^2}-1)(1+4\delta^2+4\delta\abs{\text{cos}(\lambda)})}{4\delta^2\text{sin}(\lambda)^2}\;.
\end{equation}
To give the HCRB, we define the following function
\begin{equation}
\beta=1-\frac{(-1+\text{e}^{2\delta^2})\abs{\text{cos}(\lambda)}}{2\delta}\\
\end{equation}
Theorem 1 of Ref.~\cite{suzuki2016explicit} gives the HCRB as
\begin{equation}
\label{eqn:HCRBdiffforms}
\mathcal{C}_\text{H}=
\begin{cases}
\mathcal{C}_\text{R}& \text{ if }  \beta\leq0\\
\gamma & \text{ if }  \beta\geq0
\end{cases}
\end{equation}
where 
\begin{equation}
\gamma=\frac{-1+e^{4\delta^2}+8e^{2\delta^2}\delta^2+\text{cos}(2\lambda)(-1+e^{2\delta^2})^2}{8\delta^2\text{sin}(\lambda)^2}\;.
\end{equation}
There is a smooth transition between the different solutions to the HCRB in Eq.~\eqref{eqn:HCRBdiffforms} when $\beta$ changes sign. 

Using Refs.~\cite{gill2005state} and \cite{nagaoka2005generalization}, we can also solve the NCRB. For a qubit model the NCRB is given by
\begin{equation}
\mathcal{C}_\text{N}=(\text{Tr}[J_\text{S}^{-1/2}])^2\;,
\end{equation}
where $J_\text{S}$ is the SLD Fisher information matrix, which can be calculated from the SLD operators. This gives a NCRB of
\begin{equation}
\label{eq:NCRBanal}
\mathcal{C}_\text{N}=\frac{\text{e}^{2\delta^2}}{\text{sin}^2(\lambda)}\big(1+\frac{1-\text{e}^{-2\delta^2}}{4\delta^2}+\frac{\sqrt{1-\text{e}^{-2\delta^2}}}{\delta}\big)\;.
\end{equation}

\subsection{Proof of a strict inequality between the NCRB and HCRB}
\label{apen:strictineqHN}
We now wish to show that the two bounds derived in the previous section obey a strict inequality, $\mathcal{C}_\text{N}>\mathcal{C}_\text{H}$. Consider first when $\beta\leq0$, we can write
\begin{equation}
\delta\leq \frac{1}{2}(-1+\text{e}^{2\delta^2})\abs{\text{cos}(\lambda)}\;.
\end{equation}
The difference between the two bounds is given by
\begin{equation}
\mathcal{C}_\text{N}-\mathcal{C}_\text{H}=\frac{4(\text{e}^{2\delta^2}\sqrt{1-\text{e}^{-2\delta^2}}+\delta-(-1+\text{e}^{2\delta^2})\abs{\text{cos}(\lambda)})}{4\delta\text{sin}(\lambda)^2}\;.
\end{equation}
Note that $\mathcal{C}_\text{N}-\mathcal{C}_\text{H}>0$ is equivalent to 
\begin{equation}
\sqrt{1-\text{e}^{-2\delta^2}}+\text{e}^{-2\delta^2}\delta-(1-\text{e}^{-2\delta^2})\abs{\text{cos}(\lambda)})>0\;,
\end{equation}
where we have simply removed terms which are guaranteed to be positive for $\delta>0$. We use the substitution $x=\sqrt{1-\text{e}^{-2\delta^2}}$ and $a=\abs{\text{cos}(\lambda)}$. Then we can rewrite the above expression as
\begin{equation}
\begin{split}
&x+\text{e}^{-2\delta^2}\delta-ax^2\\
=&x(1-ax)+\text{e}^{-2\delta^2}\delta\;.
\end{split}
\end{equation}
We note that $f(x)=x(1-ax)$ is guaranteed to be positive for $0<x<1$ (which is satisfied for $\delta>0$) and $a=\abs{\text{cos}(\lambda)}$. As $\text{e}^{-2\delta^2}\delta$ is also guaranteed to be positive, we can conclude that in this case there is a non-zero gap between the NCRB and HCRB.

We now consider the second possibility when $\beta\geq0$.
\begin{equation}
\label{eq:cond2}
\begin{split}
2\delta&\geq (-1+\text{e}^{2\delta^2})\abs{\text{cos}(\lambda)}\\
&\geq(-1+\text{e}^{2\delta^2})\text{cos}(\lambda)\;.
\end{split}
\end{equation}
The difference between the two bounds in this case is given by
\begin{equation}
\begin{split}
\mathcal{C}_\text{N}-\mathcal{C}_\text{H}=\frac{-1}{8\delta^2\text{sin}(\lambda)^2}\bigg(&1+\text{e}^{4\delta^2}-2\text{e}^{2\delta^2}(1+4\sqrt{1-\text{e}^{-2\delta^2}}\delta)\\ 
&+(2\text{cos}(\lambda)^2-1)(-1+\text{e}^{2\delta^2})^2\bigg)\;.
\end{split}
\end{equation}
As above, we drop any terms with definite positive sign, and using Eq.~\eqref{eq:cond2}, we find that the following is a sufficient condition for $\mathcal{C}_\text{N}-\mathcal{C}_\text{H}>0$
\begin{equation}
8\delta(\text{e}^{2\delta^2}\sqrt{1-\text{e}^{-2\delta^2}}-\delta)>0\;.
\end{equation}
As before, we shall attempt to prove that this term is strictly positive. Using the substitution $x=2\delta^2$, we see that our aim is to show
\begin{equation}
\text{e}^{x}\sqrt{1-\text{e}^{-x}}>\sqrt{\frac{x}{2}}\;.
\end{equation}
As both sides of this inequality are positive (for $\delta>0$), we can square both sides, so that we wish to show
\begin{equation}
\label{eq:cond1grad1}
\text{e}^{2x}(1-\text{e}^{-x})-\frac{x}{2}>0\;,
\end{equation}
for $x>0$. This quantity evaluated at $x=0$ gives 0, hence our aim is simply to show that Eq.~\eqref{eq:cond1grad1} is monotonically increasing for $x>0$. The derivative of Eq.~\eqref{eq:cond1grad1} is \mbox{$-\frac{1}{2}-\text{e}^{x}+2\text{e}^{2x}$}, which is positive for $x>0$. Hence, we also have \mbox{$\mathcal{C}_\text{N}>\mathcal{C}_\text{H}$} when $\beta\geq0$.

\subsection{Optimal estimator matrices for HCRB and NHCRB}
\label{apensubsecOptXPb1}
In this section, we present explicit solutions for the matrices that optimise the HCRB and NHCRB for the simultaneous estimation of phase and phase diffusion. The density matrix is given by Eq.~\eqref{eq:exrho}. For simplicity and clarity, we focus on the specific parameters of interest used in the main text, $\lambda=\pi/2$ and $\phi=0$. Retaining the dependence on $\phi$ for now (until the derivatives are evaluated) gives:
\begin{equation}
S_{\phi,\delta}=\frac{1}{2}\begin{pmatrix}
1&\mathrm{e}^{-\mathrm{i}\phi-\delta^2}\\
\mathrm{e}^{\mathrm{i}\phi-\delta^2}&1
\end{pmatrix}\;.
\end{equation}
Taking the derivative with respect to the parameters of interest $\phi$ and $\delta$ gives:
\begin{equation}
\frac{\partial S_{\phi,\delta}}{\partial \phi}=\frac{\mathrm{i}}{2}\begin{pmatrix}
0&-\mathrm{e}^{-\mathrm{i}\phi-\delta^2}\\
\mathrm{e}^{\mathrm{i}\phi-\delta^2}&0
\end{pmatrix}\;,
\end{equation}
and
\begin{equation}
\label{equation:ddelta}
\frac{\partial S_{\phi,\delta}}{\partial \delta}=\begin{pmatrix}
0&-\delta \mathrm{e}^{-\mathrm{i}\phi-\delta^2}\\
-\delta \mathrm{e}^{\mathrm{i}\phi-\delta^2}&0
\end{pmatrix}\;.
\end{equation}
We now evaluate all of these terms at $\phi=0$ to find:
\begin{equation}
S_{\delta}=\frac{1}{2}\begin{pmatrix}
1&\mathrm{e}^{-\delta^2}\\
\mathrm{e}^{-\delta^2}&1
\end{pmatrix}\;,
\end{equation}
\begin{equation}
\frac{\partial S_{\delta}}{\partial \phi}=\frac{\mathrm{i}}{2}\begin{pmatrix}
0&-\mathrm{e}^{-\delta^2}\\
\mathrm{e}^{-\delta^2}&0
\end{pmatrix}\;,
\end{equation}
and
\begin{equation}
\frac{\partial S_{\delta}}{\partial \delta}=\begin{pmatrix}
0&-\delta \mathrm{e}^{-\delta^2}\\
-\delta \mathrm{e}^{-\delta^2}&0
\end{pmatrix}\;.
\end{equation}

Next, we write down a general form for the optimisers for the HCRB
\begin{equation}
X_\phi=\begin{pmatrix}
a_\phi&b_\phi+\mathrm{i}c_\phi\\
b_\phi-\mathrm{i}c_\phi&d_\phi
\end{pmatrix}\;,
\end{equation}
and
\begin{equation}
X_\delta=\begin{pmatrix}
a_\delta&b_\delta+\mathrm{i}c_\delta\\
b_\delta-\mathrm{i}c_\delta&d_\delta
\end{pmatrix}\;.
\end{equation}
where $a_i, b_i, c_i$ and $d_i$ are real. There are 6 unbiased conditions ($\tr{SX_i}=0$ and $\tr{S_iX_j}=\delta_{i,j}$). Using these conditions to eliminate the variables $a_i, b_i$ and $c_i$ we arrive at:
\begin{equation}
X_\phi=\begin{pmatrix}
-d_\phi&-\mathrm{i}\mathrm{e}^{\delta^2}\\
\mathrm{i}\mathrm{e}^{\delta^2}&d_\phi
\end{pmatrix}\;,
\end{equation}
and
\begin{equation}
X_\delta=\begin{pmatrix}
-d_\delta+\frac{1}{\delta}&-\frac{\mathrm{e}^{\delta^2}}{2\delta}\\
-\frac{\mathrm{e}^{\delta^2}}{2\delta}&d_\delta
\end{pmatrix}\;.
\end{equation}
Evaluating the first term in the HCRB gives:
\begin{equation}
\begin{split}
\label{eq:apen1sttermHCRB}
&\text{tr}\{S(X_\phi.X_\phi+X_\delta.X_\delta)\}\\
=&d_\phi^2+d_\delta^2-\frac{d_\delta}{\delta}+\frac{\mathrm{e}^{2\delta^2}(1+4\delta^2)}{4\delta^2}\;.
\end{split}
\end{equation}
The second term in the HCRB is:
\begin{equation}
\begin{split}
\label{eq:apen2ndtermHCRB}
&|\text{tr}\{\sqrt{S}(X_\phi.X_\delta-X_\delta.X_\phi)\sqrt{S}\}|\\
=&\left|-2d_\delta+\frac{1}{\delta}\right|\;.
\end{split}
\end{equation}
Therefore we can write
\begin{equation}
\mathcal{C}_\text{H}=\min_{d_\phi,d_\delta}d_\phi^2+d_\delta^2-\frac{d_\delta}{\delta}+\frac{\mathrm{e}^{2\delta^2}(1+4\delta^2)}{4\delta^2}+\left|2d_\delta-\frac{1}{\delta}\right|\;.
\end{equation}
We can separate out the term $d_\phi^2$, which is minimised when $d_\phi=0$. We then wish to differentiate the remaining expression with respect to $d_\delta$, and find the value of $d_\delta$ which makes the gradient vanish. However, the optimal value of $d_\delta$ depends on the sign of $\mathrm{i}(2d_\delta-\frac{1}{\delta})$. In general, this is quite a difficult optimisation, but observe that the $d_\delta$ value which minimises Eq.~\eqref{eq:apen1sttermHCRB} is
\begin{equation}
\label{eq:optddelta}
d_\delta=\frac{1}{2\delta}\;,
\end{equation}
which also minimises Eq.~\eqref{eq:apen2ndtermHCRB}. Therefore, this value of $d_\delta$ must be optimal. This gives a HCRB of
\begin{equation}
\mathcal{C}_\text{H}=\frac{-1+\mathrm{e}^{2\delta^2}(1+4\delta^2)}{4\delta^2}\;,
\end{equation}
in agreement with Eq.~\eqref{eqn:HCRBdiffformsmain} in the main text. The final optimal matrices are
\begin{equation}
\label{EqXphiopt}
X_\phi=\begin{pmatrix}
0&-\mathrm{i}\mathrm{e}^{\delta^2}\\
\mathrm{i}\mathrm{e}^{\delta^2}&0
\end{pmatrix}\;,
\end{equation}
and
\begin{equation}
\label{EqXdeltaopt}
X_\delta=\begin{pmatrix}
\frac{1}{2\delta}&-\frac{\mathrm{e}^{\delta^2}}{2\delta}\\
-\frac{\mathrm{e}^{\delta^2}}{2\delta}&\frac{1}{2\delta}
\end{pmatrix}\;.
\end{equation}

\subsection{Resolution of apparent conflict when $\delta\to0$}
\label{apen:conflict}
When taking the limit of Eq.~\eqref{eqn:HCRBdiffforms} as $\delta\to0$ we find
\begin{equation}
\lim_{\delta\to0}\mathcal{C}_\text{H}=\frac{1}{\text{sin}(\lambda)^2}(1+\frac{1}{2})\;.
\end{equation}
Taking the same limit for the NCRB, Eq.~\eqref{eq:NCRBanal}, we find
\begin{equation}
\lim_{\delta\to0}\mathcal{C}_\text{N}=\frac{1}{\text{sin}(\lambda)^2}(1+\frac{1}{\sqrt{2}})^2\;.
\end{equation}
We observe that as $\delta\to0$, $\mathcal{C}_\text{N}\to(1+2\sqrt{2}/3)\mathcal{C}_\text{H}$, contrary to what is expected, $\mathcal{C}_\text{N}\to\mathcal{C}_\text{H}$. Using the trade-off relation derived in Ref.~\cite{vidrighin2014joint}, we observe the same relationship numerically. However, as noted in the main text, this is not a true conflict, as the optimisation involved in the two bounds does not commute with taking the limit $\delta\to0$.

Using the parameterisation in the main text, the derivative of $S_{\phi,\delta}$ with respect to $\delta$, $S_\delta$, vanishes as $\delta\to0$. This gives rise to singular behaviour, however this is not a true singularity as it can be corrected by a reparameterisation of the model. With the parameterisation given in Eq.~\eqref{eq:exrho} and setting $\phi$ to 0, $S_\delta$ is given by
\begin{equation}
S_\delta=-2\text{e}^{-\delta^2}\delta\text{cos}(\lambda/2)\text{sin}(\lambda/2)\begin{pmatrix}
0&1\\
1&0
\end{pmatrix}\;,
\end{equation}
which vanishes as $\delta\to0$. However, if we use the parameterisation $\delta\to\sqrt{\tilde{\delta}}$, $S_{\tilde{\delta}}$ is given by
 \begin{equation}
S_{\tilde{\delta}}=-\text{e}^{-\tilde{\delta}}\text{cos}(\lambda/2)\text{sin}(\lambda/2)\begin{pmatrix}
0&1\\
1&0
\end{pmatrix}\;,
\end{equation}
which does not vanish as $\delta\to0$. Using this parameterisation, we can compute the HCRB and NCRB for estimating $\tilde{\delta}$ and $\phi$ as $\tilde{\delta}\to0$. Using the same techniques as in Appendix~\ref{apen:analyticHol} we find that at $\tilde{\delta}=0$, $\mathcal{C}_\text{H}=\mathcal{C}_\text{N}=1/(\text{sin}(\lambda)^2)$.

\vspace{2cm}
\bibliography{NH_ratio_bib}
\end{document}